\documentclass[10pt]{article}
\pdfoutput=1
\usepackage{jpub}
\usepackage[utf8]{inputenc}
\usepackage{amsmath,amssymb,amsbsy,amstext,amsthm,simplewick,amsfonts}
\usepackage{graphicx, placeins}
\usepackage{rotating}
\usepackage{subcaption}
\usepackage{lscape}
\usepackage{cancel}
\usepackage{wrapfig}
\usepackage{upgreek}
\usepackage{bm} 
\usepackage{framed}
\usepackage{bbm}
\usepackage{textcomp}
\usepackage{mathtools}
\usepackage{stackengine}
\setstackEOL{\\}
\usepackage{multirow}
\usepackage{pifont}

\usepackage{tikz}
\tikzset{every picture/.style={line width=0.75pt}} 
\usetikzlibrary{matrix,shapes,fit,tikzmark,calc}
\usepackage{tikz-cd, tikz-feynman} 
\tikzset{
	myblob/.style={
		shape = ellipse (2cm and 1cm)
	}
}

\usepackage{adjustbox}
\usepackage{makecell}
\usepackage{tcolorbox}
\usepackage{physics}
\usepackage{empheq}
\usepackage[normalem]{ulem}
\usepackage{enumitem}
\usepackage{braket} 
\usepackage{array}
\usepackage{dsfont}
\usepackage{ulem}
\usepackage{titlesec}
\titleformat{\section}{\normalfont\fontsize{12}{16}\bfseries}{\thesection}{1em}{}

\definecolor{blue3}{RGB}{31,119,180}
\definecolor{red3}{RGB}{214,39,40}
\definecolor{orange3}{RGB}{255,127,14}
\definecolor{green3}{RGB}{44,160,44}

\usepackage{hyperref}
\hypersetup{colorlinks=true,linkcolor=teal,citecolor=orange3,urlcolor=green3,pdfencoding=auto}

\begin{document}

\title{Causality and quasi-normal modes in the GREFT}

\author{Scott Melville
}

\affiliation{Queen Mary University of London, Mile End Road, London, E1 4NS, U.K.}

\date{today}
\abstract{
The General Relativity Effective Field Theory (GREFT) introduces higher-derivative interactions to parameterise the gravitational effects of massive degrees of freedom which are too heavy to be probed directly. The coefficients of these interactions have recently been constrained using causality: both from the analytic structure of 4-point graviton scattering and the time delay of gravitational waves on a black hole background. In this work, causality is used to constrain the quasi-normal mode spectrum of GREFT black holes. Demanding that quasi-normal mode perturbations decay faster in the GREFT than in General Relativity---a new kind of causality condition which stems from the analytic structure of 2-point functions on a black hole background---leads to further constraints on the GREFT coefficients. The causality constraints and compact expressions for the GREFT quasi-normal mode frequencies presented here will inform future parameterised gravitational waveforms, and the observational prospects for gravitational wave observatories are briefly discussed. 
}


\setcounter{tocdepth}{3}
\maketitle


\section{Introduction}

A quantum theory of gravity has remained elusive for two reasons. 
The first is phenomenological: experimentally we either probe the small-scale small-curvature regime (e.g. in particle colliders) or the large-scale large-curvature regime (e.g. solar system, cosmology), but to see quantum and gravitational effects simultaneously would require both small scales and  large curvatures.
The second reason is theoretical: unlike the fundamental forces of the Standard Model, General Relativity (GR) is not a renormalisable field theory, and this typically leads to a loss of predictivity at high energies. 
Reconciling these difficulties and uncovering a complete quantum mechanical description of gravity has been a central aim of theoretical physics for the past century.  \\

Fortunately, these two difficulties also suggest a way forward. 
Since gravitational phenomena are typically observed on large scales (in the low-energy, or `IR', regime), they are well described by an Effective Field Theory (EFT). 
An EFT description of gravity also resolves the theoretical issues surrounding renormalisation\footnote{
For instance the cosmological constant is not renormalised in the EFT: only when embedding this EFT into a UV completion is there any fine-tuning or ``naturalness'' issue.  
}, since an EFT is renormalisable at any finite order in its derivative expansion. 
The goal of this work is to better understand how the physical principle of causality can be used as a guide when constructing and applying gravitational EFTs.  \\

The interpretation of General Relativity as an Effective Field Theory goes back several decades \cite{Donoghue:1995cz, Burgess:2003jk} and is by now widely known. 
In this framework, the Einstein-Hilbert action of GR is extended by all possible higher-derivative interactions consistent with the symmetries of the problem: namely diffeomorphism invariance (plus any flavour/gauge symmetries of the matter sector). 
This produces an EFT extension of GR, also known as the ``EFT of gravity'' or the ``General Relativity Effective Field Theory'' (GREFT). This latter title best highlights the many parallels with the Standard Model Effective Field Theory (SMEFT), which is an analogous extension of the Standard Model by all possible interactions which are higher-order in derivatives and fields. \\

Just as the SMEFT was developed for model-agnostic searches for BSM physics at colliders, the GREFT can be viewed as a parameterised framework in which to search for new physics beyond General Relativity.
This is particularly important for gravitational wave (GW) astronomy. 
The number of black hole (BH) or neutron star binary mergers detected by gravitational wave observatories is now at least 90 \cite{KAGRA:2021vkt}, and is forecast to rise to thousands in the coming observing runs \cite{KAGRA:2013rdx}. 
These gravitational waves were created by compact objects in very high-curvature environments, and therefore open an exciting new window into the gravitational Universe \cite{Berti:2015itd,LIGOScientific:2016lio,LIGOScientific:2018dkp, LIGOScientific:2019fpa, Barausse:2020rsu, LIGOScientific:2020tif, LIGOScientific:2021sio}. 
The GREFT framework has been used to study the inspiral \cite{Endlich:2017tqa, Brandhuber:2019qpg, AccettulliHuber:2020dal}, merger \cite{Cayuso:2023xbc} and ringdown \cite{Cardoso:2018ptl, Cardoso:2019mqo, McManus:2019ulj, deRham:2020ejn, Cano:2020cao, Cano:2023jbk} phase of a binary merger, and compared with existing GW data in \cite{Sennett:2019bpc, Silva:2022srr}. 
With future observing runs and new GW observatories planned for the coming years, developing this framework both theoretically and phenomenologically will allow for the most precise tests of GR and its possible extensions.  \\

The GREFT action is made up of two components: a gravitational sector and a coupling to matter. 
In four spacetime dimensions, all parity-preserving interactions in the gravitational sector with up to eight derivatives can be written as \cite{Endlich:2017tqa},
 \begin{align}
  S_{\rm grav} = \int d^4 x \, \sqrt{-g} \, \frac{M_P^2}{2} \, \left[ R + \frac{b_1}{\Lambda^4} R^{(3)} + \frac{c_1}{\Lambda^6} \left( R^{(2)} \right)^2 + \frac{c_2}{\Lambda^6} \left( \tilde{R}^{(2)} \right)^2   \right]
  \label{eqn:intro_GREFT}
 \end{align}
where $R^{(n)}$ denotes the following contraction of $n$ Riemann tensors,
\begin{align}
 R^{(2)} &= R^{\mu\nu}_{\phantom{\mu\nu}  \alpha \beta} R^{\alpha \beta}_{\phantom{\mu\nu}  \mu \nu}   \; , 
& \tilde{R}^{(2)} &= \frac{1}{2} R^{\mu\nu}_{\phantom{\mu\nu}  \alpha \beta} \epsilon^{\alpha \beta}_{ \phantom{\alpha \beta} \rho \sigma} R^{\rho \sigma}_{\phantom{\mu\nu}  \mu \nu}  \; , 
 &R^{(3)} &= R^{\mu\nu}_{\phantom{\mu\nu}  \alpha \beta} R^{\alpha \beta}_{\phantom{\mu\nu}  \rho \sigma}  R^{\rho \sigma}_{\phantom{\mu\nu}  \mu \nu}  \; ,
\end{align}
where $\epsilon_{\mu\nu\alpha\beta}$ is the antisymmetric Levi-Civita tensor. 
The constant coefficients $\{b_1, c_1, c_2 \}$ encode the underlying UV physics, i.e. different high energy theories (string theory\footnote{
For instance, comparing the 4-point EFT amplitude to the superstring amplitude \cite{Kawai:1985xq} gives the following GREFT coefficients at this order in derivatives \cite{Bern:2021ppb}:
\begin{align}
\text{Superstring:} \qquad \qquad b_1 = 0 \;, \qquad c_1 = c_2 =  \frac{ - \psi^{(2)} (1) }{ 8 M^6 } \approx  +  \frac{ 0.3 }{ M^6 }
\end{align}
where $\psi^{(n)}$ is the polygamma function and $M^2 = 4/\alpha'$ is related to the mass of the lowest-lying state beyond the EFT. 
}, loop quantum gravity, etc.) will match onto different choices of these coefficients.  
The common energy scale $\Lambda$ is used to track the EFT's regime of validity, since \eqref{eqn:intro_GREFT} can be viewed as an expansion in powers of $\nabla_\mu / \Lambda$ (which is therefore expected to break down when length/time scales become order $1/\Lambda$).
The basis of interactions in \eqref{eqn:intro_GREFT} captures all physics involving only gravity\footnote{
For observables involving matter fields, the GREFT is only complete once certain non-minimal couplings are included. 
}: for instance graviton scattering amplitudes, as well as the physics of single black holes (e.g. their effective horizon, quasi-normal modes and BH-GW scattering). 
The grand ambition of the GREFT is to use gravitational observations to fix (or at least constrain) the coefficients $\{ b_1/\Lambda^4 , c_1 / \Lambda^6 , c_2 / \Lambda^6 , ...\}$, and then use this information to infer properties of the underlying high-energy quantum theory. \\

This work aims to answer two related questions:
\begin{itemize}

\item[(i)] Given the recent causality constraints that have been placed on the GREFT coefficients, what are the phenomenological consequences for GREFT black holes and in particular their quasi-normal mode (QNM) spectrum? 

\item[(ii)] Given a conjectured causality/stability property of black hole quasi-normal modes, what further constraints can be placed on the GREFT coefficients? 

\end{itemize}
The first is important for future analysis which fits GREFT coefficients to data, since it will inform more accurate GW templates for the GREFT and hence lead to stronger, more reliable constraints from experiment.
The second is also important for phenomenology---since limiting the parameter space with theoretical priors before fitting to data can lead to qualitatively different results---but mainly it complements a growing theoretical effort to characterise the space of consistent EFTs.
There have been many recent advances in this direction: including UV/IR sum rules \cite{Adams:2006sv, deRham:2022hpx}, the swampland conjectures \cite{Palti:2019pca, Vafa:2005ui}, and numerical bootstrap techniques \cite{Kruczenski:2022lot}. 
These different approaches all leverage some physical property of the underlying UV physics, usually causality and unitarity, to place constraints on the EFT (and hence on IR phenomenology).
This work will show that, at least in the context of the GREFT \eqref{eqn:intro_GREFT}, existing causality constraints lead to BH solutions being ``more stable'' (in the sense that quasi-normal mode perturbations decay faster) than in GR. This property will be referred to as ``QNM causality''. 
In fact, turning this around and \emph{requiring} QNM causality will lead to bounds on the GREFT coefficients which align remarkably well with existing constraints and in some cases are even stronger.  
This opens up the possibility of placing qualitatively new constraints on the parameter space of gravitational EFTs.  \\

\begin{figure}
\centering
\includegraphics[width = 0.6 \textwidth]{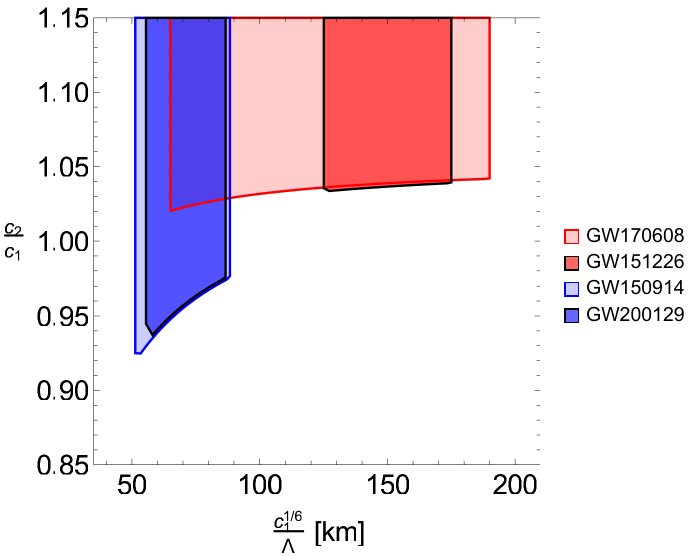}
\caption{Observational constraints on the GREFT parameter space from recent GW events (assuming $b_1$ is subdominant and that $c_2$ has little effect on these waveforms since it is spin-suppressed). The coloured regions are strongly disfavoured. Red indicates that only the inspiral phase was used \cite{Sennett:2019bpc}, and hence the EFT remains a valid description until a much higher $1/\Lambda$. Blue indicates that a full inspiral-merger-ringdown template was used \cite{Silva:2022srr}, and hence produce a stronger lower bound on $1/\Lambda$. The causality bounds derived in this work lead to the lower bounds on $c_2/c_1$ required for each event to resolvable within the EFT's regime of validity.}
\label{fig:obs}
\end{figure}

Concretely, this work will compare three different notions of causality and how they constrain the GREFT:
\begin{itemize}

\item[-] \emph{UV causality}. This is a property of 4-particle scattering amplitudes in the Minkowski vacuum, which in the low-energy EFT take the following form:
\begin{align}
 \mathcal{A}_{\rm EFT} (s, t) = \mathcal{A}_{\rm IR} (s,t) + \sum_{a,b} g_{ab} s^a t^b \; ,  
\end{align}
where $s$ and $t$ are the usual Mandelstam variables, $\mathcal{A}_{\rm IR}$ represents the pole and branch cut contributions from the light fields present in the EFT, and the coefficients $g_{ab}$ characterise the unknown UV physics. 
``UV causality'' is the assumption that the underlying amplitude in the full UV theory is analytic in the complex $s$-plane (up to the normal branch cut thresholds required by unitarity) \cite{Eden:1966dnq}, which in practice means that each $g_{ab}$ can be written explicitly as a contour integral of this underlying UV amplitude. Further assumptions about the UV, such as unitarity, locality and Lorentz invariance, place bounds on this UV integral and therefore the $g_{ab}$ coefficients. 
The simplest such bound is \cite{Adams:2006sv},  
\begin{align}
 g_{2n,0} > 0 
  \qquad \text{(UV causality)} \; , 
 \label{eqn:pos_intro}
\end{align}
where $n \geq 1$ in the absence of gravity and $n\geq 2$ in gravitational EFTs\footnote{
The difficulty in applying positivity bounds to $g_{20}$ in a gravitational theory is that the tree-level exchange of a graviton (responsible for the classical Newtonian potential) leads to the IR divergent pole term $s^2/t$ in the amplitude. 
There has been much recent progress in tackling this issue, for instance by assuming a particular Regge growth \cite{Tokuda:2020mlf, Alberte:2021dnj, deRham:2022gfe}, scattering at finite impact parameter \cite{Caron-Huot:2021rmr}, compatification \cite{Bellazzini:2019xts, Alberte:2020bdz,Alberte:2020jsk} or considering EFT observables which do not contain a graviton pole \cite{Chowdhury:2021ynh, Caron-Huot:2022ugt, Caron-Huot:2022jli}. 
}.
The importance of causality in these dispersion relation arguments was first highlighted in \cite{Adams:2006sv}, where they presented simple examples in which a violation of \eqref{eqn:pos_intro} goes hand-in-hand with superluminal propagation on certain non-trivial backgrounds.

\item[-] \emph{IR causality}.
The constraints from causality on non-trivial backgrounds has recently been explored further \cite{CarrilloGonzalez:2022fwg, CarrilloGonzalez:2023cbf}, and in general can lead to seemingly independent bounds on the EFT coefficients.
However, implementing causality in a gravitational context is subtle.
For waves passing by a compact object, the Shapiro time delay of GR is consistent with causality (i.e. it leads to a time delay rather than a time advance in the arrival time of the waves), and so it would seem that any perturbatively small EFT corrections on top of this would also be perfectly causal. 
One solution, developed in a recent series of papers \cite{deRham:2019ctd, deRham:2020zyh, Chen:2021bvg, deRham:2021bll, Chen:2023rar}, is the condition of ``IR causality'': namely that the small EFT corrections do not introduce a time advance relative to the GR time delay.
This means that if the total time delay is split into two parts,
\begin{align}
  t_{\rm GREFT} =  t_{\rm GR} + \delta t_{\rm EFT}
\end{align} 
where $t_{\rm GR}$ is the usual Shapiro time-delay from the Einstein-Hilbert term, then the shift induced by the EFT corrections must not lead to any time advance---at least, not one that is resolvable given the uncertainty principle. This leads to the condition\footnote{
Constraints from other notions of causality applied to the asymptotic time delay can be found in \cite{Camanho:2014apa} and \cite{Bellazzini:2021shn}.
}:
\begin{align}
- \omega \, \delta t_{\rm EFT}  \lesssim 1   \qquad \left( \text{IR causality} \right)
\label{eqn:intro_IR_causality}
\end{align}
where $\omega$ is the frequency of the wave experiencing the time advance.
To calculate $\delta t_{\rm EFT}$ for waves propagating on a black hole spacetime, a WKB approximation is often invoked in which the impact parameter is much larger than the size of the black hole (since if the classical trajectory stays sufficiently far from the horizon, then the fraction of the wave absorbed into the black hole can be neglected).  \\

\item[-] \emph{QNM causality}. 
The new causality condition put forward in this work is related to the time dependence of black hole quasi-normal modes, which may be parameterised in terms of an oscillatory frequency $\omega$ and a decay lifetime $\tau$:
\begin{align}
 \delta g_{\mu\nu} \sim e^{- \frac{t}{\tau} } e^{ i \omega t} \; ,
\end{align}
where the dependence on other data (angular momentum, overtone number, parity) is left implicit.
For the black hole to be stable, the decay rate $1/\tau$ must be positive. 
However, implementing this in the GREFT is subtle because the leading GR contribution from the Einstein-Hilbert term gives a positive contribution to $\tau$ and a small EFT correction would never flip this sign without become non-perturbatively large.
Instead, one should split the decay time into two parts,
\begin{align}
 \tau_{\rm GREFT} = \tau_{\rm GR} + \delta  \tau_{\rm EFT} \; , 
\end{align}
where $\delta \tau_{\rm EFT}$ is the contribution from EFT interactions which vanishes as $\Lambda \to \infty$. 
``QNM causality'' is then the condition that $\delta \tau_{\rm EFT}$ contributes positively to the decay rate---or at least, any decrease in the decay rate should not be resolvable given the uncertainty principle. This leads to the condition\footnote{
Notice that $\delta \tau_{\rm EFT} < 0$ corresponds to a positive contribution to the decay rate $1/(\tau_{\rm GR} + \delta \tau_{\rm EFT})$ and therefore $\delta \tau_{\rm EFT} > 0$ is the problematic sign. 
},
\begin{align}
 \omega \, \delta \tau_{\rm EFT}  \lesssim 1  \qquad \left( \text{QNM causality} \right) \; . 
 \label{eqn:intro_QNM_causality}
\end{align}
Unlike IR causality (large impact parameter scattering), this condition is naturally formulated near the horizon and so probes the theory in a qualitatively different region.   \\

\end{itemize}

Here are three rough arguments for the QNM causality condition. 
The first argument is that causality usually implies that any singularity in the response function is in the lower half of the complex plane. 
The quasi-normal mode frequencies in GR certainly lie in the lower half plane. 
Since the heavy physics, when decoupled from gravity, should push these points into the lower-half plane, then in the full gravitational theory they should be pushing the QNM frequencies even deeper into the complex plane. 
The second argument is that it is the straightforward analogue of the IR causality logic but applied to the stability of the BH solution. If it were possible to decouple the effects of the Einstein-Hilbert term (at the level of perturbations, but retaining the BH background), then the QNM causality condition would simply become the condition that the background is stable (just as the IR causality condition in the decoupling limit becomes the usual condition that there is no resolvable time advance). 
The third argument is that, at least in simple theories like \eqref{eqn:intro_GREFT}, it appears that causality constraints from scattering amplitudes (UV causality) and from considering the time delay of scattered waves (IR causality) both impose constraints on the EFT which limit or completely remove any positive $\delta \tau_{\rm EFT}$ from the QNM spectrum. 
Rather than a numerical coincidence, this seems further evidence that the condition \eqref{eqn:intro_QNM_causality} should be viewed as yet another avatar of causality in gravitational field theories. \\

Below is a short summary of the main results, namely the constraints on the GREFT coefficients from each of these three causality conditions. 
It is followed in section~\ref{sec:GREFT} by a short description of the GREFT and its simplest (stationary, spherically symmetric) black hole solutions. 
Then in section~\ref{sec:IR_causality}, the time delay of waves passing a GREFT black hole is computed and used to place IR causality bounds on the coefficients $\{ b_1, c_1, c_2 \}$. 
Section~\ref{sec:QNM_causality} calculates the quasi-normal mode spectrum of a GREFT black hole and compares the condition of QNM causality with IR and UV causality. 
In light of these causality bounds, the observational prospects for the GREFT is discussed briefly in section~\ref{sec:obs}, before some concluding remarks are given in section~\ref{sec:disc}.

\subsection{Summary of main results}
Assuming that the GREFT is under perturbative control up to a maximum $\left( \omega / r \right)_{\rm max} \equiv \epsilon_\omega \Lambda^2$, the IR causality constraint \eqref{eqn:intro_IR_causality} and QNM causality constraint \eqref{eqn:intro_QNM_causality} are found to imply the following lower bounds on the quartic GREFT coefficients:
\begin{align}
&\text{IR causality:} \qquad &c_1 \epsilon_{\omega}^3 &\gtrsim - 1.8 \times 10^{-4} \; ,   &c_2 \epsilon_{\omega}^3 &\gtrsim - 4.2 \times 10^{-5} \nonumber \\
&\text{QNM causality:} \qquad &c_1 \epsilon_{\omega}^3 &\gtrsim - 1.2 \times 10^{-4} \; ,   &c_2 \epsilon_{\omega}^3 &\gtrsim - 3.0 \times 10^{-5}
\label{eqn:intro_c_lower}
\end{align}
Remarkably, both QNM causality and IR causality impose nearly identical bounds. 
Choosing $\Lambda$ so that $\epsilon_{\omega} \sim \mathcal{O} (1)$ (i.e. so that it accurately reflects the true EFT cut-off), both conditions give $c_{1, 2} \gtrsim - \mathcal{O} \left( 10^{-4} \right)$. 
This is consistent with the sharp bounds from UV causality \cite{Bellazzini:2015cra},
\begin{align}
& \text{UV causality:}  \qquad &c_{1} &> 0 \;, \qquad &c_2 &> 0
 \end{align}
The small negative contribution in \eqref{eqn:intro_c_lower} is from the uncertainty in resolving wave-like behaviour.  

For this value of $\epsilon_\omega$, there is also an upper bound on each Wilson coefficient.
This bound depends on the size $r_s = 2 GM$ of the black hole. 
What matters phenomenologically is how large $1/\Lambda$ can be relative to $r_s$, and these contraints imply that,
\begin{align}
&\text{IR causality:} \qquad &\left(  \frac{1}{ G M \Lambda} \right)^6 &\lesssim \begin{cases}
\frac{ 7 \times 10^{-2}  }{c_1}   & \text{if } \frac{c_2}{c_1} \to 0   \\
\frac{1}{  13.581 c_1 - 12.896 c_2  }    &\text{if }  \mathcal{O} \left( 10^{-5} \right) \lesssim \frac{c_2}{c_1} \lesssim \mathcal{O} \left( 10^0 \right)  \;  \\
\text{no bound}   &\text{ if } \frac{c_2}{c_1} > 1.1  
 \end{cases}   \nonumber \\[12pt]
& \text{QNM causality:} \qquad &\left(  \frac{1}{ G M \Lambda} \right)^6 &\lesssim \begin{cases}
\frac{ 3 \times 10^{-3}  }{c_1}   & \text{if } \frac{c_2}{c_1} \to 0   \\
 \frac{ 1 }{ 2.862 c_1 - 2688 c_2 }   &\text{if } \mathcal{O} (10^{-4} ) \lesssim \frac{c_2}{c_1} \lesssim \mathcal{O} (10^{-3} )    \\
 \text{no bound}  &\text{if } \frac{c_2}{c_1} > 1.1 \times 10^{-3}
 \end{cases}
\end{align}  
This upper bound from IR causality when $c_2 = 0$ agrees with that derived in \cite{deRham:2021bll} up to an unimportant numerical factor. 
Both notions of causality lead to qualitatively similar upper bounds, with QNM causality the stronger at low $c_2$ and IR causality the stronger at large $c_2$. 
These bounds can be used to infer which of the binary mergers detected by the LIGO-Virgo network are within the EFTs regime of validity, and hence identify which regions of parameter space can be excluded by current gravitational wave data (see Figure \ref{fig:obs}). 
There is also an interesting similarity with recent UV causality bounds \cite{Caron-Huot:2022ugt}, 
\begin{align}
 \text{UV causality:}   \qquad  4 ( c_1 - c_2 ) \leq   12.3 \log \left( \frac{M}{m_{\rm IR}} \right)  - 13.5  
\end{align}
where $m_{\rm IR}$ is an IR cut-off responsible for regulating the graviton pole. Conceptually, this also implies that $c_2$ must be sufficiently greater than $c_1$ in order to avoid issues with causality.

Furthermore, there is a two-sided bound on the cubic coefficient,
\begin{align}
&\text{IR causality:} \qquad  &-2.8 \times 10^{-5} \lesssim  b_1  \epsilon_s^4 &\lesssim + 2.8 \times 10^{-5}   \nonumber \\
& \text{QNM causality:}  \qquad &  - 2.9 \times 10^{-5}  \lesssim  b_1 \epsilon_s^4 &\lesssim  +2.4 \times 10^{-5}
\label{eqn:b1_intro}
\end{align}
QNM and IR causality again impose similar constraints on the GREFT coefficient, namely $|b_1| \epsilon_s^4 \lesssim \mathcal{O} \left( 10^{-5} \right)$.
This is consistent with the causality bounds in \cite{Camanho:2014apa} from eikonal resummation of the 3-point function, and extends the one-sided bound from IR causality in \cite{deRham:2021bll} to a two-sided bound.
It also implies that the $b_1$ interaction can not be the leading phenomenological effect in gravitational wave templates, at least not if the EFT is to resolve $\epsilon_\omega \approx 1$. Sacrificing some resolving power, either by lowering $\epsilon_\omega$ or assuming a weaker perturbativity condition $\omega < \epsilon_\omega' \Lambda$, could render $|b_1| \epsilon_s^4 \lesssim \mathcal{O} ( 10^{-2} )$ and potentially relevant for phenomenology. In either case, a cubic interaction with comparable quartic interactions (which is what arises from generic UV completions) would describe a range of black hole backgrounds with no causality issues.

\begin{figure}
\centering
\includegraphics[width=0.8\textwidth]{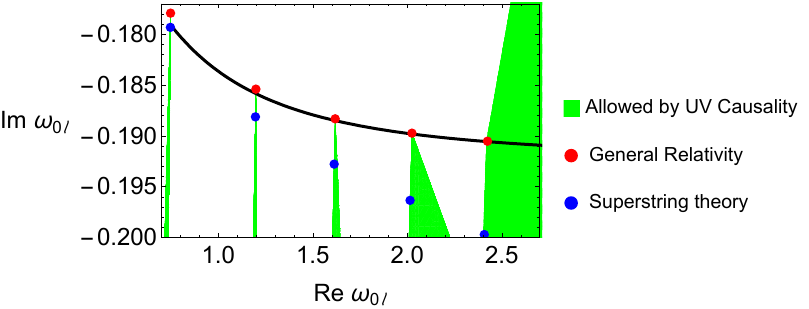}
\caption{
QNMs in the complex plane. Red points show the fundamental quasi-normal mode frequencies of GR in units of $r_s$, which from left to right correspond to $\ell = 2, 3, 4, 5, 6$. The black line is the large $\ell$ WKB approximation \eqref{eqn:QNM_GR}. The green region shows how each mode can be shifted by higher-curvature interactions, given the simplest positivity constraints $c_1 > 0$ and $c_2 > 0$ from UV causality. Once $\ell \geq 6$, it becomes possible to shift the frequency towards the real axis and decrease the decay rate of black holes. The additional bounds from IR causality prevent this shift from being resolvable within the EFT's regime of validity. As a concrete example, the blue points show the frequencies in superstring theory for a black hole with $\alpha'^3/(GM)^6 \approx 0.1$ (i.e. $c_1 \epsilon_s^6 = c_2 \epsilon_s^6 \approx 5 \times 10^{-4}$).   
}
\label{fig:QNM_pos}
\end{figure}

Finally, the UV causality bound \cite{Bern:2021ppb, Caron-Huot:2022ugt},
\begin{align}
\text{UV causality:} \qquad | b_1 |^2 <  \frac{2}{9} c_1
\end{align}
also implies that cubic interactions in the EFT must come with quartic interactions. 
While qualitatively similar, this bound is different from \eqref{eqn:b1_intro} in two key respects. 
Firstly, since it follows from scattering in the Minkowski vacuum it has no dependence on $r_s$: it is therefore a constraint on the theory space, whereas \eqref{eqn:b1_intro} could be viewed instead as a constraint on the space of backgrounds which this EFT can reliably describe. 
Secondly, it is quadratic in $b_1$ relative to $c_1$, whereas the inclusion of quartic interactions in \eqref{eqn:b1_intro} would produce a linear bound on the combination $b_1 + c_1/(G M \Lambda)^2$.
So in some respects, \eqref{eqn:b1_intro} is closer to the upper bound on $| b_1|^2$ alone which was obtained in \cite{Caron-Huot:2022ugt} by including an IR cut-off $m_{\rm IR}$ or in \cite{Haring:2023zwu} by imposing unitarity for a finite $n_{\rm max}$ of heavy states.

The main conclusion is that this new ``QNM causality'' condition---that EFT corrections do not lead to resolvable increase in the decay time of black hole perturbations---leads to constraints on the GREFT coefficients which closely parallel existing causality constraints. 
This imposes causality in a qualitatively distinct regime, near to the horizon of a black hole, and in future it will be interesting to explore its consequences for other EFT interactions (e.g. including matter fields) and black hole backgrounds (e.g. spinning black holes).    

Finally, Table~\ref{tab:QNM_delta} records the first nine fundamental QNM frequencies of the simplest GREFT black hole (the first three of which agree with the previous results of \cite{Cardoso:2018ptl, deRham:2020ejn}).  
Equations (\ref{eqn:first_QNM}--\ref{eqn:last_QNM}) provide analytic approximations for these frequencies which become exact in the limit of large angular momentum $\ell$. 
Independently of any causality constraint, this QNM data for any $\ell$ and overtone number $n$ will be useful for constructing ringdown templates in future gravitational wave analyses.

\section{The General Relativity Effective Field Theory}
\label{sec:GREFT}

This section briefly reviews the GREFT action \eqref{eqn:intro_GREFT} and properties of its black hole solutions.

\paragraph{Power counting.}
Any low-energy EFT is defined by three considerations: (i) the light degrees of freedom it should describe, (ii) any symmetries or other constraints which limit the allowed interactions, and (iii) a power counting scheme which can be used to estimate the size of each interaction. 
In the case of gravity, (i) is the spacetime metric $g_{\mu\nu}$ (plus any matter fields) and (ii) is diffeomorphism invariance  (plus any matter symmetries). 
Choosing which power counting scheme to use is less straightforward, however a scheme must be chosen in order to reliably determine which contributions to keep in any given observable (and to ensure radiative stability against quantum corrections). 
In fact, \cite{Serra:2022pzl} recently stressed the importance of power counting in gravitational EFTs in the context of causality bounds.

This work will adopt the following power counting:
\begin{align}
 S_{\rm GREFT} [ g_{\mu\nu} , \phi ] = \int d^4 x \, \sqrt{-g} \left\{ 
 M_P^2 \Lambda^2 \mathcal{L}_{\rm grav} \left[  \frac{R}{\Lambda^2} , \frac{\nabla}{\Lambda}  \right] +  \Lambda^4 \mathcal{L}_{\rm matt} \left[ \frac{\phi}{\Lambda} , \frac{R}{\Lambda^2} , \frac{\nabla}{\Lambda}   \right]
 \right\}
 \label{eqn:GREFT_power_counting}
\end{align}
where $R$ is the Riemann tensor of $g_{\mu\nu}$, $\phi$ denotes all of the dynamical matter fields and $\Lambda \ll M_P$ is related to the mass of the heavy fields which have been integrated out. 
This is clearly radiatively stable, and has the feature that matter loops give a small correction to the gravitational sector (so in that sense it is weakly coupled). 

The gravitational sector contains the Einstein-Hilbert term, plus corrections built from increasing numbers of Riemann tensors and their derivatives\footnote{
The cosmological constant has been set to zero so that the background spacetime is asymptotically flat. 
Notice that while the natural value $\sim M_P^2 \Lambda^2$ from \eqref{eqn:GREFT_power_counting} is typically much larger than the observed dark energy density in our Universe (the infamous cosmological constant problem), since it is not renormalised in this low-energy EFT there is no obstacle to simply fixing its value to be zero (or to its observed value, providing the black holes are of a sufficiently small size and separation that the resulting cosmological expansion can be neglected). See e.g. \cite{Goon:2016ihr} for a modern account of this non-renormalization.
}, 
\begin{align}
\mathcal{L}_{\rm grav} = \frac{R}{2 \Lambda^2} 
+ \sum_{n=2} \frac{ \mathcal{L}_n [R] }{ \Lambda^{2n} } 
\end{align}
A complete minimal basis for the $\mathcal{L}_n$ has been constructed in \cite{Ruhdorfer:2019qmk, Li:2023wdz} using Hilbert series methods.
The goal is ultimately to construct a set of interactions which:
\begin{itemize}

\item[(i)] respects the symmetries of GR, i.e.  diffeomorphism invariance, so are built only from the Riemann tensor and its covariant contractions with $g_{\mu\nu}$. 

\item[(ii)] affects local observables\footnote{
While local observables may not exist in a fully quantum theory of gravity, in the low-energy EFT regime considered here QFT is a good description and local operators certainly exist.
}, i.e. does not contain total derivatives,

\item[(iii)] does not contain any redundancies, i.e. a minimal set of operators which cannot be reduced to anything simpler by field redefinitions.  

\end{itemize}
\noindent These conditions lead to \eqref{eqn:intro_GREFT} as the most general parity-preserving theory with up to eight derivatives. 
Notice that, unlike the SMEFT, the operators have been grouped according to the number of derivatives they contain rather than their total mass dimension. 
For example, $R^{(3)}$ contains 6 derivatives but leads to interactions on a Ricci-flat background which start at mass dimension 9  (of the form $(\partial^2 h)^3$ for metric perturbations $h_{\mu\nu}$).


\paragraph{Expansion parameter.}
The size of the EFT corrections to any observable can be estimated by simple power counting arguments.
In GR, the Einstein-Hilbert action $\sqrt{-g} R \sim h \partial^2 h$ provides the kinetic term for metric perturbations $h$ and the coupling to matter sources $h\sim GM/r$ around any mass $M$. 
Since the first-order correction from the cubic interaction must contain a factor of $b_1/\Lambda^4$, the only dimensionless combination which can appear is the ratio $b_1 / ( G M \Lambda )^4$. The quartic interactions will similarly produce corrections proportional to $c_1 / (G M \Lambda)^6$ and $c_2/(G M \Lambda)^6$. 
The expansion parameter,
\begin{align}
 \epsilon_s \equiv \frac{1}{G M \Lambda } \equiv \frac{ 8 \pi M_P^2}{ M \Lambda}
\end{align}  
will therefore control the EFT corrections around an object of mass $M$, and the expectation is that each observable $\mathcal{O}$ in the GREFT will admit an expansion of the form,
\begin{align}
 \mathcal{O}_{\rm GREFT} = \mathcal{O}_{\rm GR} \left( 1 + \epsilon_s^4 \delta_3 \mathcal{O} + \epsilon_s^6 \delta_4 \mathcal{O}  + ... \right)
 \label{eqn:eps_expansion}
\end{align}
where $\delta_3 \mathcal{O}$ and $\delta_4 \mathcal{O}$ are the relative contributions from the cubic and quartic interactions to $\mathcal{O}$.

\paragraph{Black hole background.}
To see this in action, consider the most general spherically symmetric solution for the background metric, which takes the form,
\begin{align}
 \bar{g}_{\mu\nu} d x^\mu dx^\nu  = - f_t (r) d t^2 + \frac{1}{f_r (r )} d r^2 + f_{\Omega} (r) r^2 d \Omega^2 \; ,
 \label{eqn:background_metric}
\end{align}
where $d \Omega^2 = d \theta^2 + \sin^2 \theta d \phi^2$ is the angular element in 3 spatial dimensions. 
Coordinates can be chosen so that $f_{\Omega} = 1$. The remaining functions $f_r$ and $f_t$ are most easily found by substituting \eqref{eqn:background_metric} into the GREFT action and then solving the equations of motion,
\begin{align}
\frac{\delta S_{\rm GREFT} [ \bar{g}_{\mu\nu} ]}{ \delta g_{\mu\nu} (x) }  = 0 \;\qquad \Rightarrow \qquad  \frac{\delta}{\delta f_t}  \left( S_{\rm GREFT} [ \bar{g}_{\mu\nu} ] \right) = \frac{\delta}{\delta f_t}   \left( S_{\rm GREFT} [ \bar{g}_{\mu\nu} ] \right)  = 0
\end{align}
perturbatively in $1/\Lambda$. 
This gives \cite{Cardoso:2018ptl, deRham:2020ejn}, 
\begin{align}
 f_t &= 1 - \frac{r_s}{r} + b_1 \epsilon_s^4 \left( + \frac{5 r_s^7}{r^7} \right)  + c_1 \epsilon_s^6 \left(  + \frac{2 r_s^9}{r^9} - \frac{11 r_s^{10}}{8 r^{10}} \right) + ...  \; , \nonumber \\ 
 f_r &= 1 - \frac{r_s}{r} + b_1 \epsilon_s^4 \left( + \frac{ 54 r_s^6 }{r^6} - \frac{49 r_s^7}{r^7} \right) + c_1 \epsilon_s^6 \left(  + \frac{9 r_s^9}{r^9} - \frac{67 r_s^{10} }{8 r^{10}} \right)  + ...  \; ,
\end{align}
where $r_s = 2 GM$ is the usual Schwarschild radius.
In GR, $M$ is the mass of the black hole and $r_s$ defines the horizon of this black hole solution. 
In the GREFT, the $\mathcal{O} ( \epsilon_s )$ corrections to the background metric result in the horizon moving to $r = \tilde{r}_s$, where
\begin{align}
 \tilde{r}_s  = r_s \left(  1 - 5 b_1 \epsilon_s^4 - \frac{5}{8}  c_1 \epsilon_s^6  + ...    \right) \; ,
 \label{eqn:rH}
\end{align}
is the radial coordinate at which $f_t$ and $f_r$ both vanish. 
The $+...$ indicate higher-order corrections in both the couplings (e.g. $\mathcal{O} (b_1^2 )$) and in the parameter $\epsilon_s$. 
Note that the $c_2$ interaction has no affect on the background solution at this order.

\paragraph{Perturbations.}
Small perturbations about this background solution, $g_{\mu\nu} = \bar{g}_{\mu\nu} + h_{\mu\nu}$,  
are described by the linearised equation of motion,
\begin{align}
 \int d^4 x' \; \frac{\delta^2 S_{\rm GREFT} [ \bar{g}_{\mu\nu} ]}{  \delta g_{\mu\nu} (x) \delta g_{\alpha\beta} (x') } \; h_{\alpha \beta} (x')   = 0  \; . 
 \label{eqn:h_eom_prop}
\end{align}
In GR, this equation can be brought into the form
\begin{align}
\left(  \frac{\partial^2}{\partial r _*^2 } + \omega^2 - f (r) V_{\ell}^{\pm} (r)  \right) \Psi^{\pm}_{\ell} ( \omega , r ) = 0  
\label{eqn:GR_master_eom}
\end{align}
where $\Psi^{+}_{\ell}$ and $\Psi^-_{\ell}$ are the dynamical parity-even and parity-odd perturbations inside $h_{\mu\nu}$ (the other components of which are then fixed by gauge conditions), $f (r) = 1 - r_s/r$ and the tortoise coordinate $r_* = r + r_s \log \left( \frac{r}{r_s} - 1 \right)$ obeys $\partial_{r_*} = f \partial_r$. 

\begin{figure}
\centering
\includegraphics[width=0.8\textwidth]{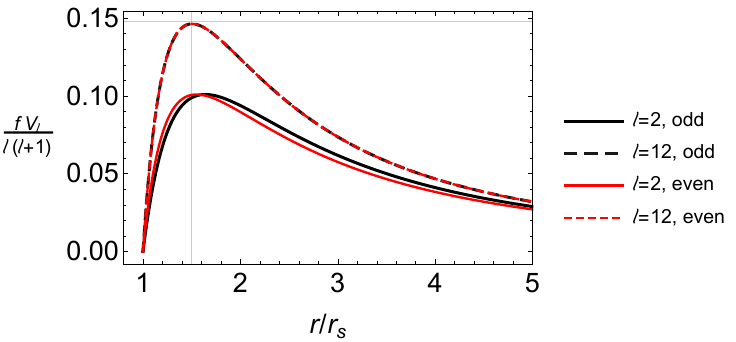}
\caption{
The effective potentials in GR, where gray lines show a maximum value of $4/27$ at $r = 3r_s/2$. 
\label{fig:GR_V}
}
\end{figure}

The master equations \eqref{eqn:GR_master_eom}  are known as the Regge-Wheeler \cite{Regge:1957td} and Zerilli \cite{Zerilli:1970se} equations. 
They are wave equations which describe the propagation of small perturbations with energy $\omega$ and angular momentum $\ell$. 
The GR potentials are,
\begin{align}
 V^-_{\ell} (r)  &= \frac{ \ell (\ell + 1 ) }{r^2} - \frac{3 r_s}{r^3} \; ,     \nonumber \\ 
 V^+_{\ell} (r) &=  \frac{Z}{r^3} + \frac{2 (r-3 r_s)}{r^3} +\frac{ r \left( 18 r_s -12 r \right)}{ Z r^3} + \frac{18 r_s^2 (r-r_s)}{Z^2 r^3}
 \label{eqn:GR_V}
\end{align}
where $Z =  ( \ell (\ell + 1) - 2) r + 3 r_s$.
The effective potential $f V_{\ell}^{\pm}$ appearing in the master equation is shown in Figure~\ref{fig:GR_V} and acts as a potential barrier between between spatial infinity and the region close to the black hole horizon.
At large $\ell$, both effective potentials coincide and reach a maximum value of $4 \ell (\ell + 1)/27$ at $r \approx 3/2$ (the height and location of the barrier).

In the GREFT, \eqref{eqn:h_eom_prop} can be brought into a similar form by carefully identifying the dynamical $\Psi^{\pm}_{\ell} (\omega, r)$ combinations of the $h_{\mu\nu}$ components (which will generally differ from the GR Regge-Wheeler and Zerilli variables by $\mathcal{O} ( \epsilon_s)$ corrections). 
It is convenient to introduce a tortoise coordinate adapted to the GREFT metric \eqref{eqn:background_metric}, namely an $r_* (r)$ defined by,
\begin{align}
 \frac{d r_*}{d r} = \frac{1}{ \sqrt{ f_t f_r } }  \; . 
\end{align}
This ``metric'' tortoise co-ordinate is useful because the master equation can be written as\footnote{
While the higher-derivative interactions would naively produce higher-derivative terms in the equation of motion, these can always be traded for terms which are higher-order in the EFT expansion \cite{Endlich:2017tqa, Cardoso:2018ptl, deRham:2019ctd, deRham:2020ejn}. From this point of view, there is no issue with unitarity \cite{Knorr:2023usb} or constructing black hole solutions \cite{Bueno:2023jtc} in such EFT extensions of GR .
}, 
\begin{align}
\left[  \frac{\partial^2}{\partial r _*^2 } + \frac{\omega^2}{c_s^2 (r)} - \sqrt{ f_t (r) f_r (r) } \left(  V^{\pm}_{\ell} (r) + \delta V_{\ell}^{\pm} ( \omega, r)  \right)  \right] \Psi^{\pm}_{\ell} ( \omega , r ) = 0  
\label{eqn:master_eom}
\end{align}
where $c_s^2 (r)$ is the physical sound speed of the metric fluctuations relative to matter\footnote{
Since the corrections to the potential will generically depend on $\omega$, the split into $c_s^2 (r)$ and $\delta V_{\ell}^{\pm} (r, \omega)$ is ambiguous: the condition that the potential contains no $\omega^2$ term resolves this ambiguity. 
}. 
This master equation was recently calculated in \cite{Cardoso:2018ptl} for the quartic interactions and in \cite{deRham:2020ejn} for the cubic interactions, where the sound speed was found to be:
\begin{align}
 c_s^2 (r) =  1   -   f \left(  
+ \frac{ 9 r_s^5 }{ r^5 } b_1 \epsilon_s^4 
  + \frac{63 r_s^8 }{ r^8} c_1 \epsilon_s^6  
 \right)  + ... 
 \label{eqn:cs}
\end{align}
where $f = 1 - r_s/r$. 
As shown in \cite{deRham:2020ejn}, the location of the horizon and the sound speed there is the same for all perturbations\footnote{
Co-ordinates could therefore be chosen so that $c_s ( \tilde{r}_s ) = 1$ and $\tilde{r}_s = 1$. While this simplifies expressions within a given theory, it makes comparing different theories difficult since generally $\tilde{r}_s$ is a function of other model parameters beyond just the black hole mass (e.g. the EFT coefficients). 
}. 
The potentials $\delta V_{\ell}^{\pm} (\omega, r)$ are somewhat lengthy and are listed in Appendix~\ref{app:master}. 

Note that the extension of the tortoise co-ordinate beyond GR is not unique. There is also the natural choice,
\begin{align}
\tilde{r}_* = r + \tilde{r}_s \text{log} \left( \frac{r}{\tilde{r}_s} - 1 \right)
\end{align}
which is the usual GR co-ordinate adapted to the new horizon. 
This ``horizon'' tortoise co-ordinate is useful because the master equation parallels that of GR with the simple replacement $r_s \to \tilde{r}_s$.
Explicitly,
\begin{align}
\left[  \frac{\partial^2}{\partial \tilde{r} _*^2 } + \frac{\omega^2}{ \tilde{c}_s^2 } - \tilde{f} (r) \left(  \tilde{V}^{\pm}_{\ell} (r) +  \delta \tilde{V}_{\ell}^{\pm} ( \omega, r)   \right)  \right] \tilde{\Psi}^{\pm}_{\ell} ( \omega , r ) = 0  
\label{eqn:master_eom_H}
\end{align}
where $\tilde{f} (r) = 1 - \tilde{r}_s/r$ and the constant factor of $\tilde{c}_s  = \lim_{r \to \tilde{r}_s} \sqrt{f_t f_r}/f$ is included for later convenience\footnote{
In short, while $\omega^2 \left( \frac{1}{ \tilde{c}_s^2}  -1 \right)$ could be absorbed into the potential, keeping it separate will ensure that the $\omega^2$ term in the potential is regular at the horizon.
}.
A tilde will be used to denote that the EFT corrections to the horizon have been included in an object: for instance
\begin{align}
 \tilde{V}_{\ell}^- (r) = \frac{j}{r^2} - \frac{3 \tilde{r}_s }{r^3}  = V_{\ell}^- (r)  + \frac{3 r_s}{r^3} \left(   5 b_1 \epsilon_s^4 + \frac{5}{8} c_1 \epsilon_s^6 + ...   \right) 
\end{align}
and similarly $\tilde{V}_{\ell}^+$ is defined by \eqref{eqn:GR_V} with $r_s \to \tilde{r}_s$. 
The master equations \eqref{eqn:master_eom_H} and \eqref{eqn:master_eom} are equivalent descriptions of the same physics although some calculations will be easier in one of the two coordinates.  
The explicit relation between $\tilde{\Psi}_{\ell}^\pm$ and $\Psi_{\ell}^{\pm}$ is,
\begin{align}
\tilde{\Psi}^{\pm}_{\ell}  = \Psi_{\ell}^{\pm} \left( \frac{\sqrt{f_t f_r}}{f} \right)^{1/2} \equiv \Psi_{\ell}^{\pm} \left( 1 + \epsilon_s^4 \delta_3 \mathcal{N} + \epsilon_s^6 \delta_4 \mathcal{N} \right)^{1/2} \; ,
\label{eqn:N_def}
\end{align}  
and the relation between their potentials is given in the Appendix~\ref{app:master}. 
Since the ``metric'' and ``horizon'' tortoise co-ordinates differ only by $\mathcal{O} ( \epsilon_s)$ corrections, both \eqref{eqn:master_eom} and \eqref{eqn:master_eom_H} correctly reduce to the GR equations when $\epsilon_s \to 0$.  

\eqref{eqn:master_eom} (or \eqref{eqn:master_eom_H}) describes the classical evolution of small perturbations about a GREFT black hole. The remainder of this work will focus on analysing this equation in different settings. 
Notice that, from a QFT perspective, the object $\frac{\delta^2 S_{\rm GREFT} [ \bar{g}_{\mu\nu} ]}{ \delta g_{\alpha\beta} (x') \delta g_{\mu\nu} (x) }$ appearing in \eqref{eqn:h_eom_prop} is the inverse propagator for the metric on this black hole background \eqref{eqn:background_metric}.
Since the consequences of causality are relatively well understood for the propagator (e.g. the $+i \epsilon$ prescription and resulting analytic structure in the complex $\omega$ plane), one should expect causality to place analogous constraints on the classical problem \eqref{eqn:master_eom}. 
Section~\ref{sec:IR_causality} below recounts the classical connection between causality and no resolvable time advance, and then section~\ref{sec:QNM_causality} will return to this connection with analyticity of the propagator.

\paragraph{Regime of validity.}
As can be seen from the expression \eqref{eqn:rH} for $\tilde{r}_s$, the condition $\epsilon_s \ll 1$ is needed for the EFT to reliably capture the black hole horizon. This is simply the requirement that $r_s \gg \frac{1}{\Lambda}$ so that at any $r \gtrsim r_s$ the derivative expansion in spatial derivatives (powers of $\partial_i /\Lambda$) is under control. 
But there is also a derivative expansion in time derivatives (power of $\partial_t/\Lambda$), which means that the EFT will break down if $\omega$ is too large. The naive condition $\omega \ll \Lambda$ implies that the EFT is valid in the range\footnote{
The factors of 2 in \eqref{eqn:w_max_weak} and \eqref{eqn:w_max_strong} are an unfortunate consequence of defining $\epsilon_s = 1/(G M \Lambda)$ rather than $1/(r_s \Lambda)$, but are retained to facilitate comparison with \cite{deRham:2021bll} in which the maximum considered energy saturates \eqref{eqn:w_max_strong}. 
},
\begin{align}
\epsilon_s  \ll  \frac{2}{  \omega r_s } \qquad \qquad \text{(weak condition)} \; . 
\label{eqn:w_max_weak}
\end{align} 
However, it was recently argued in \cite{deRham:2020zyh, Chen:2021bvg} that this EFT enjoys a much larger regime of validity if one focusses on linear perturbations in the purely gravitational sector. In that case, any $\nabla_\mu = ( \partial_t , \partial_i )$ will either act on the background and produce $\left( 0 , \frac{1}{r} \hat{\mathbf{r}} \right)$, or it will act on the fluctuations and produce the null momentum $\left( \omega ,  \omega \hat{\mathbf{p}} \right)$. 
It is the Lorentz-invariant contraction of these two 4-vectors which must be less than $\Lambda^2$ for validity, which implies the EFT is valid in the range\footnote{
Focussing on processes for which $\hat{\mathbf{p}} \cdot \hat{\mathbf{r}}$ is suppressed can lead to an even wider range of validity. 
The maximum $\epsilon_s$ for which the EFT captures GWs from a quasi-circular orbit is given in \eqref{eqn:pow_count_pheno}.
},
\begin{align}
\epsilon_s^2  \ll  \frac{2 r}{r_s} \times \frac{2}{ \omega r_s }  \qquad \qquad \text{(strong condition)}  \; . 
\label{eqn:w_max_strong}
\end{align} 
Since $ \frac{1}{\sqrt{ \omega r_s }}  \gg \frac{1}{\omega r_s}$, this condition for validity is ``stronger'' in the sense that all observables which are under perturbative control in the range \eqref{eqn:w_max_strong} are automatically under control in the range \eqref{eqn:w_max_weak}, but not vice versa (for instance scattering involving multiple fluctuations is typically only perturbative in the range \eqref{eqn:w_max_weak} but not \eqref{eqn:w_max_strong}).

\paragraph{Redundant potentials.}
Finally, note that the potentials themselves are not physical observables.
For instance, take one of the redundant cubic operators,
\begin{align}
 \mathcal{L}_3  \supset \frac{ b_3}{16}  R R^{(2)}  \; . 
 \label{eqn:L_redundant}
\end{align}
This affects the master equation in two ways. 
Firstly, it shifts the background by \cite{deRham:2020ejn}:
\begin{align}
 \delta f_t &= \frac{ b_3 \epsilon_s^4 }{2} \left( - \frac{6}{x^6} + \frac{9}{x^7}   \right)  \; , 
 & \delta f_r &= \frac{ b_3 \epsilon_s^4 }{2} \left( + \frac{36}{x^6} - \frac{33}{x^7}   \right)  \; , 
  & \delta \tilde{r}_s &= - \frac{3}{2}  b_3 \epsilon_s^4 \, r_s \; . 
  \label{eqn:df_redundant}
\end{align}
It also shifts the potentials by,
\begin{align}
 \delta V^-_{\ell}  &\supset \frac{ b_3 \epsilon_s^4 }{r_s^2} \left[  \frac{ 36 r_s^9 }{  r^9 } - \ell (\ell + 1) \frac{21 r_s^8}{2 r^8}  \right]  \; , \label{eqn:V_redundant} \\ 
   \delta V^+_{\ell}  &\supset \frac{ b_3 \epsilon_s^4 }{ r_s^2 } \left[     -\frac{21 Z r_s^8}{2 r^9} 
   + \frac{r_s^8 \left( 135 r_s-42 r \right) }{2 r^9}
   +\frac{9 r_s^9 (16 r - 27 r_s )}{r^9 Z} 
   +\frac{r_s^{10} \left( 351 r_s -297 r\right) }{r^9 Z^2}
   +\frac{162 r_s^{11} (r-r_s)}{r^9 Z^3} \right] \; . 
   \nonumber 
\end{align}
However, this redundant interaction can be converted into a universal tidal interaction, which does not affect the potential in the master equation for metric fluctuations at this order.
Consequently, \eqref{eqn:V_redundant} must be a trivial addition to the potential which exactly compensates \eqref{eqn:df_redundant} and does not affect any physical observable derived from the master equations. 
This fact can be used as a useful consistency check of the results below: the quasi-normal modes and time delays associated with \eqref{eqn:L_redundant} indeed vanish to the precision of the numerical methods used.
A more detailed discussion of the redundancies in the black hole potential is given in \cite{Kimura:2020mrh}---in particular, ``null constraints'' like  $\left(  \frac{36 r_s^9}{r^9} - \ell (\ell+1) \frac{21}{2}  \frac{r_s^8}{r^8}  \right) \approx 0$ can be used to improve the numerical accuracy with which quasi-normal modes are determined.

\section{Theoretical constraints}

When small perturbations propagate on a black hole background, the black hole presents an effective potential barrier \eqref{eqn:master_eom}. 
Waves can scatter and get reflected from this barrier, which introduces a time delay in their transit relative to Minkowski spacetime with no black hole. 
In GR, this is the familiar Shapiro time delay. 
Waves can also ``tunnel'' through this barrier: a phenomenon captured by the bound state energies of the potential. 
These are the so-called ``quasi-normal modes'' (QNM) of the black hole: the characteristic frequencies which determine the gravitational waveform emitted by a post-merger ringdown. 
In GR, the QNM are well understood and have been computed numerically up to 10 decimal places.  
Since the GREFT interactions modify the effective potential, they change both the time delay experienced by reflected waves and also the quasi-normal mode frequencies of tunnelling waves.  
In this section, causality will be used to place restrictions on the change in both of these observables, and hence constrain the GREFT coefficients which appear in the effective potential.

\subsection{IR causality}
\label{sec:IR_causality}

The goal of this subsection is to define the time delay experienced by a gravitational wave scattering from the black hole at large impact parameter, and hence use causality (positivity of this time delay) to place constraints on the GREFT coefficients $b_1, c_1$ and $c_2$.

\paragraph{Time delay.}
Consider a gravitational wave (GW) incident from spatial infinity with fixed parity $P$, angular momentum $\ell$ and an energy $\omega^2 < \text{max} \left[  \sqrt{f_t f_r} \left(  V^{P}_{\ell} +  \delta V^P_{\ell}  \right)  c_s^2 \right]$ so that the region near the black hole horizon is classically forbidden by the potential barrier. 
In terms of the total effective energy
$W^P_{\ell} = \frac{\omega^2}{c_s^2} - \sqrt{ f_t f_r} \left( V_\ell^P + \delta V_\ell^P  \right)$, the classical point of closest approach $r_t$ is determined by the condition $W_{\ell}^P ( \omega,  r_t ) = 0$ (where the kinetic and potential energies exactly balance, since this is the turning point of the classical trajectory)\footnote{
Note that $r_t$ therefore depends on $\ell, P$ and $\omega$, but this dependence is kept implicit to avoid cluttered notation.
}. 
At $r < r_t$, the energy $W_{\ell}^P$ is negative and demanding that the fluctuation decays exponentially in this region gives the WKB solution, 
\begin{align}
 \Psi^{P}_{\ell} ( \omega , r ) &\propto  \frac{1}{ ( - W_{\ell}^P )^{1/4} } \exp \left(  - \int_r^{r_t} dr \sqrt{ - W_{\ell}^P } \right)     &&( r < r_t )   
 \label{eqn:Psi_WKB}
\end{align}
This can be extended to the classically allowed region using the WKB connection formula. 
On the other hand, far from the black hole we conventionally parameterise the fluctuation as a superposition of the original incident wave and a phase-shifted reflected wave, 
\begin{align}
 \Psi^P_{\ell} ( \omega, r) &\propto  \left( 
 e^{+ i \omega r + 2 i \delta^P_{\ell} (\omega)} - e^{-i \omega r + i \pi \ell} 
 \right) \;\; && ( r \gg r_t )
\end{align}
which is an exact solution to the spherically symmetric Schrodinger problem on Minkowski. 
The phase shift $\delta_{\ell}^P (\omega)$ can then be determined from $W_{\ell}^P$ by matching onto the WKB solution. 

Crucially, this phase shift determines the so-called ``Eisenbud-Wigner time delay'' experienced by the GW\footnote{
Note that \eqref{eqn:Tl_def} defines the time delay experienced by a wave with fixed angular momentum. Taking $\partial_\omega$ at fixed $b$ instead produces the time delay at fixed impact parameter. Both are used in the literature, and usually differ only by an overall numerical factor.
},
\begin{align}
 t_{\ell}^P (\omega) =  2 \frac{\partial \delta^P_{\ell} (\omega)}{\partial \omega }  \; .
 \label{eqn:Tl_def}
\end{align}
If $T^P_{\ell}$ is the total time taken for the GW to travel from spatial coordinate $z=-\infty$ to $z=+\infty$, then this $t^P_{\ell}$ represents the difference between $T^P_{\ell}$ on the black hole spacetime \eqref{eqn:background_metric} and $T^P_{\ell}$ on Minkowski spacetime.

\paragraph{IR causality.}
It would be tempting to interpret causality as the condition $t_\ell^P > 0$, namely that the black hole has led to a time delay and not a time advance. This temptation stems from imagining that two observers at spatial infinity, communicating by sending light-like signals, might use such a time advance to send messages apparently backwards in time. 
However, as explained in \cite{deRham:2020zyh}, the implications of causality for $t_\ell^P$ are more subtle in two respects.
 
The first subtlety is that the usual Shapiro time delay of GR gives a large positive contribution to $t_\ell^P$ and no EFT correction could compete with this without the perturbation expansion breaking down (i.e. flipping the sign of $t_\ell^P$ would require the first EFT corrections to be larger than the GR Shapiro, so they are no longer small corrections which can be treated perturbatively). 
Since the EFT corrections are there to encode the effects of heavy degrees of freedom, a more refined statement of causality would be that these heavy degrees of freedom contribute positively to the time delay. 
Formally, this amounts to splitting the time delay into GR and EFT parts: 
\begin{align}
t_{\ell}^P = \left[ t_{\ell}^P \right]_{\rm GR} + 
\delta t_{\ell}^P \;\;\;\; 
\text{where} \;\;\;\; \delta t_\ell^P = \epsilon_s^4 \delta_3 t_{\ell}^P + \epsilon_s^6 \delta_4 t_{\ell}^P + ... 
\end{align}
and considering the EFT corrections separately. 
This is subtle because while one could measure $t_\ell^P = T_{\ell}^P |_{\text{GREFT BH} } - T_{\ell}^P |_{\text{Minkowski}} $ by sending signals far from / near to the black hole, there is no way to ever measure the correction $\delta t_\ell^P = T_{\ell}^P |_{\text{GREFT BH}}  - T_{\ell}^P |_{ \text{GR BH} }$ since there is no GR BH in a Universe described by the GREFT (in which unperturbed Schwarschild is not a stable solution to the classical equations of motion).
The requirement that the EFT corrections contribute positively to the time advance ultimately stems from dispersion relations and monotonicity arguments applied to sound speeds: the idea is that $\left[ t_{\ell}^P \right]_{\rm GR}$ represents a UV value of the time delay (in which all of the heavy particles have been integrated back in so there are no curvature corrections), and so the time delay in the IR must be larger if the heavy physics is causal. 

The second subtlety is that the EFT has a finite regime of validity, and therefore a sufficiently small time advance may not imply a resolvable acausality within the EFT\footnote{
This is actually already the case in quantum mechanics, for instance scattering a wave of speed $v$ from a hard sphere of radius $a$ leads to a small time advance, but Wigner argued that causality in this case should correspond to a time delay $> -2a/v$ \cite{Wigner:1955zz, DECARVALHO200283}, namely that it is larger than the quantum uncertainty associated with the sphere position / wave speed. 
}.
To be resolvable in an EFT context would usually mean a time interval $\gg 1/\Lambda$, where $\Lambda$ is approximately the EFT cut-off. 
However, in order for a time advance to be ``resolvable'' at a level where it would threaten causality, the time advance must exceed $1/\omega$ (the quantum uncertainty associated with a wavepacket of energy $\omega$). 
Taken together, the requirement that the heavy physics contribution does not lead to a resolvable time advance can be written as:
\begin{align}
 - \omega \, \delta t_{\ell}^P  \gtrsim  1  \qquad \qquad \text{(IR causality)} \; .
 \label{eqn:IR_causality}
\end{align}
This concept was recently introduced and explored in \cite{deRham:2019ctd, deRham:2020zyh, Chen:2021bvg, deRham:2021bll, Chen:2023rar}, where it was referred to as ``IR causality''. 
Unlike the positivity bounds from UV causality, it is not a numerically precise bound since the $\mathcal{O} (1)$ constant depends on the details of the scattering process (which determine the uncertainty that must be overcome in order to have a resolvable time advance).

\paragraph{Computing the time delay.}
In the WKB approximation described above, \cite{deRham:2020zyh} recently derived a compact expression for the shift in the time delay induced by the EFT interactions\footnote{
The WKB approximation for the full $t_\ell$ is,
\begin{align}
 t_{\ell}^P ( \omega ) = 2 \int_{r_{t*} }^{\infty} d r_* \left( \frac{\partial}{\partial \omega} \left( \sqrt{ W^P_{\ell} (r, \omega) }  \right)  - 1 \right) - 2 r_{t*}
\end{align}
where $r_{t*}$ is the tortoise coordinate associated with $r_t$. This can be carefully perturbed around GR to produce \eqref{eqn:dT_from_dA}. 
},
\begin{align}
 \delta_{n} t_{\ell}^P ( \omega ) = - 2 \int_{r_t^{\rm GR}}^{\infty} d r \; \mathcal{A}_{\ell}^P \, \partial_r \left( \frac{ \delta_n \mathcal{A}_{\ell}^P }{ \partial_r \mathcal{A}_{\ell}^P  }  \right)
\label{eqn:dT_from_dA}
\end{align}
where the integrand is constructed from,
\begin{align}
\mathcal{A}_\ell^P (\omega, r)  &= \frac{\omega}{ f \sqrt{\omega^2 - f  V_\ell^P } }  \nonumber \\ 
 \delta_n \mathcal{A}_\ell^P ( \omega, r ) &=  
 \mathcal{A}_\ell^P \left[ 
 \frac{ \delta_n W_\ell^P }{2 ( \omega^2 - f V_\ell^P ) } - \frac{1}{2 \omega} \frac{\partial  }{\partial \omega} \delta_n W_{\ell}^P - \delta_n \mathcal{N}
 \right]
\end{align}
and $f = 1 - \frac{r_s}{r}$, $\delta_n \mathcal{N}$ is given in \eqref{eqn:N_def} and $\delta W_\ell^P$ is given by,
\begin{align}
 W_\ell^P = f V_{\ell}^P + \delta W_\ell^P \; .
\end{align}
This can be evaluated numerically to find the time delay for a fixed $\omega$, $\ell$ and parity.
Note that the turning point $r_t^{\rm GR}$ in \eqref{eqn:dT_from_dA} is defined by the GR condition,
\begin{align}
 \omega^2 - f (r_t^{\rm GR} ) V_{\ell}^P (r_t^{\rm GR} ) = 0 \; .
\end{align}
A useful parameterisation of the $\omega$ is therefore,
\begin{align}
 \omega^2 = \gamma \, \max_r \left( 
  f (r) V_{\ell}^P (r)
  \right)  \; , 
  \label{eqn:g_def}
\end{align}
where $\gamma$ determines the turning point $r_t^{\rm GR}$ and $ 0 < \gamma < 1$ is the range of $\omega$ over which there is a classically forbidden interior region and the above WKB method can be applied. 
At large $\ell$, the turning point is $\left( r_t^{\rm GR} \right)^2 \approx \ell (\ell +1)/\omega^2 \approx \frac{27}{4} \gamma r_s^2 + \mathcal{O} \left( \gamma^2 \right)$, and so holding $\gamma$ fixed amounts to holding $r_t$ fixed\footnote{
At least in the large $\ell$ regime: at fixed $\gamma$ the turning point $r_t^{\rm GR}$ will depend on $\ell$ for small values of $\ell$.
}.

\begin{figure}[b]
\includegraphics[width=0.5 \textwidth]{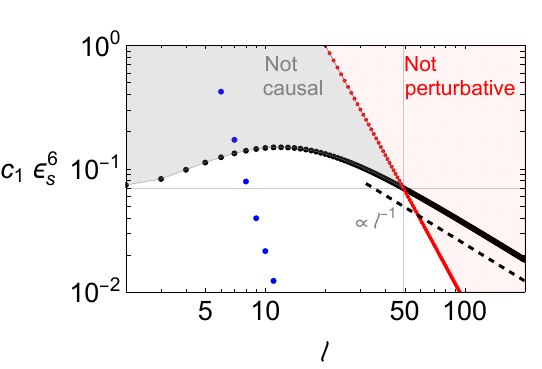}
\includegraphics[width=0.5 \textwidth]{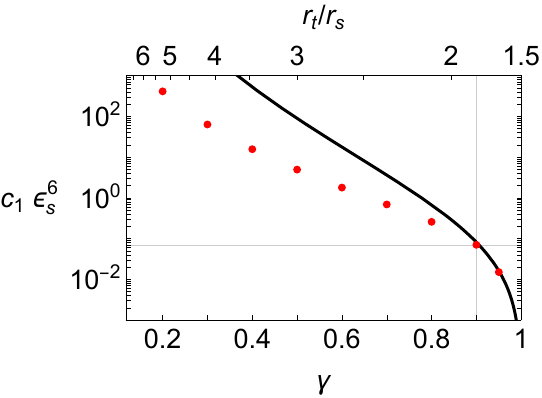}
\caption{
IR causality constraints on the quartic GREFT interaction $c_1 ( R^{(2)} )^2$. \emph{Left.} When $c_1 > 0$, the time advance for parity-odd modes with energy $\gamma = 0.9$ and angular momentum $\ell$ becomes resolvable when $c_1 \epsilon_s^6$ exceeds the black points. 
Red points show the regime of perturbative validity implied by the strong condition~\eqref{eqn:eps_max_strong} with $c_1 \epsilon_\omega^3 \approx 1$. The $\ell = 49$ mode gives the largest time advance while remaining perturbative, and leads to the IR causality bound $c_1 \epsilon_s^6 \lesssim 0.07$.
The blue points show the weak condition~\eqref{eqn:eps_max_weak} for comparison (which leads to $c_1 \epsilon_s^6 \lesssim 0.14$). 
\emph{Right.} Repeating this procedure for different $\gamma$ gives the maximum values for $c_1 \epsilon_s^6$ shown by red points. At large $\gamma$ it agrees well with the analytic result \eqref{eqn:c1_dt_bound}, shown by black line. The disagreement at low $\gamma$ is due to the $\ell$ dependence of $1/(\omega \delta_4 t_\ell^-)$: for $\gamma \lesssim 0.9$, the largest time advance occurs for $\ell \approx 3$ rather than large $\ell$ and this leads to a stronger constraint than \eqref{eqn:c1_dt_bound} for those values of $\gamma$. 
}
\label{fig:c1_dt}
\end{figure}

As an aside, note that when deriving the WKB solution \eqref{eqn:dT_from_dA}, the waves absorbed by the black hole have been neglected. 
This is a good approximation provided that the classical turning point $r_t$ is much greater than the position of the potential barrier ($\approx 3 r_s/2$ at large $\ell$). 
Since $\gamma$ is related to the turning point by,
\begin{align}
 \gamma = \frac{ 27 r_s^2 ( r_t^{\rm GR} -r_s ) }{ 4 \left( r_t^{\rm GR} \right)^3 } + \mathcal{O} \left( \frac{1}{\ell^2} \right) \; , 
 \label{eqn:g_to_rt}
\end{align}     
the limit $\gamma \to 1$ corresponds to $r_t \to 3 r_s/2$ and these absorbed waves become important. 
Following \cite{deRham:2021bll}, the fiducial value of $\gamma = 0.9$ was used in Figure~\ref{fig:c1_dt}. It is worth remarking that this fixes $r_t \approx 1.87 r_s$ in the large $\ell$ regime which dominates the causality bound, and so even if \eqref{eqn:IR_causality} could be made numerically precise there would remain an $\mathcal{O} (1)$ theoretical uncertainty from the neglected absorbed waves.

\paragraph{$c_1$ time advance.}
For instance, consider the effect of a single GREFT interaction $c_1 ( R^{(2)} )^2$ (i.e. set $b_1 = c_2 = 0$ and neglect higher-derivative terms). 
The time delay $\delta t_{\ell}^{\pm}$ is proportional to $-c_1$ for odd modes and $+ c_1$ for even modes. 
This interaction will therefore lead to a time advance if $\epsilon_s$ is too large, 
 \begin{align}
| c_1  | \epsilon_s^6 \gtrsim  \begin{cases} \frac{-1}{ \omega \delta_4 t_\ell^-}   &\text{if } c_1 > 0 \\   \frac{- 1}{ \omega \delta_4 t_\ell^+}  &\text{if } c_1 < 0 \end{cases}
 \qquad  \Rightarrow \qquad \text{resolvable time advance} \; .
 \label{eqn:c1_resolv}
 \end{align}
IR causality (forbidding such a resolvable time advance) can therefore impose an upper bound on $\epsilon_s$. This should be viewed as the range of spacetime backgrounds which this EFT can describe perturbatively. 
Just as a violation of perturbative unitarity signals that higher-order EFT corrections must become important, the violation of perturbative causality is diagnosing the range of $\epsilon_s$ over which the higher-order corrections may remain small (assuming causality). 

Expanding the integral \eqref{eqn:dT_from_dA} at large $\ell$ produces the result
\begin{align}
\frac{1}{\omega \delta_4 t_{\ell}^- (\omega )}  &= \frac{ K^-_{c_1} ( \gamma ) }{- c_1 \ell} + \mathcal{O} \left( \frac{1}{\ell^2} \right)   \nonumber \\
\frac{1}{\omega \delta_4 t_{\ell}^+ (\omega )}  &= \frac{ K^+_{c_1} ( \gamma ) }{ + c_1 \ell^3 }  + \mathcal{O} \left( \frac{1}{\ell^4} \right) 
\label{eqn:K_def}
\end{align}
where $K^{\pm}_{c_1} (\gamma)$ are positive functions of $\gamma$ given in Appendix~\ref{app:dt}. 
Figure~\ref{fig:c1_dt} shows the $-1/(\omega \delta_4 t_\ell^- (\omega) )$ upper bound obtained by numerically integrating \eqref{eqn:dT_from_dA}  for different $\ell$ at a fixed $\gamma = 0.9$, and it indeed scales as $\sim 1/\ell$ at large $\ell$.
This scaling spells trouble for the $c_1$ GREFT coefficient: regardless of its sign, scattering waves from a GREFT black hole will always produce a resolvable time advance for sufficiently large $\ell$. 
This would seem to imply that $c_1 = 0$ is the only value allowed by causality, or at least that $c_1$ cannot be the dominant interaction (i.e. other EFT interactions could correct the acausality induced by a small $c_1$).   
However, such a conclusion would be premature, since there is a finite range of $\ell$ under which the GREFT is under perturbative control. 
A finite $c_1$ would be consistent with causality as long as the time advance is only resolvable for $\ell > \ell_{\rm max}$ modes, where $\ell_{\rm max}$ is the largest angular momentum that can be reliably described by this perturbative calculation.

\paragraph{Regime of validity.}
The question becomes: what is $\ell_{\rm max}$?
As discussed above, on this GREFT background the parameter $\epsilon_s$ is also bounded by the condition that the derivative expansion is under control (this is closer in spirit to a unitarity bound). 
In order to apply the GREFT to the scattering of GWs with turning point $r_t$, it is enough for this EFT to be valid on scales $r \gtrsim r_t$ which are classically accessible\footnote{
Even if there are large corrections in the $r_s < r \ll r_t$ region close to the horizon, these affect the classically forbidden region and do not alter the WKB behaviour \eqref{eqn:Psi_WKB} near $r_t$ or the expression \eqref{eqn:dT_from_dA} for the classical time delay. 
}.
From the strong condition \eqref{eqn:w_max_strong} on the allowed range of $\omega$, suppose that the EFT is valid up to $( \omega r_s )_{\rm max} =  \epsilon_{\omega} \frac{4 r_t^{\rm GR}}{r_s} \epsilon_s^{-2}$, where $\epsilon_\omega \lesssim 1$ is a small parameter that controls the size of higher time-derivative corrections. 
In that case, the GREFT is only perturbative providing that
\begin{align}
\epsilon_s^2 <   \frac{ 4 \epsilon_{\omega}  r_t^{\rm GR} }{   \omega r_s^2 } =   \frac{  \epsilon_{\omega} }{ \ell } \left( 4 + \mathcal{O} \left(  \gamma^0 \right)  \right) + \mathcal{O} \left( \frac{1}{\ell^2} \right)  \qquad\qquad \text{(strong perturbativity)} \; ,
\label{eqn:eps_max_strong}
\end{align}
where $\epsilon_\omega$ is the small parameter that controls higher derivative corrections. 
For comparison, if the weak condition \eqref{eqn:w_max_weak} is used to instead infer an $( \omega r_s )_{\rm max} = 2 \epsilon_{\omega'} / \epsilon_s$, then the GREFT would only be perturbative for
\begin{align}
\epsilon_s^2 <   \frac{4 \epsilon_{\omega'}^2}{ (  \omega r_s )^2 } = \frac{\epsilon_{\omega'}^2}{ \ell^2 }  \ 
\frac{27}{  \gamma }  + \mathcal{O} \left( \frac{1}{\ell^2} \right) \qquad \qquad \text{(weak perturbativity)} \; . 
\label{eqn:eps_max_weak}
\end{align}
In either case, the crucial observation is that for a given EFT and spacetime background (i.e. fixed $\epsilon_s$ and $\epsilon_\omega$), waves of arbitrarily high $\ell$ at fixed $r_t$ (equivalently fixed $\gamma$) are not under perturbative control. 
Using \eqref{eqn:c1_resolv} to derive an upper bound on $\epsilon_s$ is therefore too strong, since it does not account for the finite $\ell_{\rm max}$ implied by \eqref{eqn:eps_max_strong} or \eqref{eqn:eps_max_weak}.

\paragraph{$c_1$ even modes.}
More precisely, when $c_1 < 0$ we see that the causality cut-off from parity-even modes \eqref{eqn:K_def} and the strong condition \eqref{eqn:eps_max_strong} both scale $\sim 1/\ell^3$, and therefore this interaction will lead to a problematic time advance within the EFT's regime of validity whenever $K^+ ( \gamma) <  |c_1| \left( 4 \epsilon_{\omega} r_t^{\rm GR} / \sqrt{4 \gamma / 27} \right)^3$ at large $\ell$. 
Since $K^+ ( 0.9 ) \approx 1.5$, this means that a negative $c_1$ is only consistent with causality if the higher-derivative corrections come in at a lower scale than expected: namely if the maximum $\omega$ is set by $|c_1| \epsilon_{\omega}^3 \approx 1.8 \times 10^{-4}$ rather than the expected $\epsilon_{\omega} \approx 1$.  Alternatively, if the unitarity cut-off is set by the parameterically lower weak condition \eqref{eqn:eps_max_weak}, then there is no causality issue at large $\ell$.  

Demanding that the causality cut-off does not lower the strong derivative condition from unitarity, i.e. setting $\epsilon_\omega = 1$ in \eqref{eqn:eps_max_strong}, leads to the conclusion that $c_1$ must be positive if it is the dominant interaction. 
This coincides with the positivity bound from UV causality of the $4$-point graviton amplitude.

\paragraph{$c_1$ odd modes.}
Taking $c_1 > 0$ and comparing \eqref{eqn:c1_resolv} with \eqref{eqn:eps_max_strong}, there is a problematic time advance within the EFT's regime of validity whenever,
\begin{align}
\left( \frac{4 r_t^{\rm GR}}{ \omega r_s^2}  \right)^3 \gtrsim   c_1   \epsilon_s^6  \gtrsim \frac{ K^-_{c_1} ( \gamma) }{  \ell }  
\label{eqn:c1_dt_range}
\end{align} 
at large $\ell$ (where now $c_1 \epsilon_\omega^3$ is taken to be $\mathcal{O} (1)$). 
Since $\omega^2 \approx \frac{4}{27} \gamma \ell^2$, forbidding this acausality requires that\footnote{
Note that \eqref{eqn:c1_dt_bound} follows from the range \eqref{eqn:c1_dt_range} vanishing at $\ell^2  = 648 \sqrt{3}  \left(  r_t^{\rm GR} / \sqrt{\gamma}  \right) ^3/K^-_{c_1}(\gamma)$, which must be $\gg 2^2$ in order to trust this large $\ell$ limit. 
For $\gamma = 0.9$ this is $\ell^2 \approx 49^2$ and the bound \eqref{eqn:c1_dt_bound} is indeed a good fit to the numerical result (see Figure~\ref{fig:c1_dt}). 
},
\begin{align}
 c_1 \epsilon_s^6 \lesssim  \frac{1}{18 \sqrt{2}}  \left(  \frac{\gamma^3}{3} \right)^{1/4} \, \left(  \frac{ r_s K^-_{c_1} (\gamma) }{ r_t^{\rm GR}} \right)^{3/2}  
\label{eqn:c1_dt_bound}
\end{align}
At $\gamma = 0.9$, this gives $c_1 \epsilon_s^6 \lesssim 0.07 $ as the range of backgrounds which this EFT can describe without problematic acausality\footnote{
Note that \cite{deRham:2021bll} approximates $r_t^{\rm GR} \approx (\ell + \frac{1}{2} ) / \omega \approx 2.74$ at this value of $\gamma = 0.9$, which leads to $c_1 \epsilon_s^6 \lesssim 0.04$ from \eqref{eqn:c1_dt_bound}. The difference is unimportant since the cut-off is not numerically precise, given the uncertainty in (i) choosing $\epsilon_{\omega}$, (ii) the resolvability condition \eqref{eqn:IR_causality} and (iii) neglecting the absorbed waves in the WKB solution. 
}. 

Figure \ref{fig:c1_dt} shows a numerical determination of the maximum $c_1 \epsilon_s^6$ allowed by IR causality (taking into account the perturbativity bound \eqref{eqn:eps_max_strong}), which agrees well with the result \eqref{eqn:c1_dt_bound} for $\gamma \gtrsim 0.8$. For $\gamma < 0.8$, the largest resolvable time advance actually occurs at $\ell = 3$ and gives a stronger bound than \eqref{eqn:c1_dt_bound} at those values of $\gamma$.
The strongest bound overall would come from pushing $\gamma \to 1$, where $K_{c_1}^- ( \gamma) \to 0$ since the time advance diverges. This would require $c_1 \epsilon_s^6  \to 0$ to be consistent with causality (however note that the absorbed waves become important in this limit and invalidate the WKB expression \eqref{eqn:dT_from_dA} used to derive \eqref{eqn:c1_dt_bound})\footnote{
Furthermore, any finite upper bound on $c_1 \epsilon_s^6$ would similarly imply that $c_1$ has to vanish if black hole of arbitrary small mass are considered \cite{deRham:2021bll}. 
}.
Interestingly, this is the same conclusion that would follow from UV causality if the IR cut-off is pushed unreliably close to the EFT cut-off. 
%

\paragraph{$c_2$ odd modes.}
Now consider turning on the $c_2$ interaction. 
This affects only the odd parity modes, and gives a contribution to the time advance which is proportional to $+c_2$ and grows with $\ell^2$ at large $\ell$.
As with the $c_1$ contribution to the even modes, forbidding a resolvable time advance leads to a causality cut-off of the form, 
\begin{align}
 | c_2 | \epsilon_s^6 \gtrsim  \frac{-1}{\omega \delta_4 t_{\ell}^-}  = \frac{ K^-_{c_2} (\gamma) }{ \ell^3 } + \mathcal{O} \left( \frac{1}{\ell} \right)  \;\; \text{ if } c_2 < 0 \; . 
\end{align} 
Given the EFT regime of validity \eqref{eqn:eps_max_strong}, when $c_2 < 0$ there would be a problematic time advance whenever $K^-_{c_2} ( \gamma) < | c_2 | \left( 4 \epsilon_\omega r_t^{\rm GR} / \sqrt{4 \gamma / 27} \right)^3$. 
Since $K^-_{c_2} ( 0.9) \approx 0.36$, this means that a negative $c_2$ is only consistent with causality if the maximum $\omega$ is set by $|c_2| \epsilon_\omega^3 \approx 4.2 \times 10^{-5}$. 
Demanding instead that the EFT remains reliable up to $|c_2| \epsilon_{\omega}^3$ of order unity, then IR causality implies that $c_2 > 0$. This is again in line with the UV causality bounds on the 4-point scattering amplitude on Minkowski.   

At sufficiently large $\ell$, the $c_2$ contribution dominates $\delta t_{\ell}^-$ and would cure any acausality induced by $c_1$. 
However, at small $\ell$ the $c_{1,2}$ contributions are comparable. 
The smallest $c_2$ contribution is at $\ell = 2$, where 
\begin{align}
\omega \delta_4 t^-_{\ell = 2} 
=
 -13.581 c_1 + 12.896 c_2  \;\; \text{ at  } \gamma = 0.9  \; .
\end{align}
If $c_2 > 1.053 c_1$ then the overall the time delay is positive and there is no causality violation for any value of $\epsilon_s$. 
If $c_2$ is less than this, then from the $\ell = 2$ scattering of parity-odd fluctuations, IR causality is violated if\footnote{
Interestingly, $c_1 = c_2$ in superstring theory and there is such a finite upper bound on $\epsilon_s^6$ from the lowest $\ell$ modes.
},
\begin{align}
 \epsilon_s^6  \gtrsim \frac{1}{  13.581 c_1 - 12.896 c_2  } \qquad \qquad (\text{causality violated}) \; . 
 \label{eqn:eps_upper_c1c2} 
\end{align}
Note that since the time advance from $c_1$ grows with $\ell$, as $c_2$ is made smaller and smaller it is successively higher $\ell$ modes which provide the strongest constraint on $\epsilon_s^6$ (and in the $c_2 \to 0$ limit it is the largest $\ell$ compatible with perturbativity that sets the cut-off \eqref{eqn:c1_dt_bound}).

\paragraph{$b_1$ time advance.}
Finally, consider the cubic $b_1$ interaction. 
From a straightforward power counting in $\epsilon_s$, this should be the dominant correction to $\delta t_\ell^{\pm}$. 
However, for this interaction the time delay is proportional to $+ b_1$ for even modes and $-b_1$ for odd modes, and unlike the quartic interactions the scaling with $\ell$ is the same for both parities:
\begin{align}
\frac{1}{\omega \delta_3 t_{\ell}^{\pm} (\omega ) }  
= \frac{ K^{\pm}_{b_1} ( \gamma ) }{\pm b_1 \ell} 
+ \mathcal{O} \left( \frac{1}{\ell^2} \right)   \, . 
\end{align}
As a result, there is a problematic time advance within the EFT regime of validity if,
\begin{align}
\left( \frac{4 r_t^{\rm GR} }{ \omega r_s^2}  \right)^2 \gtrsim   | b_1 | \epsilon_s^4  \gtrsim  \begin{cases}  \frac{ K_{b_1}^- (\gamma)  }{ \ell } \;\; &\text{if  } b_1 > 0  \\
\frac{ K_{b_1}^+ (\gamma)  }{ \ell } \;\; &\text{if  } b_1 < 0  
\end{cases}
\end{align}
again assuming that $|b_1| \epsilon_\omega^2$ is order unity. 
Since $K^{-}_{b_1} ( 0.9) \approx K^-_{b_1} (0.9) \approx 0.11$, IR causality imposes the two-sided bound,
\begin{align}
 |b_1 | \epsilon_s^4 \lesssim 2.8 \times 10^{-5} \; . 
 \label{eqn:b1_dt_bound}
\end{align}
This is consistent with the causality bounds of \cite{Camanho:2014apa}.
It is worth emphasising that \eqref{eqn:b1_dt_bound} should be read as $|b_1| \epsilon_s^4 \lesssim \mathcal{O} \left( 10^{-5} \right)$ since the original resolvability condition \eqref{eqn:IR_causality} is not numerically precise, and may contain factors of 2, $\pi$, etc. 
Regardless of these factors, such a tight bound would render this cubic interaction unimportant for the inspiral and ringdown phenomenology probed by gravitational waves observatories (this will be discussed further in section~\ref{sec:obs} below). 

However, note that while the bound \eqref{eqn:c1_dt_bound} on $c_1 \epsilon_s^6$ does not vary much with different $\epsilon_{\omega}$ or between the strong and weak perturbativity conditions, \eqref{eqn:b1_dt_bound} would become $|b_1| \epsilon_s^4 \lesssim \mathcal{O} \left( 10^{-2} \right)$ if the weaker perturbativity condition \eqref{eqn:eps_max_weak} were used. So the cubic interactions in GREFT may still play a phenomenologically relevant role at low energies, since this EFT could in principle preserve IR causality if the cut-off in $\omega$ is much lower than the naive expectation \eqref{eqn:w_max_strong}.

\FloatBarrier
\begin{figure}[t]
\includegraphics[width=0.5 \textwidth]{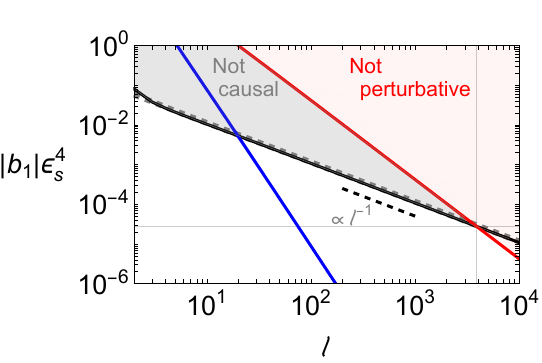}
\includegraphics[width=0.5 \textwidth]{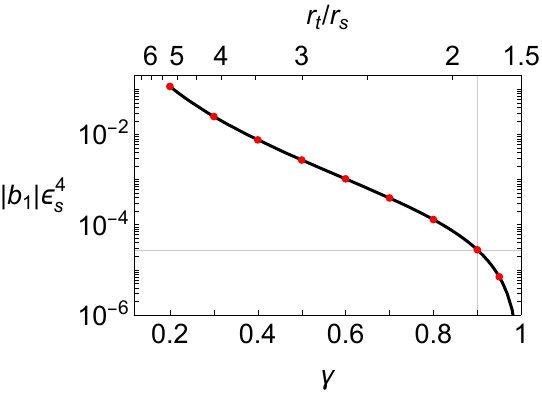}
\caption{
IR causality constraints on the cubic GREFT interaction $b_1  R^{(3)} $. 
\emph{Left.} When $b_1 > 0$ ($<0$), the time advance for parity-odd (even) modes with energy $\gamma = 0.9$ and angular momentum $\ell$ becomes resolvable when $|b_1| \epsilon_s^4$ exceeds the solid (dashed) black line. At these large values of $\ell$ the even and odd results overlap. 
The red line shows the regime of perturbative validity implied by the strong condition~\eqref{eqn:eps_max_strong} with $b_1 \epsilon_\omega^2 \approx 1$. The $\ell = 3881$ mode gives the largest time advance while remaining perturbative, and leads to the IR causality bound $|b_1| \epsilon_s^4 \lesssim 2.8 \times 10^{-5}$.
The blue line shows the weak condition~\eqref{eqn:eps_max_weak} for comparison (which leads to $|b_1| \epsilon_s^4 \lesssim 5 \times 10^{-3}$). 
\emph{Right.} Repeating this procedure for different $\gamma$ gives the maximum values for $|b_1| \epsilon_s^6$ shown by red points, which agree well with the analytic result \eqref{eqn:b1_dt_bound} shown by a black line. 
}
\label{fig:b1_dt}
\end{figure}

\FloatBarrier

As a final sanity check, note that the redundant potentials \eqref{eqn:V_redundant} give $\delta_3 t_{\ell}^{\pm} ( \omega ) \approx 0$ up to small numerical errors for the range of $\gamma$ considered in the figures (and exactly zero in the large $\ell$ limit where the integral can be performed analytically).

\subsection{QNM causality}
\label{sec:QNM_causality}

The final causality condition to be considered in this work is from the quasi-normal mode spectrum.

\paragraph{Quasi-normal modes.}
Recall that the master equation \eqref{eqn:master_eom} takes the form of the (time-independent) Schrodinger equation, 
\begin{align}
 \frac{\partial^2}{ \partial r_*^2 }  \Psi^{\pm}_{\ell}  
 =  -  W_{\ell}^{\pm}
 \Psi^{\pm}_{\ell} 
 \label{eqn:master_Schr}
\end{align}
The ``quasi-normal modes'' of the black hole refer to solutions of the following eigenvalue problem: solve \eqref{eqn:master_Schr} subject to the boundary condition,
\begin{align}
 \Psi^{\pm}_{\ell} \sim e^{\pm i \omega r_*}  \;\; \text{at} \;\; r_* \to \pm \infty \; . 
 \label{eqn:QNM_bc}
\end{align}
Physically, this condition corresponds to waves that are purely ingoing near the horizon (nothing is coming out of the black hole) and purely outgoing far from the horizon (nothing is coming in from spatial infinity).  

Solutions with these boundary conditions are only possible for a discrete set of allowed $\omega$. 
These are labelled in general by $\{ n, \ell, m \}$, namely the overtone number, the total angular momentum, and the angular momentum about one axis. 
Since both Schwarszchild and the GREFT extension \eqref{eqn:background_metric} are spherically symmetric, the allowed quasi-normal mode frequencies $\omega_{n \ell}$ for this black hole will be independent of $m$. 
The QNM frequencies are important phenomenologically because they describe how the black hole responds to a transient perturbation: for instance immediately after a binary merger, the final black hole is formed in a perturbed state which then relaxes back to \eqref{eqn:background_metric} with $\Psi_{\ell}^{\pm} \sim e^{i \omega_{n\ell} t}$ describing the characteristic gravitational wave emitted during the ringdown. See \cite{Berti:2009kk} for a review.

In GR, the QNM frequencies coincide for odd and even perturbations\footnote{
Despite the differences between the potentials \eqref{eqn:GR_V}, this isospectrality can be understood by noticing that they follow from the same superpotential \cite{Berti:2009kk}. This can be made manifest by treating odd/even perturbations in a unified manner \cite{Clarkson:2002jz}. 
}. The GREFT higher-derivative interactions typically break this coincidence and produce different allowed frequencies for the parity odd/even fluctuations, which can be written in the form:
\begin{align}
 \omega_{n \ell}^{\pm}  &=  \omega_{n \ell}^{\rm GR}  + \delta \omega_{n \ell}^{\pm} \qquad \text{where} \qquad   \delta \omega_{n \ell}^{\pm} = \epsilon_s^4 \delta_3 \omega_{n \ell}^{\pm} + \epsilon_s^6 \delta_4 \omega_{n \ell}^{\pm} + \mathcal{O} \left( \epsilon_s^8 \right) \; . 
 \label{eqn:dw_def}
\end{align}

\paragraph{QNM causality.}
In the absence of gravity, causality implies that any singularity of the linear response function must appear in the lower-half of the complex plane.
The argument is well-known, and follows from the fact that the classical response function $G(t)$ (which describes how a degree of freedom $\mathcal{O}(t)$ will respond to a source, namely $\mathcal{O} (t) = \int dt' G(t - t') J(t')$), must obey
\begin{align}
\text{Causal boundary conditions} \qquad \Rightarrow \qquad G (t) = 0 \;\; \text{ for } t < 0 \;.
\label{eqn:bc_causal}
\end{align}
Physically, this condition corresponds to the requirement that if a source is turned on at time $t_0$, it can only affect the system at times $t > t_0$. 
The response function in the frequency domain\footnote{
Note that opposite conventions for the sign of $\omega$ in the Fourier transform would lead to the opposite conclusion of analyticity for $\text{Im} \omega < 0$. When discussing QNM frequencies below, a particular sign convention has been chosen for $\omega_{n \ell}$ so that $\text{Im} \, \omega_{n \ell}^{\rm GR} < 0$.
},
\begin{align}
 G ( \omega  ) = \int_{-\infty}^{\infty} d t \, G ( t ) \, e^{i \omega t} \; , 
\end{align}
is therefore analytic in the upper-half of the complex $\omega$ plane, since if this integral converges on the real $\omega$ axis then it must also converge for any $\text{Im} \, \omega > 0$ because in that region the integrand is strictly smaller (i.e. $| e^{ i \omega t} | < 1$ for $t > 0$). 
The limited support of $G( t)$ in the time domain (i.e. that a source can only affect its future) corresponds to analyticity (i.e. no singular points) in the upper-half of the complex $\omega$-plane. 

The response function for fluctuations on this black hole background would be defined by solving
\begin{align}
\left( \frac{d^2}{dr_*^2} +  W_{\ell}^{\pm} ( \omega, r) \right)  G ( \omega , r ) \propto \delta ( r )
\end{align}
subject to appropriate boundary conditions. 
Since the quasi-normal mode frequencies $\omega_{n \ell}^{\pm}$ correspond to zeroes of the differential operator on the left-hand-side, they can be viewed as poles of $G ( \omega , r )$. 
It is therefore tempting to conclude that causality might impose some constraint on the location of these poles in the complex plane. 
For instance, $\text{Im} \, \omega_{n \ell}^{\pm} < 0$ would be the straightforward analogue of the non-gravitational analyticity of a causal response function. 
However, much like the time advance of the previous subsection, this is subtle for a number of reasons: 

\begin{itemize}

\item[(i)] The GR contribution to each QNM frequency already obeys $\text{Im} \, \omega_{n\ell} < 0$, and no small EFT correction will ever change this sign (i.e. a sufficiently large EFT correction would invalidate perturbation theory, at least naively). 
This is the analogue of EFT corrections to the Shapiro time delay in $t_{\ell}^{\pm}$. 
Learning from that example, a more refined statement would be that the EFT contribution alone, $\delta \omega_{n \ell}^{\pm}$ in \eqref{eqn:dw_def}, must shift the QNM frequencies deeper into the lower-half of the complex plane (since in the absence of gravity this is the direction which would be consistent with causality). 
Since $\delta \omega_{n \ell}^{\pm}$ is not directly measurable, this condition is again to be viewed in the same spirit as the monotonicity theorems / dispersion relations: as one integrates out heavy physics and replaces them with GREFT coefficients, this process should shift the QNMs in a particular direction if this heavy physics is causal. 

\item[(ii)] Since the EFT has a finite resolving power, a sufficiently small change in the QNM frequency should not present any pathology. 
The most basic requirement for the EFT to ``resolve'' a frequency change would be that $\delta \omega \ll 1/\Lambda$. 
However, again learning from the earlier time advance discussion, this is likely not sufficient to present a problem for causality (essentially due to the quantum uncertainty associated with any measurement of this frequency). 
A more reliable condition would be to define the characteristic lifetime of each QNM perturbation as,
\begin{align}
 \tau_{n \ell}^{\pm} = - \frac{2 \pi }{ \text{Im} \, \omega_{n \ell}^{\pm} } \; ,
\end{align} 
so that the change in the lifetime induced by the EFT interactions,
\begin{align}
 \delta \tau_{n \ell}^{\pm} \equiv - \frac{2 \pi }{ \text{Im} \, \omega_{n \ell}^{\pm} } + \frac{2 \pi }{ \text{Im} \, \omega_{n \ell}^{\rm GR} } 
 = \frac{2 \pi}{ \text{Im} \, \omega_{n \ell}^{\rm GR} } \left(  
\epsilon_s^4 \frac{ \text{Im} \, \delta_3 \omega_{n \ell}^{\pm} }{ \text{Im} \,  \omega_{n \ell}^{\rm GR} } +  \epsilon_s^6 \frac{ \text{Im} \, \delta_4 \omega_{n \ell}^{\pm} }{ \text{Im} \,  \omega_{n \ell}^{\rm GR} }  + \mathcal{O} \left( \epsilon_s^8 \right)
 \right)
\end{align}
is deemed resolvable if,
\begin{align}
\text{Re}\left[ \omega_{n \ell}^{\pm}  \right] \times \delta \tau_{n \ell}^{\pm} \; \gtrsim \; 1 \; .  
\label{eqn:dtau_resolv}
\end{align}
In this language, a QNM frequency shifting deeper into the lower-half of the complex plane corresponds to $\delta \tau_{n \ell}^{\pm} < 0$, i.e. shorter lifetimes for the QNMs. 
QNM causality is then the requirement that perturbations decay faster than in GR, namely that the GREFT black hole is ``more stable'' than a Schwarszchild black hole in GR. The condition \eqref{eqn:dtau_resolv} then reflects the point at which one could determine whether a wave of oscillatory frequency $\text{Re} \, \omega_{n \ell}^{\pm}$ has experienced a change in decay time of $\delta \tau_{n \ell}^{\pm}$.

\item[(iii)] A further subtlety, which seems unique to the QNM problem, is that the boundary conditions \eqref{eqn:QNM_bc} are not the standard `causal' boundary condition \eqref{eqn:bc_causal} used in basic proofs of analyticity.  
This will be discussed further in section~\ref{sec:disc}. 
While there are some rough arguments for this proposal, and at least for the GREFT it coincides with existing constraints from UV and IR causality, there is no rigorous proof. 
It would therefore be wise to interpret the following bounds as an interesting conjecture, which may hopefully be proven (or disproven) in the near future. 

\end{itemize}

\paragraph{Computing QNMs.}
There are now two separate motivations to carefully compute the QNM frequencies in the GREFT. 
Firstly, they inform the ringdown portion of the inspiral-merger-ringdown templates that are required to accurately test GR using gravitational wave measurements (see section~\ref{sec:obs}).
Secondly, they can be used to check when the ``QNM causality'' condition is violated and hence place theoretical constraints on the GREFT coefficients.

There are a number of ways to compute the QNMs from a given $W_{\ell}^{\pm}$. 
Below, two complementary techniques are used:
\begin{itemize}

\item[(1)] The recent parameterised approach of \cite{Cardoso:2019mqo, McManus:2019ulj}, which is particularly well-suited to numerically determine the shift induced by the EFT in low $\ell$ frequencies.
This approach parameterises the potential in the ``horizon'' master equation~\eqref{eqn:master_eom_H} with the following large $r$ expansion, 
\begin{align}
 \delta \tilde{V}_{\ell}^{\pm} ( \omega, r )
 = \frac{1}{ \tilde{r}_s^2 } \sum_{k=0}^{\infty} \alpha_{k \ell}^{\pm} ( \omega ) \left( \frac{ \tilde{r}_s }{r}  \right)^k \; .
 \label{eqn:alpha_def}
\end{align}
The shift in the QNM frequencies at leading order in $\delta \tilde{V}_{\ell}^{\pm}$ is then\footnote{
Note that including the $\tilde{c}_s^2$ rescaling of the frequency improves the convergence of this series approximation, since it effectively removes a term of the form $\omega^2/\tilde{f}$ from $\delta \tilde{V}_{\ell}^{\pm}$, which would otherwise populate every $\alpha_{k \ell}^{\pm}$ coefficient.
},
\begin{align}
\frac{ \tilde{r}_s \omega_{0 \ell}^{\pm} }{ \tilde{c}_s} = \tilde{r}_s \, \tilde{\omega}^{\rm GR}_{0 \ell} + \sum_{k=0}^{\infty}  e^{\pm}_{k \ell}  \, \alpha^{\pm}_{k \ell} ( \tilde{\omega}^{\rm GR}_{0 \ell}  )  \; , 
\end{align}
where $\tilde{\omega}_{0 \ell}^{\rm GR}$ is the QNM frequency of a black hole of size $\tilde{r}_s$ in GR, which obeys $\tilde{r}_s \tilde{\omega}_{n \ell}^{\rm GR} = r_s \omega_{n \ell}^{\rm GR}$. 
In terms of the GR Schwarszchild radius $r_s = 2 GM$ and corresponding $\omega_{0\ell}^{\rm GR}$, the final result is\footnote{
Note that since $\tilde{c}_s - 1$ and $\tilde{\omega}^{\rm GR}_{n \ell} - \omega_{n \ell}^{\rm GR}$ both $\sim \mathcal{O} ( \delta V)$, such factors can be ignored in the sum since $\alpha$ is already $\mathcal{O} \left( \delta V \right)$.
},
\begin{align}
 r_s \, \delta \omega_{0 \ell}^{\pm} =  \left(  \frac{ \tilde{c}_s r_s }{ \tilde{r}_s } -1 \right) \, r_s \,  \omega^{\rm GR}_{0 \ell} + \sum_{k=0}^{\infty} e^{\pm}_{k \ell}  \, \alpha^{\pm}_{k \ell} ( \omega^{\rm GR}_{0 \ell}  )   \; , 
 \label{eqn:QNM_para}
\end{align}
Lists of the ``basis'' coefficients $e_{k \ell}$ have been computed and provided online \cite{Cardoso:2019mqo}. 
 
\item[(2)] The popular WKB approach \cite{Iyer:1986np, Guinn:1989bn, Konoplya:2003ii, Konoplya:2004ip, Matyjasek:2017psv, Konoplya:2019hlu}, which provides simple analytic approximations that become exact at large $\ell$. 
This approach makes use of the potential and its derivatives evaluated at its maximum (i.e. the location of the potential barrier)\footnote{
Specifically, $r_{\rm max}$ is defined so that $\tilde{W}_{\ell,1}^{\pm} = 0$ (and thus depends on both $\ell$ and the parity). 
},  
\begin{align}
 \tilde{W}^{\pm}_{\ell, k} ( \omega )  \equiv   \left. \frac{ \partial^k   }{\partial \tilde{r}_*^k } \left[ 
 \tilde{f} \tilde{V}_{\ell}^{\pm} + \tilde{f} \delta \tilde{V}_{\ell}^{\pm} 
 \right] \right|_{ r_{\rm max} } 
\end{align}
to construct an approximation to the QNM frequency.
The zeroth (second) derivative determines $\text{Re} \, \omega_{n \ell}^{\pm}$ ($\text{Im} \, \omega_{n \ell}^{\pm}$) exactly in the large $\ell$ limit,
\begin{align}
 \text{Re} \left[ \frac{ \omega_{0 \ell}^{\pm} }{ \tilde{c}_s } \right] &=  j \sqrt{ \tilde{W}_{\ell,0}^{\pm} } + \mathcal{O} \left( j^0 \right) \;,  
 & \text{Im} \left[ \frac{ \omega_{0 \ell}^{\pm} }{ \tilde{c}_s } \right] &= - N \sqrt{ 
 \frac{ - \tilde{W}_{\ell,2}^{\pm} }{8 \tilde{W}_{\ell,0}^{\pm} } 
 } + \mathcal{O} \left( j^{-1} \right) \;,  
\end{align} 
where $j = \ell (\ell +1)$ and $N= 2n+1$.
The third and fourth (fifth and sixth) derivatives determine the subleading correction to $\text{Re} \, \omega_{0 \ell}^{\pm}$ ($\text{Im} \, \omega_{0 \ell}^{\pm}$)---details can be found in Appendix~\ref{app:QNM}.
To extract the shift in the QNM frequencies due to the EFT interactions, one can expand these functions of $\tilde{W}_{\ell, k}^{\pm}$ to linear order in $\delta \tilde{V}_{\ell}^{\pm}$ (remembering to account for the fact that $r_{\rm max}$ also depends on $\delta \tilde{V}^{\pm}_{\ell}$), which gives an approximation of the form, 
\begin{align}
 \frac{   \omega_{0 \ell}^{\pm}  }{ \tilde{c}_s } = 
 \tilde{\omega}_{0 \ell}^{\rm GR}
 + \sum_{k=0} \beta_{k \ell}^{\pm} \; \delta \tilde{W}_{\ell, k}^{\pm} ( \tilde{\omega}_{0 \ell}^{\rm GR} )  + \mathcal{O} \left( \, \left( \delta V_\ell^{\pm} \right)^2 \, \right)
\end{align}
where $\beta_{k\ell}^{\pm}$ are constant coefficients fixed by the GR potential. 
At zeroth order in $\delta \tilde{V}_{\ell}^{\pm}$, one finds the QNMs of a black hole of size $\tilde{r}_s$ in GR, which obey $\tilde{r}_s \tilde{\omega}_{n \ell}^{\rm GR} = r_s \omega_{n \ell}^{\rm GR}$. 
The final WKB solution is then,
\begin{align}
r_s \delta \omega_{0 \ell}^{\pm} = \left( \frac{ \tilde{c}_s r_s }{ \tilde{r}_s} - 1 \right) \, r_s \, \omega_{0 \ell}^{\rm GR} + \sum_{k=0} \beta_{k \ell}^{\pm} \; \delta \tilde{W}_{\ell, k}^{\pm} ( \omega_{0 \ell}^{\rm GR} ) \; . 
\label{eqn:QNM_WKB}
\end{align} 

\end{itemize}
Comparing the two approaches in their region of overlap $\ell \approx 10$ is a useful consistency check, and together they provide accurate results for every $\omega_{n \ell}$ in the GREFT at this order in derivatives. 
While it is typically the low $\ell$ and $n$ modes that matter most for phenomenology\footnote{
The reason is simple: low $\ell$ modes are typically excited with largest initial amplitude and then have the slowest decay (i.e. $|\text{Im} \, \omega_{n\ell}^{\rm GR}|$ increases with $n$ and $\ell$).
}, from the theoretical point of view 
there could be useful causality constraints coming from QNMs of any $n$ and $\ell$. 
In fact, since the strongest IR causality constraints come from large $\ell$, one might naively expect the same to be true for QNM causality. 
This expectation will be borne out by the $c_1$ potential for odd modes: while the shift in the fundamental QNM frequencies for $\ell = 2,3, 4$ were computed numerically in \cite{Cardoso:2018ptl} and are all consistent with causality, the computation below shows that once $\ell \geq 6$ the correction to the imaginary part of $\omega_{0 \ell}$ \emph{changes sign} and the high $\ell$ modes can lead to a resolvable violation of causality if $c_1 \epsilon_s^6$ is too large.

\paragraph{GR QNMs.}
For later comparison, recall that the WKB approximation for the GR QNM frequencies at large $\ell$ is,
\begin{align}
 \text{Re} \left[ \omega_{n \ell}^{\rm GR} \right] &= 
\frac{2 }{ 3 \sqrt{3} } \sqrt{j}  -  \frac{ 493 + 15 N^2 }{ 648 \sqrt{3} } \frac{1}{\sqrt{j}}  + \mathcal{O} \left( \frac{1}{j^{3/2} } \right)  \nonumber \\ 
 \text{Im} \left[ \omega_{n \ell}^{\rm GR} \right] &= 
 - \frac{ N }{ 3 \sqrt{3} } +  \frac{ N (6599 - 235 N^2) }{ 46656 \sqrt{3} } \frac{1}{j} + \mathcal{O} \left( \frac{1}{j^2} \right) \; . 
 \label{eqn:QNM_GR}
\end{align}
were $j = \ell (\ell+1)$ and $N = 2n + 1$. 
The frequencies at low $\ell$ have been determined numerically to high precision: the first few significant figures are listed in Table~\ref{tab:QNM_delta}.
They agree well with the WKB approximation even at relatively low $\ell$. 
Note that at high $\ell$, the QNM frequency $\omega_{n \ell}^{\rm GR}$ satisfies \eqref{eqn:g_def} with $\gamma = 1$. 
These QNM solutions are therefore probing the regime in which the time delay calculation breaks down, and also the regime in which causality constraints are expected to be the strongest.

\paragraph{GREFT QNMs.}
In the GREFT, the QNM frequencies, $\omega_{n \ell}^{\pm}$ are shifted relative to GR by the EFT interactions. This relative shift takes the form,
\begin{align}
\text{Re} \left[ \omega^{\pm}_{n \ell} \right] &= \text{Re} \left[ \omega_{n \ell}^{\rm GR} \right] \left( 
1 + \epsilon_s^4 \delta_3 \mathcal{R}^{\pm}_{n \ell} + \epsilon_s^6 \delta_4 \mathcal{R}^{\pm}_{n\ell} + ...
\right)   \nonumber \\ 
\text{Im} \left[ \omega_{n \ell}^{\pm} \right] &= \text{Im} \left[ \omega_{n \ell}^{\rm GR} \right] \left( 
1 + \epsilon_s^4 \delta_3 \mathcal{I}^{\pm}_{n \ell} + \epsilon_s^6 \delta_4 \mathcal{I}^{\pm}_{n\ell} + ...
\right)   
\label{eqn:QNM_EFT_exp}
\end{align}
where $\epsilon_s = 1/(G M \Lambda)$ as before. 
Using the parameterised approach \eqref{eqn:QNM_para} with the potentials given in Appendix~\ref{app:master} produces the numerical values given in Table~\ref{tab:QNM_delta}. 
For $\ell = 2, 3, 4$, these agree with the values previously found in \cite{deRham:2020ejn}\footnote{
Note that \cite{deRham:2020ejn} uses an expansion parameter $\mu = \left( \frac{1}{r_s \Lambda} \right)^4 = 2^{-4} \epsilon_s^4$, so their quoted values are a factor of $2^4$ larger.
}.
Using the WKB approach \eqref{eqn:QNM_WKB} gives the exact shift in the large $\ell$ limit (given in the final column of Table~\ref{tab:QNM_delta}), as well as the first correction in $1/j$ (given in (\ref{eqn:first_QNM}--\ref{eqn:last_QNM})). 
Note that the subleading corrections for the $c_1$ odd potential are large for $\ell \leq 9$ and are needed for good agreement. 
For instance, $\delta_4 \mathcal{I}_{0,9}^- = - 0.2278 c_1$ from the parameterised approach, while the WKB result for this $\ell$ and $n$ gives,
\begin{align}
\delta_4 \mathcal{I}_{0, 9}^-  = c_1 \left(  -0.3674 + 0.1399 +  \mathcal{O} \left( \frac{1}{ (90 )^2 }  \right)  \right) \approx 0.2275 c_1
\label{eqn:c1_subleading_large}
\end{align}
and agrees with the numerics up to the stated error from subleading $1/j^2$ corrections.

\paragraph{Implications of IR causality.}
One of the main observations in this work is that the sign of the $\delta \mathcal{I}^{\pm}_{n\ell}$ shifts in the imaginary part of the QNM frequencies is closely correlated with the sign of the time advances computed in the previous subsection. 
Consequently, imposing the constraints from IR causality (no resolvable time advance) leads to a constraint on the direction in which the QNMs can be shifted. 

For instance, the fastest growing contributions at large $j$ (at this order in $\epsilon_s$) to the decay lifetimes of parity-odd and parity-even black holes are\footnote{
Note that $ \text{Re} \left[  \omega_{n \ell}^{\rm GR} \right] \delta_n \tau^{\pm}_{\ell} = 4 \pi \sqrt{j} \,\delta_n \mathcal{I}_\ell^\pm + \mathcal{O} (j^{-1})$ at large $j$, which will be used repeatedly below.
}:
\begin{align}
 \text{Re} \left[  \omega_{n \ell}^{\rm GR} \right]  \delta \tau_{\ell}^{+}  &=  - c_1 \epsilon_s^6 \; j^{3/2} \, \frac{ 5120 \pi }{ 2187} + \mathcal{O} \left( \sqrt{j} \right) \; ,  \nonumber \\ 
  \text{Re} \left[  \omega_{n \ell}^{\rm GR} \right]  \delta \tau_{\ell}^{-}  &= - 4 c_2 \epsilon_s^6 \;  j^{3/2} \, \frac{ 5120 \pi }{ 2187} + \mathcal{O} \left( \sqrt{j} \right) \; . 
  \label{eqn:c_dtau}
\end{align}
The positivity bounds from UV causality, $c_1 > 0$ and $c_2 > 0$, immediately imply that both even and odd perturbations decay with a shorter lifetime than in GR, i.e. that the QNM frequencies are shifted deeper into the lower-half of the complex plane. 
This is also the result of applying the simplest IR causality bounds: in particular, it was also the large $j$ limit of the time delay which dominated the IR causality constraint.

While it only scales as $j^0$, the shift from the cubic $b_1$ interaction is leading order in $\epsilon_s$. 
Focussing on this $\delta_3 \mathcal{I}_\ell^{\pm}$ contribution, the shift in the QNM lifetime is proportional to $+b_1$ for the odd modes and proportional to $-b_1$ for the even modes, with approximately equal constants of proportionality:
\begin{align}
 \text{Re} \left[  \omega_{n \ell}^{\rm GR} \right]  \delta_3 \tau_{\ell}^{\pm}   \approx \mp 20 \sqrt{j}   \, b_1 \epsilon_s^4  + \mathcal{O} \left(\frac{1}{\sqrt{j}} \right) 
\label{eqn:b1_dtau}
\end{align}
So regardless of the sign of $b_1$ one of these modes will increase the decay time of perturbations relative to GR and make the black hole less stable. 
However, imposing the IR causality constraint $|b_1 | \epsilon_s^6 \lesssim \mathcal{O} \left( 10^{-5} \right)$ from \eqref{eqn:b1_dt_bound} means that this change in the decay lifetime is not resolvable (in the sense of \eqref{eqn:dtau_resolv}) until $\ell \gtrsim 10^4$.
Does such a large $\ell$ remain under perturbative control in the EFT?

\paragraph{Regime of validity.}
The perturbativity bound \eqref{eqn:w_max_strong} on time derivatives in the EFT, when specialised to the QNM problem in which $r \approx r_s$ and $|\omega| \approx \text{Re} \left[ \omega_{n \ell}^{\rm GR} \right]$, gives the upper bound,
\begin{align}
  \epsilon_s^2  &<  \frac{ 4 \epsilon_{\omega} }{ \text{Re} \left[ \omega_{n\ell}^{\rm GR} \right] r_s } 
   = \frac{ 6 \sqrt{3} \epsilon_\omega }{  j^{1/2} } + \mathcal{O} \left( \frac{1}{j^{3/2}} \right) \qquad \text{(strong perturbativity)}   \; 
   \label{eqn:eps_max_QNM}
\end{align}
In order to assess whether an apparent violation of QNM causality is really problematic for the EFT, one must check that it is occurring for a value of $\epsilon_s$ and $j$ that satisfies \eqref{eqn:eps_max_QNM}. 
For instance, assuming that both $\epsilon_\omega$ and $b_1$ are order unity, in order for the delay time \eqref{eqn:b1_dtau} from the $b_1$ interaction to be problematic, $\epsilon_s$ must lie in the range\footnote{
The exact result at large $\ell$ is $-\frac{19683}{ 67634176 \pi^2 }   \lesssim  b_1 \epsilon_s^4 \lesssim \frac{19683}{ 82011136 \pi^2 } $, but since there is an $\mathcal{O}(1)$ numerical ambiguity in the resolvability condition \eqref{eqn:dtau_resolv} these should be interpreted as rough orders of magnitude as in \eqref{eqn:b1_intro}.
}: 
\begin{align}
\left( \frac{ 6 \sqrt{3} }{  j^{1/2} } \right)^2  \gtrsim  | b_1| \epsilon_s^4  \gtrsim  \frac{1}{20 \sqrt{j}} \; .
\end{align}
Since this can only occur for $b_1 \epsilon_s^6 \gtrsim 2 \times 10^{-5}$, one can conclude that the IR causality bound $|b_1| \epsilon_s^4 \lesssim \mathcal{O} \left( 10^{-5} \right)$ effectively guarantees QNM causality for this interaction.

\paragraph{Implications of QNM causality.}
While UV or IR causality constraints imply that \eqref{eqn:c_dtau} is compatible with QNM causality, one could turn the question around and ask what is the range of $\{ c_1, c_2 \}$ for which \eqref{eqn:c_dtau} satisfies the QNM causality condition? One might expect additional information / constraining power to be contained in the QNM causality condition since it is probing the effective potential in a different regime (namely $r \approx r_s$). 

Comparing \eqref{eqn:c_dtau} with the perturbativity condition \eqref{eqn:eps_max_QNM} gives the result \eqref{eqn:intro_c_lower}\footnote{
Although the precise result obtained is $c_1 \epsilon_\omega^2 \gtrsim \frac{ 9 \sqrt{3} }{ 40960 \pi }$ and $c_2 \epsilon_\omega^2 \gtrsim \frac{ 9 \sqrt{3} }{ 163840 \pi }$, given the numerical ambiguity in \eqref{eqn:dtau_resolv} it again makes sense to report rough order of magnitude estimates.
}, namely $c_1 \epsilon_{\omega}^3 \gtrsim - 1.2 \times 10^{-4}$ and $c_2 \epsilon_{\omega}^3 \gtrsim - 3.0 \times 10^{-5}$. 
Remarkably, these show close agreement with the time delay values from scattering waves close to the black hole (at $\gamma = 0.9$). 
While that time delay calculation was subject to potentially large corrections from the absorption of waves by the black hole, the QNM calculation has no such corrections: it is exact at this order in $\epsilon_s$ and at large $\ell$. 
In one sense, the QNM causality condition can be used to place the IR causality condition on a firmer footing (it is at least corroborating evidence that corrections to \eqref{eqn:dT_from_dA} remain small at $\gamma = 0.9$). In another sense, the IR causality condition, which stems from well-understood semiclassical physics, can be viewed as evidence for the QNM causality conjecture. 

Finally, there is the question of whether the QNM causality condition is respected at finite values of $\ell$. 
Just as with the time delay, it turns out that the shift from the $c_1$ interaction violates QNM causality in the parity-odd sector, which must either be compensated by a sufficiently large $c_2$ interaction or else lead to an upper bound on $\epsilon_s^6$. 
From the numerically determined fundamental QNM frequencies, this violation occurs for all $\ell \geq 6$. When $c_2 / c_1 > 1.1 \times 10^{-3}$ all of the QNMs respect the causality condition. 
For $\mathcal{O} \left( 10^{-4} \right) \lesssim c_2/c_1 \lesssim \mathcal{O} \left( 10^{-3} \right)$, the tightest upper bound on $\epsilon_s$ comes from the $\ell = 9$ mode, which gives
\begin{align}
 \epsilon_s^6 \lesssim \frac{ 1 }{ 2.862 c_1 - 2688 c_2 } \; . 
\end{align}
As $c_2/c_1 \to 0$, it is higher and higher $\ell$ modes which give the tightest bound. 
This eventually converges to a decay time
\begin{align}
  \text{Re} \left[  \omega_{n \ell}^{\rm GR} \right]  \delta_4 \tau_{\ell}^{-}   = + \frac{ 28928 \pi c_1 }{ 19683 }  + \mathcal{O} \left( \frac{1}{ j } \right) \; .
\end{align}
when $c_2 = 0$, which would lead to a resolvable causality violation within the EFT regime of validity unless $c_1 \epsilon_s^6 \lesssim 3 \times 10^{-3}$.
This is somewhat stronger than the IR causality bound, but only by a factor of a few.

\begin{sidewaystable}
\centering
\begin{tabular}{| c c | c c c c c c c c c |}
\hline
 & & \multicolumn{9}{c |}{$\ell$} \\ 
   &    & 2 & 3 & 4 & 5 & 6 & 7 & 8 & 9 & $\infty$  \\ 
   \hline
\multicolumn{2}{ |c |}{  $\text{Re} \left[ \omega_{0\ell}^{\rm GR} \right]$ } & +0.7473 & +1.199 & +1.618 & +2.025 & +2.424 & +2.820 & +3.212 & +3.604 & $+\frac{2}{3 \sqrt{3}} \, \ell (\ell+1)$          \\
\multicolumn{2}{ |c |}{  $\text{Im} \left[ \omega_{0\ell}^{\rm GR} \right]$ } & $-$0.1779 & $-$0.1854 & $-$0.1883 & $-$ 0.1897 & $-$0.1905 & $-$0.1910 & $-$ 0.1913 & $-$ 0.1916   &  $-\frac{1}{3 \sqrt{3}} \approx -0.1925$ \\
 \hline
 \multirow{4}{*}{$b_1$} & $\delta \mathcal{R}^-_{0 \ell}$  &  +0.6585 & +0.5715 &  +0.5483 &  +0.5383 & +0.5330 & +0.5299 & +0.5279 &  +0.5265 & $+ \frac{380}{729} \approx + 0.52$ 
   \\  
  & $\delta \mathcal{I}^-_{0 \ell}$ &  +1.485 & +1.446 & +1.428 & +1.421 & +1.417 & +1.415 & +1.413 & +1.413 & $+\frac{1028}{729} \approx +1.41$   \\ 
\cline{3-11} %
  & $\delta \mathcal{R}^+_{0 \ell}$ &  $-$0.3854 & $-$0.4421 & $-$0.4519 & $-$0.4562  & $-$0.4588 & $-$0.4605  & $-$0.4617  & $-$0.4626 & $-\frac{ 340 }{729 } \approx -0.47$  \\ 
   & $\delta \mathcal{I}^+_{0 \ell}$ & $-$1.816 & $-$1.619 & $-$1.568 & $-$1.553 & $-$1.549 & $-$1.548 & $-$1.548 & $-$1.548 & $- \frac{1132}{729} \approx  -1.55$
 \\ 
\hline
\multirow{4}{*}{$c_1$} & $\delta \mathcal{R}^-_{0 \ell}$  &   $-$0.2124 & + 0.01045 & + 0.04949 & + 0.05618 & +0.05587 & + 0.05414 & +0.05234 & +0.05078 & $+\frac{832}{19683} \approx +0.042$
 \\  
  & $\delta \mathcal{I}^-_{0 \ell}$ &  +0.6476 & +0.4344 & + 0.1939 & +0.03069 & $-$0.07512 & $-$0.1453 &  $-$0.1935 & $-$0.2278  &  $-\frac{7232}{19683} \approx - 0.37$
 \\ 
\cline{3-11} %
  & $\delta \mathcal{R}^+_{0 \ell}$ & $-$0.4425 & $-$1.069 & $-$1.761 & $-$2.619 & $-$3.657 & $-$4.874 & $-$6.270 & $-$7.845  & $- \frac{ 64 }{ 729 } \ell ( \ell + 1 )$  
 \\ 
   & $\delta \mathcal{I}^+_{0 \ell}$ & +2.745 & +6.415 & +11.43 & +17.50 & +24.67 & +32.95 & +42.38 & +52.96  &  $+\frac{1280}{2187} \ell ( \ell + 1 ) $ 
 \\ 
\hline
\multirow{2}{*}{$c_2$} & $\delta \mathcal{R}^-_{0 \ell}$  &  $-$2.130 &  $-$4.229 & $-$6.986 & $-$10.46 & $-$14.65 & $-$19.55  & $-$25.16 & $-$31.47 &  $- \frac{ 256 }{ 729 } \ell (\ell_1 ) $   \\  
  & $\delta \mathcal{I}^-_{0 \ell}$ & +15.84 &  + 30.47 & + 49.54 &  +73.14 & +101.4 & +134.198 & +171.7 &  + 213.9 &  $ +\frac{ 5120 }{ 2187} \ell (\ell + 1) $   \\ 
\hline
\end{tabular}
\caption{The fundamental quasi-normal mode frequencies $\omega_{0 \ell}$ for a Schwarszchild black hole in General Relativity for different angular momenta $\ell$, together with the relative shifts \eqref{eqn:QNM_EFT_exp} for parity even/odd modes induced by the three leading GREFT interactions.}
\label{tab:QNM_delta}
\end{sidewaystable}

\section{Observational prospects}
\label{sec:obs}

This section discusses the possibility of measuring the GREFT coefficients with gravitational wave detectors. It will focus on the transient signals from binary mergers that are now routinely measured by the LIGO-Virgo-KAGRA network, and in particular the recent results of \cite{Sennett:2019bpc, Silva:2022srr} which place observational constraints on the GREFT \eqref{eqn:intro_GREFT} with $b_1 = c_2 = 0$ and $c_1 = 1$.
These are two-sided constraints which disfavour a particular range of $\Lambda$. 
The reason for a two-sided bound is the following.
For large enough $\Lambda$, the EFT correction to the waveform is suppressed and there is good agreement with the GR waveform. 
But as $\Lambda$ is lowered, the EFT corrections become larger and at some point become resolvable by GW observatories and hence excluded by the data. 
But once $\Lambda$ is sufficiently low, the cut-off where the EFT breaks down becomes comparable to the scales of the GW event and so there is no longer any constraint because the EFT cannot be applied to this particular data: measurements at a different scale would be required in order to constrain the theory in this regime.   
Since the causality bounds derived above place a further restriction on the range of $\Lambda$ and $M$ for which the EFT is valid, they will change one side of this two-sided bound and impact the range of parameter space which can be reliably probed using any given GW event.

\paragraph{Power counting.}
The size of the EFT corrections to the GWs emitted by a binary can be estimated by simple power counting arguments.
Recall that the Einstein-Hilbert action $\sqrt{-g} R \sim h \partial^2 h$ provides the kinetic term for metric perturbations and the coupling to matter sources $h \sim GM/r$. 
The relative correction to GR from the cubic and quartic EFT interactions can then be estimated as, 
\begin{align}
  \frac{ \tfrac{1}{\Lambda^4} \sqrt{-g} R^3 }{ \sqrt{-g} R } &\sim \frac{ \tfrac{1}{\Lambda^4} ( \partial^2 h )^3 }{ h \partial^2 h } \sim \frac{1}{ (\Lambda r )^4 }  \frac{G M}{r} \equiv \epsilon_3   \nonumber \\ 
  \frac{ \tfrac{1}{\Lambda^6} \sqrt{-g} R^4  }{ \sqrt{-g} R } &\sim \frac{ \tfrac{1}{\Lambda^6} ( \partial^2 h )^4 }{ h \partial^2 h } \sim \frac{1}{ (\Lambda r )^6 } \left( \frac{G M}{r} \right)^2  \equiv \epsilon_4 
\label{eqn:pow_count_pheno}
\end{align}
where here $R^n$ denotes any contraction of $n$ Riemann tensors and $\partial \sim 1/r$ is a typical length scale of the system (e.g. the orbital separation in the case of a binary). 
A more refined version of \eqref{eqn:eps_expansion} is therefore:
\begin{align}
 \mathcal{O}_{\rm GREFT} = \mathcal{O}_{\rm GR} \left( 1 + \epsilon_3 \delta \mathcal{O}_3 + \epsilon_4 \delta \mathcal{O}_4  + ... \right)
\end{align}
Since $GM/r \sim v^2$ for bound orbits by the virial theorem, these corrections are formally 5PN and 8PN in the post-Newtonian expansion, 
\begin{align}
 \epsilon_3 &\sim \epsilon_s^4 \left( \frac{v^2}{c^2} \right)^{5}  \;\; \Rightarrow \;\; \text{5PN}  \nonumber \\ 
 \epsilon_4 &\sim \epsilon_s^6 \left( \frac{v^2}{c^2}  \right)^{8}  \;\; \Rightarrow \;\; \text{8PN} \; . 
\end{align}
However, as can be seen in \eqref{eqn:pow_count_pheno}, if the orbital separation of the binary $r \sim 1/\Lambda$, then this acts as a very large numerical coefficient which can make $\epsilon_3$ and $\epsilon_4$ important already at 1PN and 2PN respectively. 
While an EFT with $\epsilon_s \gg 1$ will not resolve the black hole horizon reliably, what matters in the context of GW phenomenology is whether the EFT is under control for the orbital separation of the binary. 
This separation decreases with time as the objects lose energy through GW emission, leading to the characteristic `chirp' in the GW frequency $f$. 
The EFT can therefore capture the portion of the signal for which,
\begin{align}
 \left( G M \pi f \right)^{2/3} \lesssim \frac{1}{\epsilon_s}  \; ,
 \label{eqn:f_max}
\end{align}
where $M$ is the chirp mass.

In summary, if the scale $\Lambda$ in the EFT is such that that $\epsilon_s < 1$ and the black hole horizons are resolved, then the effect of these corrections on the binary inspiral will be at least 5PN (or 8PN) suppressed. However, if $\Lambda$ is much lower so that $\epsilon_s \gg 1$, then the EFT corrections to the inspiral part of the waveform can be as large as 1PN (or 2PN)\footnote{
There is actually a cancellation that takes place for the $b_1$ cubic interaction which means it does not contribute until one higher PN order than naive power counting would suggest \cite{Endlich:2017tqa}.
} and remain under perturbative control until the frequency exceeds \eqref{eqn:f_max}.   
Of course, taking $\epsilon_s \gg 1$ must be done with care, since this will usually lead to the EFT breaking down at some finite distance from the horizon. The above causality arguments are one way to quantify this breakdown.

\paragraph{Implementing causality bounds.}
In the previous section, causality led to a number of upper bounds on $\epsilon_s$. Denoting the strongest of these as $\epsilon_s^{\rm caus}$, then the GREFT may only be reliably applied to portions of the GW signal for which\footnote{
In practice, the frequency $f$ in \eqref{eqn:eps_max_obs} characterises the lowest frequency at which the GW is detected: from the bounds given in \cite{Sennett:2019bpc}, these are approximately $f \approx 58$ Hz for GW151226 and $f \approx 48$ Hz for GW170608. The red regions in Figure~\ref{fig:obs} uses $\epsilon_s < \frac{1}{( 4 G M \pi f )^{2/3}}$ so that the upper bound on $1/\Lambda$ coincides with that of \cite{Sennett:2019bpc} with $f_{\rm high} = 0.25 f_{\Lambda}$. 
}:
\begin{align}
\epsilon_s \lesssim \epsilon_s^{\rm max} = \begin{cases}
\text{min} \left(  \frac{1}{ (G M \pi f )^{2/3} } , \; \epsilon_s^{\rm caus}    \right)   &\text{for inspiral},  \\
 \text{min} \left(  1 ,  \; \epsilon_s^{\rm caus}    \right)    &\text{ for ringdown}  .
\end{cases}
\label{eqn:eps_max_obs}
\end{align}
Any GW event which is consistent with GR (i.e. consistent with $\frac{1}{\Lambda} = 0$) will therefore lead to constraints on the GREFT scale $\Lambda$ of the form, 
\begin{align}
  \frac{1}{\Lambda} \in \left[ d_{\rm min} , G M \epsilon_s^{\rm max}   \right]  \; , 
\end{align}
where $d_{\rm min}$ is the smallest $1/\Lambda$ at which the EFT corrections become excluded by the event's waveform and the upper bound corresponds to the event leaving the EFT's regime of validity (at which point that particular GW event can no longer be used to constrain the EFT, since it is not probing \eqref{eqn:intro_GREFT} but rather its UV completion). A good estimate for $M$ would be the total mass of the binary (for inspiral measurements) or the mass of the final remnant (for ringdown measurements). 

The results of \cite{Sennett:2019bpc, Silva:2022srr}, which considered the GREFT \eqref{eqn:intro_GREFT} with $b_1 = c_2 = 0$ and $c_1 = 1$, strongly disfavour\footnote{
More precisely, \cite{Silva:2022srr} excludes this region at the 90$\%$ credible level and \cite{Sennett:2019bpc} finds a Bayes factor $\log B \lesssim -5$ in this region.
} the regions,
\begin{align}
 \frac{1}{\Lambda} &\in  \left[ 65 , \; 27.6 \epsilon_s^{\rm max}   \right] \; \text{km} \qquad &\text{(GW170608 \cite{Sennett:2019bpc})}  \nonumber \\ 
  \frac{1}{\Lambda} &\in \left[ 125 , \; 31.9 \epsilon_s^{\rm max}  \right] \; \text{km} \qquad &\text{(GW151226 \cite{Sennett:2019bpc})}  \nonumber \\ 
  \frac{1}{\Lambda} &\in \left[ 51.3 , \; 88.4 \epsilon_s^{\rm max}  \right] \; \text{km}  \qquad &\text{(GW150914 \cite{Silva:2022srr})}  \nonumber \\
    \frac{1}{\Lambda} &\in \left[ 55.5 , \; 86.7 \epsilon_s^{\rm max}  \right]\; \text{km}  \qquad &\text{(GW200129 \cite{Silva:2022srr})}    
    \label{eqn:d_ranges}
\end{align}
Notice that \cite{Sennett:2019bpc} chose two of lightest events since their constraining power was coming only from the inspiral part of the waveform, while \cite{Silva:2022srr} chose two of the loudest events since they adopted an inspiral-merger-ringdown approach which was also sensitive to finite-size effects.

Crucially, the cut-off $\epsilon_s^{\rm caus} \approx 0.38$ from QNM causality seems to render all four of these GW events outside of the EFT's regime of validity.  By contrast, IR causality alone requires only that $\epsilon_s^{\rm caus} \approx 0.64$, in which case the two events of \cite{Silva:2022srr} are perhaps marginal.
Such conclusions are of course sensitive to the precise $\mathcal{O} (1)$ numerical factor in the bounds. 
But the point is that a low cut-off from causality on $\epsilon_s$ can severely limit the constraining power of GW events by pushing them outside of the EFT's regime of validity 

One resolution is to include additional EFT interactions which can relax the causality bounds. 
For example, under the assumption that adding a $c_2$ interaction does not lead to any significant changes in the emitted waveform (which is a good approximation if the spins of the black holes are small), then \eqref{eqn:d_ranges} would translate simply into constraints on the $\{ c_1, c_2 \}$ parameter space. 
Once $c_2 \gtrsim 10^{-3} c_1$ the QNM causality bound disappears, and once $c_2 \gtrsim c_1$ the IR causality bound also disappears. 
This is shown in Figure~\ref{fig:obs}. 

Finally, thanks to the results of \cite{Emond:2019crr, AccettulliHuber:2020oou, Brandhuber:2019qpg, AccettulliHuber:2020dal} and \cite{deRham:2020ejn}, the imprint of the cubic GREFT interaction in the inspiral and ringdown is well understood.
The events GW150914 and GW200129 exclude the following ranges at the 90$\%$ credible level when $b_1$ is the dominant interaction \cite{Silva:2022srr}:
\begin{align}
  \frac{b_1^{1/4}}{\Lambda} &\in \left[ 38.2 , \; 88.4 \epsilon_s^{\rm max}  \right] \; \text{km}  \qquad &\text{(GW150914)}  \nonumber \\
    \frac{b_1^{1/4}}{\Lambda} &\in \left[ 42.5 , \; 86.7 \epsilon_s^{\rm max}  \right]\; \text{km}  \qquad &\text{(GW200129)}    
\end{align}
However, given the tight constraints on $\epsilon_s$ from causality, these two events are not likely to be resolved by the EFT unless this interaction is accompanied by a comparable higher-derivative interaction. It would be interesting to explore this further in future.

\section{Discussion}
\label{sec:disc}

\paragraph{Summary.}
The General Relativity Effective Field Theory (GREFT) parameterises the effects of heavy degrees of freedom beyond GR in a systematic way that is amenable to perturbation theory and can be compared with GW data. 
The GREFT coefficients have recently been constrained using causality in both the UV (via scattering amplitude analyticity) and in the IR (via semiclassical time delay on a black hole background).
This work has put forward quasi-normal modes as a new way to implement causality constraints on this EFT. 
In contrast to existing causality bounds from the Eisenbud-Wigner time delay, which are sensitive only to large impact parameters $r \gg 3/2 r_s$, the new QNM constraints probe a different regime near to the horizon $r \approx r_s$.
In some cases they lead to even stronger upper bounds on the GREFT parameters, but for the most part they reproduce almost exactly the constraints from IR causality. 
This suggests that QNM causality is a reliable condition which may be used to constrain other gravitational EFTs. 
In light of these findings, there are a number of future directions to explore (some of which are listed below). 

\paragraph{Higher-order interactions.}
Consider adding to \eqref{eqn:intro_GREFT} a single higher-order interaction\footnote{
Of course, there are a number of other contractions which would naturally enter at this order and it seems unlikely (barring some very finely tuned UV physics) that only this single operator would be generated in the EFT. 
But this example neatly demonstrates the power of considering causality constraints on non-trivial backgrounds, since it can in principle place constraints on interactions of arbitrarily high order. 
}
\begin{align}
 \mathcal{L}_{2q} \supset  C_q \left( R^{(2)} \right)^q
\end{align}
for a fixed integer $q$. 
The black hole background for this interaction is given in \cite{Cardoso:2018ptl}. 
Remarkably, QNM causality implies that $C_q$ must be positive, regardless of the value of $q$. 
To see this, notice that the highest derivative of the metric fluctuation to appear in the master equation is\footnote{
Note that several seemingly four-derivative terms can be discarded since $h_{\mu\nu}$ is transverse and traceless, and the leading order equation of motion can be used to replace any $\Box h_{\mu\nu} $ with terms that have fewer derivatives.
},
\begin{align}
 \frac{\delta }{\delta g_{\mu\nu}} \left(  \sqrt{-g} \mathcal{L}_{2q} \right) \supset  \sqrt{-g}  C_q \left( R^{(2)} \right)^{q-2} \left[
 16 q (q-1 )  R^{\mu \rho \nu \sigma} R^{\alpha \tau \beta \upsilon} \nabla_{\rho} \nabla_{\sigma} \nabla_{\tau} \nabla_{\upsilon} h_{\alpha \beta}
 \right]
\end{align}
This does not contribute to the potential for odd modes, but for the even modes it gives a contribution $\epsilon_s^{ 4q - 2 } \delta_{2q} V_{\ell}^+$ to the potential, where
\begin{align}
 \delta_{2q} V^+_{\ell} \supset -  C_q \left( \frac{12 r_s^6}{r^6} \right)^{q - 2}  \frac{9 j^2}{4 r_s^2} \left( \frac{r_s}{r} \right)^{10} \; . 
\end{align}
All other contributions from this interactions are $\mathcal{O} ( j )$ or smaller in the large $j$ limit. 
The shift in the QNM is then,
\begin{align}
 \delta_{2q} \mathcal{R}_{0 \ell}^{+}  &= - C_q \frac{8^q j }{ 729 } + \mathcal{O} ( j^0 ) \; ,   \nonumber \\
 \delta_{2q} \mathcal{I}_{0 \ell}^{+} &= + C_q \frac{ 8^q ( q + 3  ) ( q + 6  )  }{ 4374 } \; j + \mathcal{O} ( j^0 ) \; . 
\end{align}
In order for the lifetime of the large-$j$ QNMs to decrease, $C_q$ must be positive. 
This positivity would not have been easy to prove using UV causality, since such interactions do not contribute to the $4$-point scattering amplitudes until a high loop order. 
It would be interesting to further explore the master equation and effective potential for perturbations around this background to see what further constraints QNM causality / IR causality might impose.
Studying interactions that contain derivatives of the Riemann tensor \cite{Aguilar-Gutierrez:2023kfn}, or general contractions of the type $( R^{(n)} )^q$, would also be interesting.
The main point here is that higher-point EFT interactions can be constrained by considering non-trivial backgrounds \cite{Chandrasekaran:2018qmx,Melville:2022ykg,Davis:2021oce, Serra:2023nrn}, and QNM causality is a further example of this.

\paragraph{Other black hole backgrounds.}
This work has focussed on the simplest (stationary, spherically symmetric) black hole spacetime in the GREFT, but the GREFT corrections to other black hole backgrounds are known and it would be interesting to investigate whether QNM causality for those black holes places additional constraints on the EFT. 
The obvious next step would be to study the Kerr background that describes a spinning black hole \cite{Cano:2019ore}, since phenomenologically speaking $c_2$ has a significantly larger effect on spinning objects. 
One key limitation of Figure \ref{fig:obs} is that it assumes the black holes have low enough spin that the $c_2$ interaction can be neglected: a more refined analysis in future would establish causality bounds for spinning black holes and remove this assumption.   

Given the recent progress in calculating Schwarschild-de Sitter quasi-normal modes\footnote{
Note that other interactions beyond \eqref{eqn:intro_GREFT} may become non-redundant on backgrounds which are not Ricci-flat.
}, it would also be interesting to investigate how QNM causality might constrain cosmological field theories and compare this with other recent approaches to implementing causality bounds in a cosmological setting \cite{Melville:2019wyy, Grall:2021xxm,deRham:2021fpu,Davis:2021oce, Melville:2022ykg, Salcedo:2022aal, Agui-Salcedo:2023wlq, CarrilloGonzalez:2023rmc}. 

Finally, it would be interesting to combine QNM causality with a model-agnostic approach like that of \cite{Glampedakis:2019dqh, Silva:2019scu}, which could place constraints on the ringdown waveform without the need to specify the underlying field content.

\paragraph{Parity-violating interactions.}
At eighth order in derivatives, \eqref{eqn:intro_GREFT} is not the most general EFT one could consider. 
There is also a single parity-violating cubic interaction and a single parity-violating quartic interaction. 
These have interesting effects on the QNM spectrum\footnote{
Parity-violation also leads to interesting effects in the GW propagation \cite{Callister:2023tws}.
}: in particular they can mix odd and even perturbations and lead to ``running'' of the QNM frequencies \cite{McManus:2019ulj,Cano:2023jbk}. 
It would therefore be interesting to study how to apply QNM causality in that case and whether it can constrain the parity-violating GREFT coefficients.

\paragraph{Including matter fields.}
In GR, the QNM spectrum for perturbations of a spin-$s$ matter field on a black hole background are known. One future direction would be to investigate how these are shifted by interactions like \eqref{eqn:intro_GREFT}, or (perhaps better) whether the effect of a matter self-interaction\footnote{
To constrain interactions that are non-linear in the matter fields would require those fields to acquire a non-trivial background. 
While exact scalar-tensor black hole solutions are known in only a handful of cases, it may be enough to consider just the leading corrections on top of a Schwarszchild background. 
} like $\lambda (\nabla \phi )^4$ on the scalar QNMs might connect QNM causality with the prototypical positivity bound from UV causality, $\lambda > 0$. 
There is no shortage of interactions beyond \eqref{eqn:intro_GREFT} which could be considered next, where the constraints from QNM causality could be compared with both experimental constraints and other theory constraints like UV/IR causality. 

In particular, \cite{Langlois:2021aji, Roussille:2023sdr} have recently developed a first-order formalism to determine the quasi-normal modes of modified gravity theories in which the perturbation equations are not of second-order Schrodinger form.
Exploring how the QNM frequencies are shifted by EFT corrections in that framework (and to what extent QNM causality overlaps with other causality constraints) would be a natural way to extend this discussion to a much wider range of theories.

\paragraph{Connection with other conjectures.}
This work has focussed on connecting quasi-normal modes with the constraints from UV and IR causality. 
It would be interesting to explore what restrictions are placed by other consistency conditions such as the swampland conjectures (and how these interface with QNM causality). 
For example the Weak Gravity Conjecture has a close connection with UV causality \cite{Hamada:2018dde, Bellazzini:2019xts, Arkani-Hamed:2021ajd} and was recently connected to the Love numbers of GREFT black holes in \cite{DeLuca:2022tkm}.
Furthermore, positivity of the time delay is just one of an infinite family of causality bounds \cite{Bellazzini:2022wzv}, and it would be interesting to investigate whether they also have a QNM counterpart.

\paragraph{Proving the conjecture.}
Finally, it is worth bearing in mind that no rigorous proof has been given for QNM causality. 
For one thing, QNMs are found using different boundary conditions to the causal propgators (for which analyticity is immediate). An important next step would be to establish whether EFT corrections must always obey the QNM causality condition \eqref{eqn:intro_QNM_causality}, for instance by connecting more carefully with the familiar analyticity arguments. 
In fact, from a holographic perspective the quasi-normal modes of a black hole in asymptotically AdS spacetime correspond precisely to poles in the two-point function of the boundary hydrodynamics theory \cite{Berti:2009kk}, so that could provide a setting in which to tackle this question.

\paragraph{Acknowledgements.}
S.M. thanks the other members of the Gravitational Wave Initiative at Queen Mary for useful comments. 
This work was supported by a UKRI Stephen Hawking Fellowship (EP/T017481/1).

\appendix
\section{Master equations for black hole perturbations}
\label{app:master}

This appendix collects some details about the master equations \eqref{eqn:master_eom} and \eqref{eqn:master_eom_H}. 
The GREFT potentials used throughout to calculate the time delay and QNMs are listed in (\ref{eqn:first_dV}-\ref{eqn:last_dV})

\paragraph{EFT Perturbation Theory.}
The algebra involved in computing the master equations can quickly become laborious if not done carefully.
To illustrate the basic idea, consider the toy problem of a scalar field $\phi$ described by the action $S_0 + \epsilon S_1$, where $S_1$ is a small correction to be treated perturbatively in $\epsilon$. 
The first step is to split $\phi$ into a background plus small perturbations,
\begin{align}
 \phi = \bar{\phi} + \varphi \; . 
\end{align}
Both are affected by the interactions,
\begin{align}
 \bar{\phi} &= \bar{\phi}_0 + \epsilon \bar{\phi}_1 + \mathcal{O} ( \epsilon )  \nonumber \\ 
 \varphi &= \varphi_0 + \epsilon \varphi_1 + \mathcal{O} ( \epsilon )   \; . 
\end{align}
The leading-order background $\bar{\phi}_0$ is determined first, by solving:
\begin{align}
 \frac{\delta S_0 }{\delta \phi } \big|_{\phi = \bar{\phi}_0} = 0  \; .
\end{align}
Then one should solve:
\begin{align}
\frac{\delta S_0 }{ \delta \phi } \big|_{\bar{\phi}_1} +  \frac{ \delta S_1 }{\delta \phi } \big|_{\bar{\phi}_0} = 0
\end{align}
for $\bar{\phi}_1$, the first correction to the background (e.g. for the GREFT, this produces \eqref{eqn:background_metric}). 
Then moving on to the small perturbations, they satisfy,
\begin{align}
\frac{ \delta^2 S_0 }{ \delta \phi^2} \big|_{ \bar{\phi}_0 } \; \varphi_0 = 0 
\end{align}
in the absence of the $\epsilon$ correction. 
To find the leading correction $\varphi_1$ to the perturbations induced by the $\epsilon$ correction, one must finally solve,
\begin{align}
 \frac{ \delta^2 S_0 }{ \delta \phi^2} \big|_{ \bar{\phi}_0 } \; \varphi_1  
 +  
 \frac{ \delta^2 S_0 }{ \delta \phi^2} \big|_{ \bar{\phi}_1 } \; \varphi_0  
+
 \frac{ \delta^2 S_1 }{ \delta \phi^2} \big|_{ \bar{\phi}_0 } \; \varphi_0 
 = 0  \; . 
\end{align}
It is an equation of this form (with $\phi$ replaced by $g_{\mu\nu}$) which leads to the master equations for $\Psi_{\ell}^{\pm}$. 
In particular, note that to first order in $\epsilon$ one may use the leading-order equation of motion to remove higher-derivative terms, which ensures that the final equation for the perturbations will be at most second-order (despite the variation of the original Lagrangian initially generating higher-order derivatives). 



\paragraph{Converting between potentials.}
By rescaling the perturbation,
\begin{align}
\tilde{ \Psi} = \left(  \frac{ \sqrt{f_r f_t} }{f}  \right)^{1/2} \Psi
\end{align}
then the master equation \eqref{eqn:master_eom} is recast as \eqref{eqn:master_eom_H} to leading order in $\epsilon$. 
The potentials are related by:
 \begin{align}
\tilde{V}  &= \frac{ \tilde{f} }{\sqrt{f_r f_t}} V  + \frac{1}{2 \tilde{f}} \frac{d^2}{d r_*^2} \left( \frac{ \sqrt{f_r f_t} }{ \tilde{f} }  \right)  + \frac{\omega^2}{ \tilde{f} } \left[ \frac{1}{ \tilde{c}_s^2}  - \frac{ \tilde{f}^2 }{  c_s^2  f_r f_t } \right] \; . 
\end{align} 
Note that since $c_s^2 ( \tilde{r}_s ) = 1$ at the horizon, the final square bracket vanishes at $r= \tilde{r}_s$.
In the GREFT \eqref{eqn:intro_GREFT}, $\tilde{c}_s$ is given by,
\begin{align}
   \frac{ \tilde{c}_s r_s}{ \tilde{r}_s }  =& \;\; 1 + 2 b_! \epsilon_s^4 +  \frac{1}{2} c_1 \epsilon_s^6 
\end{align}
which is useful when evaluating \eqref{eqn:QNM_para}. 

The potentials $\delta V_{\ell}^{\pm}$ and $\delta \tilde{V}_{\ell}^{\pm}$ are given below for each of the interactions in \eqref{eqn:intro_GREFT} and \eqref{eqn:L_redundant}. They take the form:
\begin{align}
 \delta V_\ell^{\pm} (\omega, r) &= \epsilon_s^4 \delta_3 V_{\ell}^{\pm} ( \omega, r) + \epsilon_s^6 \delta_4 V_{\ell}^{\pm} ( \omega, r)  + ... \nonumber \\
 \delta \tilde{V}_\ell^{\pm} (\omega, r) &=  \epsilon_s^4 \delta_3 \tilde{V}_{\ell}^{\pm} ( \omega, r) + \epsilon_s^6 \delta_4 \tilde{V}_{\ell}^{\pm} ( \omega, r) + ...  
\end{align}
Some important features are:
\begin{itemize}

\item $b_1$ contributes like $\ell^0$ to both odd and even modes but with opposite signs,

\item $c_1$ contributes like $\ell^2$ to even modes and like $\ell^0$ for odd modes, also with opposite signs.Note that the $\ell^0$ contribution to the time advance $\delta t_{\ell}^-$ is further surpressed since the LO WKB approximation vanishes, and relatedly the subleading QNM correction is large (see \eqref{eqn:c1_subleading_large}). 

\item $c_2$ contributes like $\ell^2$ to odd modes and has no effect on the even modes or the background. 

\end{itemize}

\subsection*{Cubic interaction $b_1$}

The Riemann-cubed interaction in the GREFT modifies the master equation \eqref{eqn:master_eom} by \cite{deRham:2020ejn},
\begin{align}
\frac{ r_s^2 \delta_3 V^-_{\ell} }{b_1} =& j \left(\frac{45}{x^7}-\frac{855}{16 x^8}\right)-\frac{270}{x^7}+\frac{5175}{8
   x^8}-\frac{2979}{8 x^9}  \label{eqn:first_dV}  \\
\frac{ r_s^2 \delta_3 V^+_{\ell} }{b_1} =& + Z
\left(\frac{405}{16 x^9}-\frac{45}{2 x^8}\right)
     +\frac{90}{x^7}-\frac{783}{4 x^8}+\frac{1755}{16 x^9}   \nonumber  \\
     &
     +\frac{-\frac{432}{x^7}+\frac{963}{x^8}-\frac{4347}{8
   x^9}}{Z}
+ \frac{\frac{567}{x^7}-\frac{9369}{8 x^8}+\frac{4851}{8
   x^9}}{Z^2}
   +\frac{-\frac{243}{x^7}+\frac{1863}{4 x^8}-\frac{891}{4
   x^9}}{Z^3}   \nonumber  
\end{align}
where $x = r/r_s$ is a dimensionless radial coordinate and both $j = \ell (\ell +1)$ and $ Z = (j -2 ) x + 3$ depend on the angular momentum\footnote{
Note that $\lambda = \ell ( \ell + 1 ) - 2$ is often used instead of $j$ because it is the eigenvalue of the Laplace-Beltrami operator on the 2-sphere acting on a tensor-type spherical harmonic. 
}. 
Once converted into the ``horizon'' tortoise coordinate \eqref{eqn:master_eom_H}, the odd potential becomes:
\begin{align}
\frac{ r_s^2 \, \delta_3 \tilde{V}_{\ell}^- }{ b_1 } =& 
  +  \omega^2 r_s^2 \left(-\frac{7}{8
   x^2}-\frac{3}{2 x^3}-\frac{17}{8 x^4}-\frac{47}{4 x^5}-\frac{1}{4
   x}+\frac{3}{8}\right) \nonumber \\
&+ j \left(\frac{5}{16 x^3}+\frac{5}{16 x^4}+\frac{5}{16 x^5}+\frac{5}{16
   x^6}+\frac{725}{16 x^7}-\frac{877}{16 x^8}\right) \nonumber \\ 
  &-\frac{5}{4 x^3}-\frac{45}{32 x^4}-\frac{25}{16
   x^5}-\frac{55}{32 x^6}-\frac{2175}{8 x^7}+\frac{21769}{32
   x^8}-\frac{1605}{4 x^9}  \nonumber 
\end{align}
and the even potential becomes:
\begin{align}
\frac{ r_s^2 \, \delta_3 \tilde{V}_{\ell}^+ }{ b_1 } =& 
+ \omega^2 r_s^2 \left(-\frac{7}{8 x^2}-\frac{3}{2 x^3}-\frac{17}{8 x^4}-\frac{47}{4
   x^5}-\frac{1}{4 x}  +\frac{3}{8}\right)    \\ 
   &+
   \left(\frac{5}{16 x^4}+\frac{5}{16 x^5}+\frac{5}{16
   x^6}+\frac{5}{16 x^7}-\frac{355}{16 x^8}+\frac{383}{16 x^9}\right)
   Z   \nonumber \\  
   & -\frac{5}{8 x^3}-\frac{55}{32 x^4}-\frac{15}{8 x^5}-\frac{65}{32
   x^6}+\frac{1405}{16 x^7}-\frac{5313}{32 x^8}+\frac{1359}{16 x^9}  \nonumber \\ 
  &+\frac{-\frac{15}{4 x^2}+\frac{15}{2 x^3}+\frac{15}{8 x^4}+\frac{15}{8
   x^5}+\frac{15}{8 x^6}-\frac{3441}{8 x^7}+\frac{7881}{8 x^8}-\frac{4545}{8
   x^9}}{Z}  \nonumber \\ 
   &+\frac{\frac{45}{2 x^2}-\frac{225}{8
   x^3}+\frac{567}{x^7}-\frac{2403}{2 x^8}+\frac{5049}{8
   x^9}}{Z^2}
   +\frac{-\frac{135}{4 x^2}+\frac{135}{4
   x^3}-\frac{243}{x^7}+\frac{1863}{4 x^8}-\frac{891}{4
   x^9}}{Z^3}  \; ,  \nonumber 
\end{align}
from which the $\alpha_{k \ell}^{\pm}$ coefficients in \eqref{eqn:QNM_para} can be easily read off.

\subsection*{Quartic interaction $c_1$}

The $c_1$ Riemann$^4$ interaction in the GREFT modifies the master equation \eqref{eqn:master_eom} by \cite{Cardoso:2018ptl},
\begin{align}
\frac{r_s^2 \delta_4 V_\ell^-}{ c_1 } =&+ j \left(\frac{73}{x^{11}}-\frac{72}{x^{10}}\right) +\frac{2916}{x^{10}}-\frac{13509}{2 x^{11}}+\frac{15561}{4 x^{12}}  \nonumber \\ 
\frac{ r_s^2 \delta_4 V_\ell^+ }{ c_1 } =& 
-\frac{9 Z^2}{2 x^{12}} 
+ Z \left(\frac{100}{x^{12}}-\frac{81}{x^{11}}\right) 
-\frac{792}{x^{10}}+\frac{4549}{2 x^{11}}-\frac{6069}{4 x^{12}}  \nonumber \\ 
&+ \frac{\frac{3348}{x^{10}}-\frac{7776}{x^{11}}+\frac{4410}{x^{12}}}{Z}
+ \frac{-\frac{3240}{x^{10}}+\frac{6669}{x^{11}}-\frac{13599}{4 x^{12}}}{Z^2}
+ \frac{\frac{972}{x^{10}}-\frac{1836}{x^{11}}+\frac{864}{x^{12}}}{Z^3}
\end{align}
as well as shifting the sound speed \eqref{eqn:cs}.

Converting this to the ``horizon'' tortoise coordinate \eqref{eqn:master_eom_H} produces for the odd modes:
\begin{align}
\frac{ \delta_4 \tilde{V}^-_{\ell} }{ c_1  } = & 
   + \omega^2 \left(-\frac{9}{4 x^2}-\frac{7}{2
   x^3}-\frac{19}{4 x^4}-\frac{6}{x^5}-\frac{29}{4 x^6}-\frac{17}{2
   x^7}+\frac{213}{4 x^8}-\frac{1}{x}+\frac{1}{4}\right)  \nonumber \\ 
&+ \frac{j}{r_s^2} \left(\frac{5}{8 x^3}+\frac{5}{8 x^4}+\frac{5}{8 x^5}+\frac{5}{8
   x^6}+\frac{5}{8 x^7}+\frac{5}{8 x^8}+\frac{5}{8 x^9}-\frac{571}{8
   x^{10}}+\frac{545}{8 x^{11}}\right)
   \nonumber \\ 
   &- \frac{1}{r_s^2} \left( \frac{5}{2
   x^3}-\frac{45}{16 x^4}-\frac{25}{8 x^5}-\frac{55}{16 x^6}-\frac{15}{4
   x^7}-\frac{65}{16 x^8}-\frac{35}{8 x^9}+\frac{46581}{16
   x^{10}}-\frac{6512}{x^{11}}+\frac{58617}{16 x^{12}} \right)  \nonumber 
\end{align}
and for the even modes:
\begin{align}
\frac{ r_s^2 \delta_4 \tilde{V}_{\ell}^+ }{c_1} =&+ 
 \omega^2 r_s^2 \left(-\frac{9}{4 x^2}-\frac{7}{2 x^3}-\frac{19}{4
   x^4}-\frac{6}{x^5}-\frac{29}{4 x^6}-\frac{17}{2 x^7}+\frac{213}{4
   x^8}-\frac{1}{x}+\frac{1}{4}\right)
   -\frac{9 Z^2}{2
   x^{12}}  \nonumber \\ 
   &+Z \left(
   \frac{5}{8 x^4}+\frac{5}{8 x^5}+\frac{5}{8 x^6}+\frac{5}{8 x^7}+\frac{5}{8
   x^8}+\frac{5}{8 x^9}+\frac{5}{8 x^{10}}-\frac{643}{8 x^{11}}+\frac{761}{8
   x^{12}}\right)   \nonumber \\
&+ \left(   -\frac{5}{4 x^3}-\frac{55}{16
   x^4}-\frac{15}{4 x^5}-\frac{65}{16 x^6}-\frac{35}{8 x^7}-\frac{75}{16
   x^8}-\frac{5}{x^9}-\frac{12757}{16 x^{10}}+\frac{20043}{8
   x^{11}}-\frac{27669}{16 x^{12}} \right)  \nonumber \\
  &+\frac{-\frac{15}{2 x^2}+\frac{15}{x^3}+\frac{15}{4
   x^4}+\frac{15}{4 x^5}+\frac{15}{4 x^6}+\frac{15}{4 x^7}+\frac{15}{4
   x^8}+\frac{15}{4 x^9}+\frac{13407}{4 x^{10}}-\frac{30825}{4
   x^{11}}+\frac{17289}{4 x^{12}}}{Z}  \nonumber \\
   &+   \frac{\frac{45}{x^2}-\frac{225}{4
   x^3}-\frac{3240}{x^{10}}+\frac{6570}{x^{11}}-\frac{3312}{x^{12}}}{Z^2}  + \frac{-\frac{135}{2 x^2}+\frac{135}{2
   x^3}+\frac{972}{x^{10}}-\frac{1836}{x^{11}}+\frac{864}{x^{12}}}{Z^3}  
\end{align}
where $Z = (j -2 ) x + 3$.

\subsection*{Quartic interaction $c_2$}

The $c_2$ Riemann$^4$ interaction in the GREFT modifies the master equation \eqref{eqn:master_eom} for the odd modes only:
\begin{align}
\frac{ r_s^2 \delta_4 V^-_{\ell} }{c_2} = - \frac{18 j ( j -2 ) }{ x^{10} } \; . 
\end{align}
Since $c_2$ does not affect the background metric or sound speed, the $c_2$ contribution to the master equation in ``horizon'' tortoise coordinates \eqref{eqn:master_eom_H} is simply the same:
\begin{align}
\frac{ r_s^2 \delta_4 \tilde{V}^-_{\ell} }{c_2} = - \frac{18 j ( j -2) }{ x^{10} } \; . 
\end{align}

\subsection*{Redundant cubic interaction $b_3$}

For the redundant cubic operator \eqref{eqn:L_redundant}, it shifts the background by \eqref{eqn:df_redundant} and leaves the sound speed unchanged. 
Converting the potentials given in \eqref{eqn:V_redundant} into the ``horizon'' tortoise coordinates gives the following contribution to \eqref{eqn:master_eom_H}: 
\begin{align}
\frac{ r_s^2 \delta_3 \tilde{V}^-_{\ell} }{b_3}  =&  
+ \omega^2 r_s^2
   \left(-\frac{3}{x^2}-\frac{6}{x^3}-\frac{9}{x^4}-\frac{12}{x^5}+3\right)
  + j \left(\frac{3}{2 x^3}+\frac{3}{2 x^4}+\frac{3}{2 x^5}+\frac{3}{2
   x^6}+\frac{3}{2 x^7}-\frac{33}{2 x^8}\right)  \nonumber \\ 
  &- \frac{6}{x^3}-\frac{27}{4 x^4}-\frac{15}{2 x^5}-\frac{33}{4
   x^6}-\frac{9}{x^7}+\frac{591}{4 x^8}-\frac{90}{x^9}     
   \nonumber \\ 
 \frac{  r_s^2 \delta_3 \tilde{V}^+_{\ell} }{b_3}  =&  + \omega^2 r_s^2
   \left(-\frac{3}{x^2}-\frac{6}{x^3}-\frac{9}{x^4}-\frac{12}{x^5}+3\right)  
      +Z \left(\frac{3}{2 x^4}+\frac{3}{2 x^5}+\frac{3}{2 x^6}+\frac{3}{2
   x^7}+\frac{3}{2 x^8}-\frac{33}{2 x^9}\right) \nonumber \\
   &-\frac{3}{x^3}-\frac{33}{4 x^4}-\frac{9}{x^5}-\frac{39}{4 x^6}-\frac{21}{2 x^7}+\frac{441}{4
   x^8}-\frac{81}{2 x^9}
   +\frac{-\frac{18}{x^2}+\frac{36}{x^3}+\frac{9}{x^4}+\frac{9}{x^5}+\frac{9}
   {x^6}+\frac{9}{x^7}+\frac{243}{x^8}-\frac{351}{x^9}}{Z} \nonumber \\ 
   &+\frac{\frac{108}{x^2}-\frac{135}{x^3}-\frac{432}{x^8}+\frac{459}{x^9}}{Z^2}
   +\frac{-\frac{162}{x^2}+\frac{162}{x^3}+\frac{162}{x^8}-\frac{162}{x^9}}{Z^3}  \label{eqn:last_dV}
\end{align}
Amusingly, the $\alpha_{k \ell}^{\pm}$ coefficients \eqref{eqn:alpha_def} read off from this potential satisfy \cite{deRham:2020ejn},
\begin{align}
 \sum_k^{\infty} e_{k \ell}^{\pm} \alpha_{k \ell}^{\pm} (\omega^{\rm GR}_{0 \ell} ) = 0  \; , 
 \label{eqn:null_constraint}
\end{align}
up to small numerical errors, which is consistent with the expectation that this redundant potential does not affect any purely gravitational observable (like the QNMs). 
The null constraint \eqref{eqn:null_constraint} can be used to infer higher-order $e_{k \ell}$ from lower-order ones \cite{Kimura:2020mrh}, or alternatively could be used to infer $\omega_{n \ell}^{\rm GR}$ if the $e_{k \ell}^{+}$ or $e_{k\ell}^-$ were independently known.

\section{Time delay details}
\label{app:dt}

Starting from \eqref{eqn:dT_from_dA}, one can compute the time delay in the large $\ell$ limit as follows. 
First perform a large $\ell$ expansion of both the GR and EFT potentials,
\begin{align}
f V^{\pm}_{\ell} &= \frac{j}{r^2} \left( 1 - \frac{r_s}{r} \right) + \mathcal{O} \left( j^0 \right)  \nonumber \\
  \delta W_{\ell}^{\pm} ( \omega , r ) \big|_{ \omega = \sqrt{\gamma (f V)_{\rm max} }} &=  j^p \kappa^{\pm}_0 \left( \frac{r_t}{r} , \frac{r_s}{r} \right)  + \mathcal{O} \left( j^{p-1} \right) \; ,  \nonumber \\
\partial_\omega  \delta W_{\ell}^{\pm} ( \omega  , r ) \big|_{ \omega = \sqrt{\gamma (f V)_{\rm max}}} &=  j^p \kappa^{\pm}_1 \left( \frac{r_t}{r} , \frac{r_s}{r} \right)  + \mathcal{O} \left( j^{p-1} \right)
\label{eqn:dt_app_1}
\end{align}
where $\gamma$ can be replaced with $r_t$ using \eqref{eqn:g_to_rt} and $p$ is a fixed constant that depends on the interaction considered.
Then construct the integrand,
\begin{align}
\mathcal{A}_{\ell}^P \, \partial_r \left( \frac{ \delta \mathcal{A}_{\ell}^P }{ \partial_r \mathcal{A}_{\ell}^P  }  \right)
= j^{p-1} \mathcal{I}^P \left( \frac{r_t}{r} , \frac{r_s}{r} \right) + \mathcal{O} \left( j^{p-2} \right) \; .
\end{align}
For instance from the odd potential $\delta V_{\ell}^-$ from the $c_1$ interaction, the power $p = 1$ and the time delay tends to a constant at large $j$. Writing,
\begin{align}
 - 2 \omega \int_{r_t^{\rm GR}}^{\infty} dr \;  \mathcal{I}^- \left( \frac{r_t}{r} , \frac{r_s}{r} \right)  \equiv \frac{ c_1 \ell}{ K_{c_1}^- (\gamma) } \; , 
\end{align}
immediately produces \eqref{eqn:K_def}. 
The $\mathcal{I}^-$ is relatively simple and can be integrated analytically in terms of elliptic functions. 
The full expression is somewhat cumbersome, but it is plotted as the black line in Figure~\ref{fig:c1_dt}, showing good agreement with the numerical result at large $\ell$. 

Finally, note that in the eikonal limit of large $r_t \gg r_s$ ($\gamma \ll 1$), one can further expand \eqref{eqn:dt_app_1} in powers of $r_s/r_t$, 
\begin{align}
f V^{\pm}_{\ell} &= \frac{j}{r^2} \ + \mathcal{O} \left( j^0 , \frac{r_s}{r_t} \right)  \nonumber \\
  \delta W_{\ell}^{\pm} ( \omega , r ) \big|_{ \omega = \sqrt{\gamma (f V)_{\rm max} }} &=  j^p \kappa_0^{\pm} \left( \frac{r_t}{r} , 0 \right)  + \mathcal{O} \left( j^{p-1} , \frac{r_s}{r_t} \right) \; ,  \nonumber \\
\partial_\omega  \delta W_{\ell}^{\pm} ( \omega  , r ) \big|_{ \omega = \sqrt{\gamma (f V)_{\rm max}}} &=  j^p \kappa_1^{\pm} \left( \frac{r_t}{r} , 0 \right)  + \mathcal{O} \left( j^{p-1} , \frac{r_s}{r_t} \right) \; . 
\end{align}
and simplify \eqref{eqn:g_to_rt} in this limit to simply $r_t^{\rm GR} \approx \sqrt{j}/\omega \equiv b$, the Newtonian impact parameter.
The time delay can then be massaged into the usual eikonal form \cite{deRham:2020zyh},
\begin{align}
 \delta t_\ell^P \approx \frac{\partial}{\partial \omega} \left[ 
 \frac{1}{\omega} \int_b^{\infty} d r \; \frac{ \delta W_\ell^P }{\sqrt{1- \frac{b^2}{r^2} }}
 \right]    \; ,
 \label{eqn:dt_eikonal}
\end{align}
up to subleading corrections in $1/j$ and $r_s/r_t$. 
However, note that for the $\delta V_{\ell}^-$ potential from $c_1$, this eikonal limit \emph{vanishes} since,
\begin{align}
 \delta W_\ell^- = 9 \omega^2 \, \frac{ r_s^{8} \left( 7 r^2 - 8 b^2 \right) }{r^{10}}  + ...
\end{align}
integrates to zero.
As a result, the naive expectation that $\delta t_{\ell}^- \sim \omega/b^7$ in the eikonal limit overestimates the time delay, which is actually $\mathcal{O} \left( \omega / b^8 \right)$ and comes from the subleading $r_s/r_t$ corrections.
Regardless of this cancellation, \eqref{eqn:dt_eikonal} is only a good approximation at large impact parameter ($\gamma \ll 1$), and does not capture the $\gamma = 0.9$ regime studied in the main text (which requires the full $\mathcal{I}^P \left( \frac{r_t}{r} , \frac{r_s}{r} \right)$ at large $j$, or the numerical integration of \eqref{eqn:dT_from_dA} at small $j$).

\section{Quasi-normal mode details}
\label{app:QNM}

This Appendix describes how to find the QNM spectrum of fluctuations $\Psi$ which obey the Schrodinger-like equation\footnote{
Note that the $r$ and $V$ used in this Appendix differ from those in the main text.
}:
\begin{align}
\left(  \frac{d^2}{dr^2}  + \omega^2 \right) \Psi =  
V  \Psi 
\end{align}
where $V = V_{\rm GR} +  \delta V$ with $\delta V$ a small perturbation to GR.
The general procedure is:
\begin{itemize}

\item[(i)] Following \cite{Schutz:1985km,Iyer:1986np,Konoplya:2003ii, Hatsuda:2019eoj}, the WKB approximation can be used to relate $\omega^2_{n \ell}$ to $V^{(k)} (   \omega_{n \ell} , r_{\rm max} )$, derivatives of the potential evaluated at its maximum (i.e. $r_{\rm max}$ is defined by the condition $V^{(1)} = 0$). 
Explicitly, this gives,
\begin{align}
\omega_{n \ell}^2 &= V^{(0)}   - i N \sqrt{-V^{(2)}/2}   
 +\frac{1}{32} \left[  -\frac{1}{9}\frac{V^{(3)2}}{V^{(2)2}}(7+ 15 N^2)+\frac{V^{(4)}}{V^{(2)}}(1+ N^2)\right] \nonumber\\
&  - \frac{ i N}{  576 \sqrt{2} }\left[ \frac{5}{24}\frac{V^{(3)4}}{(-V^{(2)})^{9/2}}(77 + 47 N^2) 
+ \frac{3}{4}\frac{V^{(3)2}V^{(4)}}{(-V^{(2)})^{7/2}}(51+25 N^2)   \right.   \\
&\qquad\qquad \left. +\frac{1}{8}\frac{V^{(4)2}}{(-V^{(2)})^{5/2}}(67+17 N^2) 
+ \frac{V^{(3)}V^{(5)}}{(-V^{(2)})^{5/2}}(19+7 N^2)  
+ \frac{V^{(6)}}{(-V^{(2)})^{3/2}}(5+N^2)
\right] , \nonumber
\end{align}
where $N=2n+1$ with $n=0,1,2,\dots$ the overtone number.

\item[(ii)] Expand each derivative in terms of $V_{\rm GR}$ and $\delta V$ using,
\begin{align}
 V^{(k)} \left( r_{\rm max} , \omega_{n \ell} \right) = V_{\rm GR}^{(k)} ( r_{\rm max}^{\rm GR} ) + \delta V^{(k)} (  \omega^{\rm GR}_{n\ell} , r_{\rm max}^{\rm GR} ) - \frac{ V_{\rm GR}^{(k+1)} ( r_{\rm max}^{\rm GR} ) }{ V_{\rm GR}^{(2)} ( r_{\rm max}^{\rm GR}  ) } \delta V^{(1)} (  \omega^{\rm GR}_{n\ell} , r_{\rm max}^{\rm GR}  ) 
 + \mathcal{O} \left( \left( \delta V \right)^2  \right) \; , 
\end{align}
where the right-hand-side is evaluated at the GR value  $r_{\rm max}^{\rm GR}$ and the final term accounts for the difference with the full theory's $r_{\rm max}$. 

\item[(iii)] Expand each $V_{\rm GR}^{(k)}$ and $\delta V^{(k)}$ in the large $\ell$ limit,
\begin{align}
 V_{\rm GR}^{(k)} &= j  \sum_{p=0} \bar{\nu}_{k, 2p} \;  j^{-p}   \nonumber \\
\delta V^{(k)} (  \omega^{\rm GR}_{n\ell} , r^{\rm GR}_{\rm max} ) &= j^{ p } \left( \sum_{a=0}  \delta \nu_{k , 2a} ( n ) j^{-a} + i \sqrt{j} \sum_{a=0} \delta \nu_{k, 2a+1} ( n ) \, j^{-a}  \right)    
\end{align} 
where $j= \ell (\ell + 1)$ and the power $p$ depends on the interaction. 

\end{itemize}

The result of this procedure is an analytic expression for the QNM frequencies which becomes exact in the large $j$ limit. 
At zeroth order in $\delta V$ it recovers the GR frequencies \eqref{eqn:QNM_GR}. 
At first order in $\delta V$, it gives an expression for the shift in the QNM frequencies relative to GR: for instance the leading contribution to the imaginary part is:  
\begin{align}
\frac{ \text{Im} \left( \omega_{n\ell} -  \omega^{\rm GR}_{n \ell} \right) }{ \text{Im} \left(  \omega^{\rm GR}_{n\ell} \right) } &= j^{p -1} \left( 
  - \frac{ \delta \nu_{00} }{2 \bar{\nu}_{00} } 
+ \frac{\delta \nu_{20} }{2 \bar{\nu}_{20} } 
- \frac{  \bar{\nu}_{30} \delta \nu_{10} }{2 \bar{\nu}_{20}^2} 
 - \frac{ \delta \nu_{01} }{  \sqrt{-2 \bar{\nu}_{20}}}  + \mathcal{O} \left( \frac{1}{j} \right)  
 \right)
\end{align}
Both real and imaginary part at leading and next-to-leading order are given below for the GREFT potentials.

\paragraph{GREFT QNMs.}
The shift in the QNM frequency from the GREFT interactions can be written in the form \eqref{eqn:QNM_EFT_exp}. 
This WKB approach gives the following shift from the cubic $b_1$ interaction in the odd QNMs:
\begin{align}
\delta_3 \mathcal{R}_{n \ell}^- &= b_1 \left( \frac{380}{729} + \frac{ 24037 - 6303 N^2 }{ 39366 j }  + \mathcal{O} \left( \frac{1}{j^2} \right)  \right)  \; ,
 \nonumber \\
\delta_3 \mathcal{I}_{n \ell}^- &= b_1 \left( \frac{1028}{729} + \frac{  35 (535 + 358 N^2) }{ 177147 j } + \mathcal{O} \left( \frac{1}{j^2} \right) \; , \right) 
\label{eqn:first_QNM}
\end{align}
and in the even QNMs: 
\begin{align}
\delta_3 \mathcal{R}_{n \ell}^+ &= b_1 \left( 
- \frac{340}{729} + \frac{ 6217 + 8277 N^2 }{ 39366 j }
+ \mathcal{O} \left( \frac{1}{j^2} \right)  \right)  \; ,  \nonumber \\
\delta_3 \mathcal{I}_{n \ell}^+ &= b_1 \left( 
- \frac{ 1132 }{ 729 } - \frac{ 5 (-25111 + 2264 N^2) }{ 177147 j }
+ \mathcal{O} \left( \frac{1}{j^2} \right)  \right) \; .
\end{align}
Similarly, the WKB approach gives the following shift from the quartic interactions in the odd modes:
\begin{align}
\delta_4 \mathcal{R}_{n \ell}^- &= - \frac{  256 c_2 }{ 729 } j  - \frac{ 32 c_2 (-919 + 615 N^2) }{ 59049 } + \frac{832 c_1}{19683} + \frac{ 112 c_1 (1391 + 57 N^2) }{ 177147 j } + \mathcal{O} \left( \frac{c_2}{j} , \frac{c_1}{j^2} \right) 
  \\
\delta_4 \mathcal{I}_{n \ell}^- &= + \frac{5120 c_2}{ 2187} j  + \frac{8 c_2 (197531 + 21425 N^2)} {531441}  - \frac{ 7232 c_1 }{19683} - \frac{ 28 c_1 (-722441 + 5437 N^2) }{ 1594323 j } + \mathcal{O} \left( \frac{c_2}{j} ,  \frac{c_1}{j^2} \right) \; . \nonumber 
\end{align}
and in the even modes:
\begin{align}
\delta_4 \mathcal{R}_{n \ell}^+ &=  - \frac{ 64 c_1 }{ 729} j  - \frac{  8 c_1 (-1231 + 615 N^2 ) }{ 59049 } + \mathcal{O} \left( \frac{1}{j} \right) \; 
 \nonumber \\
\delta_4 \mathcal{I}_{n \ell}^+ &= +\frac{ 1280 c_1 }{ 2187} j + \frac{ 2 c_1 (99899 + 21425 N^2 ) }{ 531441 } + \mathcal{O} \left( \frac{1}{j} \right) \; ,
\label{eqn:last_QNM}
\end{align}
where $j = \ell ( \ell + 1)$ and $N = 2n+1$ as before.

%

\bibliographystyle{JHEP}
\bibliography{refs}

\providecommand{\href}[2]{#2}\begingroup\raggedright\begin{thebibliography}{10}

\bibitem{Donoghue:1995cz}
J.~F. Donoghue, \emph{{Introduction to the effective field theory description
  of gravity}},  in \emph{{Advanced School on Effective Theories}}, 6, 1995.
\newblock \href{https://arxiv.org/abs/gr-qc/9512024}{{\ttfamily
  gr-qc/9512024}}.

\bibitem{Burgess:2003jk}
C.~P. Burgess, \emph{{Quantum gravity in everyday life: General relativity as
  an effective field theory}},
  \href{http://dx.doi.org/10.12942/lrr-2004-5}{\emph{Living Rev. Rel.}
  {\bfseries 7} (2004) 5--56},
  [\href{https://arxiv.org/abs/gr-qc/0311082}{{\ttfamily gr-qc/0311082}}].

\bibitem{KAGRA:2021vkt}
{\scshape KAGRA, VIRGO, LIGO Scientific} collaboration, R.~Abbott et~al.,
  \emph{{GWTC-3: Compact Binary Coalescences Observed by LIGO and Virgo during
  the Second Part of the Third Observing Run}},
  \href{http://dx.doi.org/10.1103/PhysRevX.13.041039}{\emph{Phys. Rev. X}
  {\bfseries 13} (2023) 041039},
  [\href{https://arxiv.org/abs/2111.03606}{{\ttfamily 2111.03606}}].

\bibitem{KAGRA:2013rdx}
{\scshape KAGRA, LIGO Scientific, Virgo, VIRGO} collaboration, B.~P. Abbott
  et~al., \emph{{Prospects for observing and localizing gravitational-wave
  transients with Advanced LIGO, Advanced Virgo and KAGRA}},
  \href{http://dx.doi.org/10.1007/s41114-020-00026-9}{\emph{Living Rev. Rel.}
  {\bfseries 21} (2018) 3}, [\href{https://arxiv.org/abs/1304.0670}{{\ttfamily
  1304.0670}}].

\bibitem{Berti:2015itd}
E.~Berti et~al., \emph{{Testing General Relativity with Present and Future
  Astrophysical Observations}},
  \href{http://dx.doi.org/10.1088/0264-9381/32/24/243001}{\emph{Class. Quant.
  Grav.} {\bfseries 32} (2015) 243001},
  [\href{https://arxiv.org/abs/1501.07274}{{\ttfamily 1501.07274}}].

\bibitem{LIGOScientific:2016lio}
{\scshape LIGO Scientific, Virgo} collaboration, B.~P. Abbott et~al.,
  \emph{{Tests of general relativity with GW150914}},
  \href{http://dx.doi.org/10.1103/PhysRevLett.116.221101}{\emph{Phys. Rev.
  Lett.} {\bfseries 116} (2016) 221101},
  [\href{https://arxiv.org/abs/1602.03841}{{\ttfamily 1602.03841}}].

\bibitem{LIGOScientific:2018dkp}
{\scshape LIGO Scientific, Virgo} collaboration, B.~P. Abbott et~al.,
  \emph{{Tests of General Relativity with GW170817}},
  \href{http://dx.doi.org/10.1103/PhysRevLett.123.011102}{\emph{Phys. Rev.
  Lett.} {\bfseries 123} (2019) 011102},
  [\href{https://arxiv.org/abs/1811.00364}{{\ttfamily 1811.00364}}].

\bibitem{LIGOScientific:2019fpa}
{\scshape LIGO Scientific, Virgo} collaboration, B.~P. Abbott et~al.,
  \emph{{Tests of General Relativity with the Binary Black Hole Signals from
  the LIGO-Virgo Catalog GWTC-1}},
  \href{http://dx.doi.org/10.1103/PhysRevD.100.104036}{\emph{Phys. Rev. D}
  {\bfseries 100} (2019) 104036},
  [\href{https://arxiv.org/abs/1903.04467}{{\ttfamily 1903.04467}}].

\bibitem{Barausse:2020rsu}
E.~Barausse et~al., \emph{{Prospects for Fundamental Physics with LISA}},
  \href{http://dx.doi.org/10.1007/s10714-020-02691-1}{\emph{Gen. Rel. Grav.}
  {\bfseries 52} (2020) 81},
  [\href{https://arxiv.org/abs/2001.09793}{{\ttfamily 2001.09793}}].

\bibitem{LIGOScientific:2020tif}
{\scshape LIGO Scientific, Virgo} collaboration, R.~Abbott et~al., \emph{{Tests
  of general relativity with binary black holes from the second LIGO-Virgo
  gravitational-wave transient catalog}},
  \href{http://dx.doi.org/10.1103/PhysRevD.103.122002}{\emph{Phys. Rev. D}
  {\bfseries 103} (2021) 122002},
  [\href{https://arxiv.org/abs/2010.14529}{{\ttfamily 2010.14529}}].

\bibitem{LIGOScientific:2021sio}
{\scshape LIGO Scientific, VIRGO, KAGRA} collaboration, R.~Abbott et~al.,
  \emph{{Tests of General Relativity with GWTC-3}},
  \href{https://arxiv.org/abs/2112.06861}{{\ttfamily 2112.06861}}.

\bibitem{Endlich:2017tqa}
S.~Endlich, V.~Gorbenko, J.~Huang and L.~Senatore, \emph{{An effective
  formalism for testing extensions to General Relativity with gravitational
  waves}}, \href{http://dx.doi.org/10.1007/JHEP09(2017)122}{\emph{JHEP}
  {\bfseries 09} (2017) 122},
  [\href{https://arxiv.org/abs/1704.01590}{{\ttfamily 1704.01590}}].

\bibitem{Brandhuber:2019qpg}
A.~Brandhuber and G.~Travaglini, \emph{{On higher-derivative effects on the
  gravitational potential and particle bending}},
  \href{http://dx.doi.org/10.1007/JHEP01(2020)010}{\emph{JHEP} {\bfseries 01}
  (2020) 010}, [\href{https://arxiv.org/abs/1905.05657}{{\ttfamily
  1905.05657}}].

\bibitem{AccettulliHuber:2020dal}
M.~Accettulli~Huber, A.~Brandhuber, S.~De~Angelis and G.~Travaglini,
  \emph{{From amplitudes to gravitational radiation with cubic interactions and
  tidal effects}},
  \href{http://dx.doi.org/10.1103/PhysRevD.103.045015}{\emph{Phys. Rev. D}
  {\bfseries 103} (2021) 045015},
  [\href{https://arxiv.org/abs/2012.06548}{{\ttfamily 2012.06548}}].

\bibitem{Cayuso:2023xbc}
R.~Cayuso, P.~Figueras, T.~Fran\c{c}a and L.~Lehner, \emph{{Self-Consistent
  Modeling of Gravitational Theories beyond General Relativity}},
  \href{http://dx.doi.org/10.1103/PhysRevLett.131.111403}{\emph{Phys. Rev.
  Lett.} {\bfseries 131} (2023) 111403}.

\bibitem{Cardoso:2018ptl}
V.~Cardoso, M.~Kimura, A.~Maselli and L.~Senatore, \emph{{Black Holes in an
  Effective Field Theory Extension of General Relativity}},
  \href{http://dx.doi.org/10.1103/PhysRevLett.121.251105}{\emph{Phys. Rev.
  Lett.} {\bfseries 121} (2018) 251105},
  [\href{https://arxiv.org/abs/1808.08962}{{\ttfamily 1808.08962}}].

\bibitem{Cardoso:2019mqo}
V.~Cardoso, M.~Kimura, A.~Maselli, E.~Berti, C.~F.~B. Macedo and R.~McManus,
  \emph{{Parametrized black hole quasinormal ringdown: Decoupled equations for
  nonrotating black holes}},
  \href{http://dx.doi.org/10.1103/PhysRevD.99.104077}{\emph{Phys. Rev. D}
  {\bfseries 99} (2019) 104077},
  [\href{https://arxiv.org/abs/1901.01265}{{\ttfamily 1901.01265}}].

\bibitem{McManus:2019ulj}
R.~McManus, E.~Berti, C.~F.~B. Macedo, M.~Kimura, A.~Maselli and V.~Cardoso,
  \emph{{Parametrized black hole quasinormal ringdown. II. Coupled equations
  and quadratic corrections for nonrotating black holes}},
  \href{http://dx.doi.org/10.1103/PhysRevD.100.044061}{\emph{Phys. Rev. D}
  {\bfseries 100} (2019) 044061},
  [\href{https://arxiv.org/abs/1906.05155}{{\ttfamily 1906.05155}}].

\bibitem{deRham:2020ejn}
C.~de~Rham, J.~Francfort and J.~Zhang, \emph{{Black Hole Gravitational Waves in
  the Effective Field Theory of Gravity}},
  \href{http://dx.doi.org/10.1103/PhysRevD.102.024079}{\emph{Phys. Rev. D}
  {\bfseries 102} (2020) 024079},
  [\href{https://arxiv.org/abs/2005.13923}{{\ttfamily 2005.13923}}].

\bibitem{Cano:2020cao}
P.~A. Cano, K.~Fransen and T.~Hertog, \emph{{Ringing of rotating black holes in
  higher-derivative gravity}},
  \href{http://dx.doi.org/10.1103/PhysRevD.102.044047}{\emph{Phys. Rev. D}
  {\bfseries 102} (2020) 044047},
  [\href{https://arxiv.org/abs/2005.03671}{{\ttfamily 2005.03671}}].

\bibitem{Cano:2023jbk}
P.~A. Cano, K.~Fransen, T.~Hertog and S.~Maenaut, \emph{{Quasinormal modes of
  rotating black holes in higher-derivative gravity}},
  \href{https://arxiv.org/abs/2307.07431}{{\ttfamily 2307.07431}}.

\bibitem{Sennett:2019bpc}
N.~Sennett, R.~Brito, A.~Buonanno, V.~Gorbenko and L.~Senatore,
  \emph{{Gravitational-Wave Constraints on an Effective Field-Theory Extension
  of General Relativity}},
  \href{http://dx.doi.org/10.1103/PhysRevD.102.044056}{\emph{Phys. Rev. D}
  {\bfseries 102} (2020) 044056},
  [\href{https://arxiv.org/abs/1912.09917}{{\ttfamily 1912.09917}}].

\bibitem{Silva:2022srr}
H.~O. Silva, A.~Ghosh and A.~Buonanno, \emph{{Black-hole ringdown as a probe of
  higher-curvature gravity theories}},
  \href{http://dx.doi.org/10.1103/PhysRevD.107.044030}{\emph{Phys. Rev. D}
  {\bfseries 107} (2023) 044030},
  [\href{https://arxiv.org/abs/2205.05132}{{\ttfamily 2205.05132}}].

\bibitem{Kawai:1985xq}
H.~Kawai, D.~C. Lewellen and S.~H.~H. Tye, \emph{{A Relation Between Tree
  Amplitudes of Closed and Open Strings}},
  \href{http://dx.doi.org/10.1016/0550-3213(86)90362-7}{\emph{Nucl. Phys. B}
  {\bfseries 269} (1986) 1--23}.

\bibitem{Bern:2021ppb}
Z.~Bern, D.~Kosmopoulos and A.~Zhiboedov, \emph{{Gravitational effective field
  theory islands, low-spin dominance, and the four-graviton amplitude}},
  \href{http://dx.doi.org/10.1088/1751-8121/ac0e51}{\emph{J. Phys. A}
  {\bfseries 54} (2021) 344002},
  [\href{https://arxiv.org/abs/2103.12728}{{\ttfamily 2103.12728}}].

\bibitem{Adams:2006sv}
A.~Adams, N.~Arkani-Hamed, S.~Dubovsky, A.~Nicolis and R.~Rattazzi,
  \emph{{Causality, analyticity and an IR obstruction to UV completion}},
  \href{http://dx.doi.org/10.1088/1126-6708/2006/10/014}{\emph{JHEP} {\bfseries
  10} (2006) 014}, [\href{https://arxiv.org/abs/hep-th/0602178}{{\ttfamily
  hep-th/0602178}}].

\bibitem{deRham:2022hpx}
C.~de~Rham, S.~Kundu, M.~Reece, A.~J. Tolley and S.-Y. Zhou, \emph{{Snowmass
  White Paper: UV Constraints on IR Physics}},  in \emph{{Snowmass 2021}}, 3,
  2022.
\newblock \href{https://arxiv.org/abs/2203.06805}{{\ttfamily 2203.06805}}.

\bibitem{Palti:2019pca}
E.~Palti, \emph{{The Swampland: Introduction and Review}},
  \href{http://dx.doi.org/10.1002/prop.201900037}{\emph{Fortsch. Phys.}
  {\bfseries 67} (2019) 1900037},
  [\href{https://arxiv.org/abs/1903.06239}{{\ttfamily 1903.06239}}].

\bibitem{Vafa:2005ui}
C.~Vafa, \emph{{The String landscape and the swampland}},
  \href{https://arxiv.org/abs/hep-th/0509212}{{\ttfamily hep-th/0509212}}.

\bibitem{Kruczenski:2022lot}
M.~Kruczenski, J.~Penedones and B.~C. van Rees, \emph{{Snowmass White Paper:
  S-matrix Bootstrap}},  \href{https://arxiv.org/abs/2203.02421}{{\ttfamily
  2203.02421}}.

\bibitem{Eden:1966dnq}
R.~J. Eden, P.~V. Landshoff, D.~I. Olive and J.~C. Polkinghorne, \emph{{The
  analytic S-matrix}}.
\newblock Cambridge Univ. Press, Cambridge, 1966.

\bibitem{Tokuda:2020mlf}
J.~Tokuda, K.~Aoki and S.~Hirano, \emph{{Gravitational positivity bounds}},
  \href{http://dx.doi.org/10.1007/JHEP11(2020)054}{\emph{JHEP} {\bfseries 11}
  (2020) 054}, [\href{https://arxiv.org/abs/2007.15009}{{\ttfamily
  2007.15009}}].

\bibitem{Alberte:2021dnj}
L.~Alberte, C.~de~Rham, S.~Jaitly and A.~J. Tolley, \emph{{Reverse
  Bootstrapping: IR Lessons for UV Physics}},
  \href{http://dx.doi.org/10.1103/PhysRevLett.128.051602}{\emph{Phys. Rev.
  Lett.} {\bfseries 128} (2022) 051602},
  [\href{https://arxiv.org/abs/2111.09226}{{\ttfamily 2111.09226}}].

\bibitem{deRham:2022gfe}
C.~de~Rham, S.~Jaitly and A.~J. Tolley, \emph{{Constraints on Regge behavior
  from IR physics}},
  \href{http://dx.doi.org/10.1103/PhysRevD.108.046011}{\emph{Phys. Rev. D}
  {\bfseries 108} (2023) 046011},
  [\href{https://arxiv.org/abs/2212.04975}{{\ttfamily 2212.04975}}].

\bibitem{Caron-Huot:2021rmr}
S.~Caron-Huot, D.~Mazac, L.~Rastelli and D.~Simmons-Duffin, \emph{{Sharp
  boundaries for the swampland}},
  \href{http://dx.doi.org/10.1007/JHEP07(2021)110}{\emph{JHEP} {\bfseries 07}
  (2021) 110}, [\href{https://arxiv.org/abs/2102.08951}{{\ttfamily
  2102.08951}}].

\bibitem{Bellazzini:2019xts}
B.~Bellazzini, M.~Lewandowski and J.~Serra, \emph{{Positivity of Amplitudes,
  Weak Gravity Conjecture, and Modified Gravity}},
  \href{http://dx.doi.org/10.1103/PhysRevLett.123.251103}{\emph{Phys. Rev.
  Lett.} {\bfseries 123} (2019) 251103},
  [\href{https://arxiv.org/abs/1902.03250}{{\ttfamily 1902.03250}}].

\bibitem{Alberte:2020bdz}
L.~Alberte, C.~de~Rham, S.~Jaitly and A.~J. Tolley, \emph{{QED positivity
  bounds}}, \href{http://dx.doi.org/10.1103/PhysRevD.103.125020}{\emph{Phys.
  Rev. D} {\bfseries 103} (2021) 125020},
  [\href{https://arxiv.org/abs/2012.05798}{{\ttfamily 2012.05798}}].

\bibitem{Alberte:2020jsk}
L.~Alberte, C.~de~Rham, S.~Jaitly and A.~J. Tolley, \emph{{Positivity Bounds
  and the Massless Spin-2 Pole}},
  \href{http://dx.doi.org/10.1103/PhysRevD.102.125023}{\emph{Phys. Rev. D}
  {\bfseries 102} (2020) 125023},
  [\href{https://arxiv.org/abs/2007.12667}{{\ttfamily 2007.12667}}].

\bibitem{Chowdhury:2021ynh}
S.~D. Chowdhury, K.~Ghosh, P.~Haldar, P.~Raman and A.~Sinha, \emph{{Crossing
  Symmetric Spinning S-matrix Bootstrap: EFT bounds}},
  \href{http://dx.doi.org/10.21468/SciPostPhys.13.3.051}{\emph{SciPost Phys.}
  {\bfseries 13} (2022) 051},
  [\href{https://arxiv.org/abs/2112.11755}{{\ttfamily 2112.11755}}].

\bibitem{Caron-Huot:2022ugt}
S.~Caron-Huot, Y.-Z. Li, J.~Parra-Martinez and D.~Simmons-Duffin,
  \emph{{Causality constraints on corrections to Einstein gravity}},
  \href{http://dx.doi.org/10.1007/JHEP05(2023)122}{\emph{JHEP} {\bfseries 05}
  (2023) 122}, [\href{https://arxiv.org/abs/2201.06602}{{\ttfamily
  2201.06602}}].

\bibitem{Caron-Huot:2022jli}
S.~Caron-Huot, Y.-Z. Li, J.~Parra-Martinez and D.~Simmons-Duffin,
  \emph{{Graviton partial waves and causality in higher dimensions}},
  \href{http://dx.doi.org/10.1103/PhysRevD.108.026007}{\emph{Phys. Rev. D}
  {\bfseries 108} (2023) 026007},
  [\href{https://arxiv.org/abs/2205.01495}{{\ttfamily 2205.01495}}].

\bibitem{CarrilloGonzalez:2022fwg}
M.~Carrillo~Gonzalez, C.~de~Rham, V.~Pozsgay and A.~J. Tolley, \emph{{Causal
  effective field theories}},
  \href{http://dx.doi.org/10.1103/PhysRevD.106.105018}{\emph{Phys. Rev. D}
  {\bfseries 106} (2022) 105018},
  [\href{https://arxiv.org/abs/2207.03491}{{\ttfamily 2207.03491}}].

\bibitem{CarrilloGonzalez:2023cbf}
M.~Carrillo~Gonz\'alez, C.~de~Rham, S.~Jaitly, V.~Pozsgay and A.~Tokareva,
  \emph{{Positivity-causality competition: a road to ultimate EFT consistency
  constraints}},  \href{https://arxiv.org/abs/2307.04784}{{\ttfamily
  2307.04784}}.

\bibitem{deRham:2019ctd}
C.~de~Rham and A.~J. Tolley, \emph{{Speed of gravity}},
  \href{http://dx.doi.org/10.1103/PhysRevD.101.063518}{\emph{Phys. Rev. D}
  {\bfseries 101} (2020) 063518},
  [\href{https://arxiv.org/abs/1909.00881}{{\ttfamily 1909.00881}}].

\bibitem{deRham:2020zyh}
C.~de~Rham and A.~J. Tolley, \emph{{Causality in curved spacetimes: The speed
  of light and gravity}},
  \href{http://dx.doi.org/10.1103/PhysRevD.102.084048}{\emph{Phys. Rev. D}
  {\bfseries 102} (2020) 084048},
  [\href{https://arxiv.org/abs/2007.01847}{{\ttfamily 2007.01847}}].

\bibitem{Chen:2021bvg}
C.~Y.~R. Chen, C.~de~Rham, A.~Margalit and A.~J. Tolley, \emph{{A cautionary
  case of casual causality}},
  \href{http://dx.doi.org/10.1007/JHEP03(2022)025}{\emph{JHEP} {\bfseries 03}
  (2022) 025}, [\href{https://arxiv.org/abs/2112.05031}{{\ttfamily
  2112.05031}}].

\bibitem{deRham:2021bll}
C.~de~Rham, A.~J. Tolley and J.~Zhang, \emph{{Causality Constraints on
  Gravitational Effective Field Theories}},
  \href{http://dx.doi.org/10.1103/PhysRevLett.128.131102}{\emph{Phys. Rev.
  Lett.} {\bfseries 128} (2022) 131102},
  [\href{https://arxiv.org/abs/2112.05054}{{\ttfamily 2112.05054}}].

\bibitem{Chen:2023rar}
C.~Y.~R. Chen, C.~de~Rham, A.~Margalit and A.~J. Tolley, \emph{{Surfin'
  pp-waves with Good Vibrations: Causality in the presence of stacked
  shockwaves}},  \href{https://arxiv.org/abs/2309.04534}{{\ttfamily
  2309.04534}}.

\bibitem{Camanho:2014apa}
X.~O. Camanho, J.~D. Edelstein, J.~Maldacena and A.~Zhiboedov, \emph{{Causality
  Constraints on Corrections to the Graviton Three-Point Coupling}},
  \href{http://dx.doi.org/10.1007/JHEP02(2016)020}{\emph{JHEP} {\bfseries 02}
  (2016) 020}, [\href{https://arxiv.org/abs/1407.5597}{{\ttfamily 1407.5597}}].

\bibitem{Bellazzini:2021shn}
B.~Bellazzini, G.~Isabella, M.~Lewandowski and F.~Sgarlata,
  \emph{{Gravitational causality and the self-stress of photons}},
  \href{http://dx.doi.org/10.1007/JHEP05(2022)154}{\emph{JHEP} {\bfseries 05}
  (2022) 154}, [\href{https://arxiv.org/abs/2108.05896}{{\ttfamily
  2108.05896}}].

\bibitem{Bellazzini:2015cra}
B.~Bellazzini, C.~Cheung and G.~N. Remmen, \emph{{Quantum Gravity Constraints
  from Unitarity and Analyticity}},
  \href{http://dx.doi.org/10.1103/PhysRevD.93.064076}{\emph{Phys. Rev. D}
  {\bfseries 93} (2016) 064076},
  [\href{https://arxiv.org/abs/1509.00851}{{\ttfamily 1509.00851}}].

\bibitem{Haring:2023zwu}
K.~H\"aring and A.~Zhiboedov, \emph{{The Stringy S-matrix Bootstrap: Maximal
  Spin and Superpolynomial Softness}},
  \href{https://arxiv.org/abs/2311.13631}{{\ttfamily 2311.13631}}.

\bibitem{Serra:2022pzl}
F.~Serra, J.~Serra, E.~Trincherini and L.~G. Trombetta, \emph{{Causality
  constraints on black holes beyond GR}},
  \href{http://dx.doi.org/10.1007/JHEP08(2022)157}{\emph{JHEP} {\bfseries 08}
  (2022) 157}, [\href{https://arxiv.org/abs/2205.08551}{{\ttfamily
  2205.08551}}].

\bibitem{Goon:2016ihr}
G.~Goon, K.~Hinterbichler, A.~Joyce and M.~Trodden, \emph{{Aspects of Galileon
  Non-Renormalization}},
  \href{http://dx.doi.org/10.1007/JHEP11(2016)100}{\emph{JHEP} {\bfseries 11}
  (2016) 100}, [\href{https://arxiv.org/abs/1606.02295}{{\ttfamily
  1606.02295}}].

\bibitem{Ruhdorfer:2019qmk}
M.~Ruhdorfer, J.~Serra and A.~Weiler, \emph{{Effective Field Theory of Gravity
  to All Orders}}, \href{http://dx.doi.org/10.1007/JHEP05(2020)083}{\emph{JHEP}
  {\bfseries 05} (2020) 083},
  [\href{https://arxiv.org/abs/1908.08050}{{\ttfamily 1908.08050}}].

\bibitem{Li:2023wdz}
H.-L. Li, Z.~Ren, M.-L. Xiao, J.-H. Yu and Y.-H. Zheng, \emph{{On-shell
  operator construction in the effective field theory of gravity}},
  \href{http://dx.doi.org/10.1007/JHEP10(2023)019}{\emph{JHEP} {\bfseries 10}
  (2023) 019}, [\href{https://arxiv.org/abs/2305.10481}{{\ttfamily
  2305.10481}}].

\bibitem{Regge:1957td}
T.~Regge and J.~A. Wheeler, \emph{{Stability of a Schwarzschild singularity}},
  \href{http://dx.doi.org/10.1103/PhysRev.108.1063}{\emph{Phys. Rev.}
  {\bfseries 108} (1957) 1063--1069}.

\bibitem{Zerilli:1970se}
F.~J. Zerilli, \emph{{Effective potential for even parity Regge-Wheeler
  gravitational perturbation equations}},
  \href{http://dx.doi.org/10.1103/PhysRevLett.24.737}{\emph{Phys. Rev. Lett.}
  {\bfseries 24} (1970) 737--738}.

\bibitem{Knorr:2023usb}
B.~Knorr, \emph{{Momentum-dependent field redefinitions in Asymptotic Safety}},
   \href{https://arxiv.org/abs/2311.12097}{{\ttfamily 2311.12097}}.

\bibitem{Bueno:2023jtc}
P.~Bueno, P.~A. Cano and R.~A. Hennigar, \emph{{On the stability of Einsteinian
  Cubic Gravity black holes in EFT}},
  \href{https://arxiv.org/abs/2306.02924}{{\ttfamily 2306.02924}}.

\bibitem{Kimura:2020mrh}
M.~Kimura, \emph{{Note on the parametrized black hole quasinormal ringdown
  formalism}}, \href{http://dx.doi.org/10.1103/PhysRevD.101.064031}{\emph{Phys.
  Rev. D} {\bfseries 101} (2020) 064031},
  [\href{https://arxiv.org/abs/2001.09613}{{\ttfamily 2001.09613}}].

\bibitem{Wigner:1955zz}
E.~P. Wigner, \emph{{Lower Limit for the Energy Derivative of the Scattering
  Phase Shift}}, \href{http://dx.doi.org/10.1103/PhysRev.98.145}{\emph{Phys.
  Rev.} {\bfseries 98} (1955) 145--147}.

\bibitem{DECARVALHO200283}
C.~{de Carvalho} and H.~Nussenzveig, \emph{Time delay},
  \href{http://dx.doi.org/https://doi.org/10.1016/S0370-1573(01)00092-8}{\emph{Physics
  Reports} {\bfseries 364} (2002) 83--174}.

\bibitem{Berti:2009kk}
E.~Berti, V.~Cardoso and A.~O. Starinets, \emph{{Quasinormal modes of black
  holes and black branes}},
  \href{http://dx.doi.org/10.1088/0264-9381/26/16/163001}{\emph{Class. Quant.
  Grav.} {\bfseries 26} (2009) 163001},
  [\href{https://arxiv.org/abs/0905.2975}{{\ttfamily 0905.2975}}].

\bibitem{Clarkson:2002jz}
C.~A. Clarkson and R.~K. Barrett, \emph{{Covariant perturbations of
  Schwarzschild black holes}},
  \href{http://dx.doi.org/10.1088/0264-9381/20/18/301}{\emph{Class. Quant.
  Grav.} {\bfseries 20} (2003) 3855--3884},
  [\href{https://arxiv.org/abs/gr-qc/0209051}{{\ttfamily gr-qc/0209051}}].

\bibitem{Iyer:1986np}
S.~Iyer and C.~M. Will, \emph{{Black Hole Normal Modes: A {WKB} Approach. 1.
  Foundations and Application of a Higher Order {WKB} Analysis of Potential
  Barrier Scattering}},
  \href{http://dx.doi.org/10.1103/PhysRevD.35.3621}{\emph{Phys. Rev. D}
  {\bfseries 35} (1987) 3621}.

\bibitem{Guinn:1989bn}
J.~W. Guinn, C.~M. Will, Y.~Kojima and B.~F. Schutz, \emph{{High Overtone
  Normal Modes of Schwarzschild Black Holes}},
  \href{http://dx.doi.org/10.1088/0264-9381/7/2/006}{\emph{Class. Quant. Grav.}
  {\bfseries 7} (1990) L47}.

\bibitem{Konoplya:2003ii}
R.~A. Konoplya, \emph{{Quasinormal behavior of the d-dimensional Schwarzschild
  black hole and higher order WKB approach}},
  \href{http://dx.doi.org/10.1103/PhysRevD.68.024018}{\emph{Phys. Rev. D}
  {\bfseries 68} (2003) 024018},
  [\href{https://arxiv.org/abs/gr-qc/0303052}{{\ttfamily gr-qc/0303052}}].

\bibitem{Konoplya:2004ip}
R.~A. Konoplya, \emph{{Quasinormal modes of the Schwarzschild black hole and
  higher order WKB approach}}, {\emph{J. Phys. Stud.} {\bfseries 8} (2004)
  93--100}.

\bibitem{Matyjasek:2017psv}
J.~Matyjasek and M.~Opala, \emph{{Quasinormal modes of black holes. The
  improved semianalytic approach}},
  \href{http://dx.doi.org/10.1103/PhysRevD.96.024011}{\emph{Phys. Rev. D}
  {\bfseries 96} (2017) 024011},
  [\href{https://arxiv.org/abs/1704.00361}{{\ttfamily 1704.00361}}].

\bibitem{Konoplya:2019hlu}
R.~A. Konoplya, A.~Zhidenko and A.~F. Zinhailo, \emph{{Higher order WKB formula
  for quasinormal modes and grey-body factors: recipes for quick and accurate
  calculations}},
  \href{http://dx.doi.org/10.1088/1361-6382/ab2e25}{\emph{Class. Quant. Grav.}
  {\bfseries 36} (2019) 155002},
  [\href{https://arxiv.org/abs/1904.10333}{{\ttfamily 1904.10333}}].

\bibitem{Emond:2019crr}
W.~T. Emond and N.~Moynihan, \emph{{Scattering Amplitudes, Black Holes and
  Leading Singularities in Cubic Theories of Gravity}},
  \href{http://dx.doi.org/10.1007/JHEP12(2019)019}{\emph{JHEP} {\bfseries 12}
  (2019) 019}, [\href{https://arxiv.org/abs/1905.08213}{{\ttfamily
  1905.08213}}].

\bibitem{AccettulliHuber:2020oou}
M.~Accettulli~Huber, A.~Brandhuber, S.~De~Angelis and G.~Travaglini,
  \emph{{Eikonal phase matrix, deflection angle and time delay in effective
  field theories of gravity}},
  \href{http://dx.doi.org/10.1103/PhysRevD.102.046014}{\emph{Phys. Rev. D}
  {\bfseries 102} (2020) 046014},
  [\href{https://arxiv.org/abs/2006.02375}{{\ttfamily 2006.02375}}].

\bibitem{Aguilar-Gutierrez:2023kfn}
S.~E. Aguilar-Gutierrez, P.~Bueno, P.~A. Cano, R.~A. Hennigar and Q.~Llorens,
  \emph{{Aspects of higher-curvature gravities with covariant derivatives}},
  \href{http://dx.doi.org/10.1103/PhysRevD.108.124075}{\emph{Phys. Rev. D}
  {\bfseries 108} (2023) 124075},
  [\href{https://arxiv.org/abs/2310.09333}{{\ttfamily 2310.09333}}].

\bibitem{Chandrasekaran:2018qmx}
V.~Chandrasekaran, G.~N. Remmen and A.~Shahbazi-Moghaddam, \emph{{Higher-Point
  Positivity}}, \href{http://dx.doi.org/10.1007/JHEP11(2018)015}{\emph{JHEP}
  {\bfseries 11} (2018) 015},
  [\href{https://arxiv.org/abs/1804.03153}{{\ttfamily 1804.03153}}].

\bibitem{Melville:2022ykg}
S.~Melville and J.~Noller, \emph{{Positivity bounds from multiple vacua and
  their cosmological consequences}},
  \href{http://dx.doi.org/10.1088/1475-7516/2022/06/031}{\emph{JCAP} {\bfseries
  06} (2022) 031}, [\href{https://arxiv.org/abs/2202.01222}{{\ttfamily
  2202.01222}}].

\bibitem{Davis:2021oce}
A.-C. Davis and S.~Melville, \emph{{Scalar fields near compact objects:
  resummation versus UV completion}},
  \href{http://dx.doi.org/10.1088/1475-7516/2021/11/012}{\emph{JCAP} {\bfseries
  11} (2021) 012}, [\href{https://arxiv.org/abs/2107.00010}{{\ttfamily
  2107.00010}}].

\bibitem{Serra:2023nrn}
F.~Serra and L.~G. Trombetta, \emph{{Five-point Superluminality Bounds}},
  \href{https://arxiv.org/abs/2312.06759}{{\ttfamily 2312.06759}}.

\bibitem{Cano:2019ore}
P.~A. Cano and A.~Ruip\'erez, \emph{{Leading higher-derivative corrections to
  Kerr geometry}}, \href{http://dx.doi.org/10.1007/JHEP05(2019)189}{\emph{JHEP}
  {\bfseries 05} (2019) 189},
  [\href{https://arxiv.org/abs/1901.01315}{{\ttfamily 1901.01315}}].

\bibitem{Melville:2019wyy}
S.~Melville and J.~Noller, \emph{{Positivity in the Sky: Constraining dark
  energy and modified gravity from the UV}},
  \href{http://dx.doi.org/10.1103/PhysRevD.101.021502}{\emph{Phys. Rev. D}
  {\bfseries 101} (2020) 021502},
  [\href{https://arxiv.org/abs/1904.05874}{{\ttfamily 1904.05874}}].

\bibitem{Grall:2021xxm}
T.~Grall and S.~Melville, \emph{{Positivity bounds without boosts: New
  constraints on low energy effective field theories from the UV}},
  \href{http://dx.doi.org/10.1103/PhysRevD.105.L121301}{\emph{Phys. Rev. D}
  {\bfseries 105} (2022) L121301},
  [\href{https://arxiv.org/abs/2102.05683}{{\ttfamily 2102.05683}}].

\bibitem{deRham:2021fpu}
C.~de~Rham, S.~Melville and J.~Noller, \emph{{Positivity bounds on dark energy:
  when matter matters}},
  \href{http://dx.doi.org/10.1088/1475-7516/2021/08/018}{\emph{JCAP} {\bfseries
  08} (2021) 018}, [\href{https://arxiv.org/abs/2103.06855}{{\ttfamily
  2103.06855}}].

\bibitem{Salcedo:2022aal}
S.~A. Salcedo, M.~H.~G. Lee, S.~Melville and E.~Pajer, \emph{{The Analytic
  Wavefunction}}, \href{http://dx.doi.org/10.1007/JHEP06(2023)020}{\emph{JHEP}
  {\bfseries 06} (2023) 020},
  [\href{https://arxiv.org/abs/2212.08009}{{\ttfamily 2212.08009}}].

\bibitem{Agui-Salcedo:2023wlq}
S.~Agui-Salcedo and S.~Melville, \emph{{The Cosmological Tree Theorem}},
  \href{https://arxiv.org/abs/2308.00680}{{\ttfamily 2308.00680}}.

\bibitem{CarrilloGonzalez:2023rmc}
M.~Carrillo~Gonz\'alez, \emph{{Bounds on EFT's in an expanding Universe}},
  \href{https://arxiv.org/abs/2312.07651}{{\ttfamily 2312.07651}}.

\bibitem{Glampedakis:2019dqh}
K.~Glampedakis and H.~O. Silva, \emph{{Eikonal quasinormal modes of black holes
  beyond General Relativity}},
  \href{http://dx.doi.org/10.1103/PhysRevD.100.044040}{\emph{Phys. Rev. D}
  {\bfseries 100} (2019) 044040},
  [\href{https://arxiv.org/abs/1906.05455}{{\ttfamily 1906.05455}}].

\bibitem{Silva:2019scu}
H.~O. Silva and K.~Glampedakis, \emph{{Eikonal quasinormal modes of black holes
  beyond general relativity. II. Generalized scalar-tensor perturbations}},
  \href{http://dx.doi.org/10.1103/PhysRevD.101.044051}{\emph{Phys. Rev. D}
  {\bfseries 101} (2020) 044051},
  [\href{https://arxiv.org/abs/1912.09286}{{\ttfamily 1912.09286}}].

\bibitem{Callister:2023tws}
T.~Callister, L.~Jenks, D.~Holz and N.~Yunes, \emph{{A New Probe of
  Gravitational Parity Violation Through (Non-)Observation of the Stochastic
  Gravitational-Wave Background}},
  \href{https://arxiv.org/abs/2312.12532}{{\ttfamily 2312.12532}}.

\bibitem{Langlois:2021aji}
D.~Langlois, K.~Noui and H.~Roussille, \emph{{Black hole perturbations in
  modified gravity}},
  \href{http://dx.doi.org/10.1103/PhysRevD.104.124044}{\emph{Phys. Rev. D}
  {\bfseries 104} (2021) 124044},
  [\href{https://arxiv.org/abs/2103.14750}{{\ttfamily 2103.14750}}].

\bibitem{Roussille:2023sdr}
H.~Roussille, D.~Langlois and K.~Noui, \emph{{Numerical computation of
  quasinormal modes in the first-order approach to black hole perturbations in
  modified gravity}},  \href{https://arxiv.org/abs/2312.11986}{{\ttfamily
  2312.11986}}.

\bibitem{Hamada:2018dde}
Y.~Hamada, T.~Noumi and G.~Shiu, \emph{{Weak Gravity Conjecture from Unitarity
  and Causality}},
  \href{http://dx.doi.org/10.1103/PhysRevLett.123.051601}{\emph{Phys. Rev.
  Lett.} {\bfseries 123} (2019) 051601},
  [\href{https://arxiv.org/abs/1810.03637}{{\ttfamily 1810.03637}}].

\bibitem{Arkani-Hamed:2021ajd}
N.~Arkani-Hamed, Y.-t. Huang, J.-Y. Liu and G.~N. Remmen, \emph{{Causality,
  unitarity, and the weak gravity conjecture}},
  \href{http://dx.doi.org/10.1007/JHEP03(2022)083}{\emph{JHEP} {\bfseries 03}
  (2022) 083}, [\href{https://arxiv.org/abs/2109.13937}{{\ttfamily
  2109.13937}}].

\bibitem{DeLuca:2022tkm}
V.~De~Luca, J.~Khoury and S.~S.~C. Wong, \emph{{Implications of the weak
  gravity conjecture for tidal Love numbers of black holes}},
  \href{http://dx.doi.org/10.1103/PhysRevD.108.044066}{\emph{Phys. Rev. D}
  {\bfseries 108} (2023) 044066},
  [\href{https://arxiv.org/abs/2211.14325}{{\ttfamily 2211.14325}}].

\bibitem{Bellazzini:2022wzv}
B.~Bellazzini, G.~Isabella and M.~M. Riva, \emph{{Classical vs quantum eikonal
  scattering and its causal structure}},
  \href{http://dx.doi.org/10.1007/JHEP04(2023)023}{\emph{JHEP} {\bfseries 04}
  (2023) 023}, [\href{https://arxiv.org/abs/2211.00085}{{\ttfamily
  2211.00085}}].

\bibitem{Schutz:1985km}
B.~F. Schutz and C.~M. Will, \emph{{BLACK HOLE NORMAL MODES: A SEMIANALYTIC
  APPROACH}}, \href{http://dx.doi.org/10.1086/184453}{\emph{Astrophys. J.
  Lett.} {\bfseries 291} (1985) L33--L36}.

\bibitem{Hatsuda:2019eoj}
Y.~Hatsuda, \emph{{Quasinormal modes of black holes and Borel summation}},
  \href{http://dx.doi.org/10.1103/PhysRevD.101.024008}{\emph{Phys. Rev. D}
  {\bfseries 101} (2020) 024008},
  [\href{https://arxiv.org/abs/1906.07232}{{\ttfamily 1906.07232}}].

\end{thebibliography}\endgroup

\end{document}